\documentclass[12pt]{article}
\pdfoutput=1
\usepackage{color}
\usepackage{graphicx}

\input xy
\xyoption{all}
\xyoption{web}

\newlength{\abstractwidth}

\renewcommand{\thefootnote}{\fnsymbol{footnote}}
\renewcommand{\thanks}[1]{\footnote{#1}}
\newcommand{\starttext}{
\setcounter{footnote}{0}
\renewcommand{\thefootnote}{\arabic{footnote}}}

\newcommand{\bea}{\begin{eqnarray}}
\newcommand{\eea}{\end{eqnarray}}
\newcommand{\ee}{\end{equation}}
\newcommand{\be}{\begin{equation}}

\newcommand{\ea}{\end{array}}
\newcommand{\bac}{\begin{array}{c}}
\newcommand{\bacc}{\begin{array}{cc}}
\newcommand{\barcl}{\begin{array}{r@{}c@{}l}}
\newcommand{\brcl}{\begin{array}{rcl}}
\newcommand{\bdm}{\begin{displaymath}}
\newcommand{\edm}{\end{displaymath}}

\newcommand{\half}{\frac{1}{2}}

\def\cA{{\cal A}}
\def\cB{{\cal B}}
\def\cC{{\cal C}}

\def\cE{{\cal E}}

\def\cI{{\cal I}}

\def\cK{{\cal K}}

\def\cM{{\cal M}}
\def\cN{{\cal N}}

\def\cS{{\cal S}}

\def\bC{{\bf C}}

\def\bR{{\bf R}}

\def\Re{{\rm Re}}
\def\Im{{\rm Im}}

\def\half{ {1\over 2}}
\def\p{\partial}

\def\a{\alpha}
\def\b{\beta}

\def\ep{\varepsilon}

\def\g{\gamma}
\def\l{\lambda}
\def\L{\Lambda}
\def\o{\omega}

\def\G{\Gamma}

\def\g{\gamma}
\def\s{\sigma}
\def\t{\tau}

\def\no{\nonumber}
\def\sm{\smallskip}
\def\ti{\tilde}

\begin{document}
\starttext
\setcounter{footnote}{0}

\begin{flushright}
IGC-11/7-1
\end{flushright}

\bigskip

\begin{center}

{\Large \bf Exact half-BPS string-junction solutions} 

\vskip 0.15in

{\Large \bf in six-dimensional supergravity\footnote{This work is supported in part by NSF grants
PHY-08-55356 and PHY-07-57702.}}

\vskip 0.6in

{ \bf Marco Chiodaroli$^a$, Eric D'Hoker$^b$, Yu Guo$^b$,  Michael Gutperle$^b$}

\vskip .2in

 ${}^a$ {\sl Institute for Gravitation and the Cosmos,}\\
{\sl The Pennsylvania State University, University Park, PA 16802, USA} \\
{\tt \small mchiodar@gravity.psu.edu}

\vskip 0.2in

 ${}^b$ { \sl Department of Physics and Astronomy }\\
{\sl University of California, Los Angeles, CA 90095, USA}\\
{\tt \small dhoker, guoyu, gutperle@physics.ucla.edu; }

\end{center}

\vskip 0.2in

\begin{abstract}

We construct $SO(2,1) \times SO(3)$-invariant half-BPS solutions in six-dimensional 
(0,4) supergravity with $m$ tensor multiplets. The space-time manifold of each one of 
these solutions consists of an  $AdS_2 \times S^2$ warped over a Riemann 
surface $\Sigma$ with boundary. The most general local solution is parametrized by one 
real harmonic function, and $m+2$ holomorphic functions which are subject to a quadratic 
constraint and a hermitian inequality, both of which are manifestly $SO(2,m)$ invariant.
Imposing suitable conditions on these harmonic and holomorphic functions, we construct 
globally regular supergravity solutions with $N$ distinct  $AdS_3 \times S^3$ asymptotic 
regions and contractible $\Sigma$. These solutions have an intricate moduli space, whose 
dimension equals $2(m+1)N -m-2$ and matches the counting of three-form charge vectors 
and un-attracted scalars of the tensor multiplet. Exact explicit formulas for all supergravity
fields are obtained in terms of the moduli.

\sm

Our solutions give the near-horizon geometries for junctions of $N$ self-dual strings in six 
dimensions, and are holographic duals to CFTs defined on $N$ half-planes which share
a common interface line. For $m=5$, the solutions lift to quarter-BPS solutions for 
six-dimensional (4,4) supergravity.

\end{abstract}

\newpage

\tableofcontents

\baselineskip=16pt
\setcounter{equation}{0}
\setcounter{footnote}{0}

\newpage

\section{Introduction}
\setcounter{equation}{0}
\label{sec1}

Over the past few years, considerable progress has been made on the  construction of exact
half-BPS classical solutions to Type IIB and M-theory supergravity.  Research motivated by studies
in gauge/gravity duality \cite{Maldacena:1997re,Gubser:1998bc,Witten:1998qj}
(for reviews, see e.g. \cite{Aharony:1999ti,D'Hoker:2002aw}) has focused on solutions whose 
space-time manifold contains a (warped) factor of an $AdS$ space, and on so-called {\sl bubbling solutions}. 
By construction, all these solutions have non-trivial isometries, and an asymptotic symmetry  
which coincides with the asymptotic symmetry of the maximally supersymmetric solution. 

\sm

A general classification of such half-BPS solutions for both Type IIB and M-theory supergravity
in terms of conformal superalgebras and their subalgebras was proposed in \cite{D'Hoker:2008ix}. 
Many of the resulting cases admit immediate interpretations in terms of intersecting branes,
and are holographic duals to gauge theory states with defects of various dimensions. 
For example, Type IIB solutions which are holographic duals to states of $\cN=4$ super-Yang-Mills (SYM) theory in 
$3+1$-dimensional space-time include the following: (1) the { bubbling solutions} of  \cite{Lin:2004nb}, which 
are invariant under the superalgebra $PSU(2|2)^2$ and dual to local half-BPS operators; (2) the solutions 
of  \cite{D'Hoker:2007fq}, which are invariant under the superalgebra $OSp(4^*|4)$ and dual to straight 
Wilson lines (see also \cite{Yamaguchi:2006te,Gomis:2006sb,Lunin:2006xr}); (3) the half-BPS Janus 
solutions of \cite{D'Hoker:2007xy,D'Hoker:2007xz}, which are invariant under $OSp(4|4,\bR)$, 
are dual to a planar $2+1$-dimensional interface or defect (see also 
\cite{Bachas:2001vj,Gauntlett:2005ww,Clark:2005te,D'Hoker:2006uv,Gomis:2006cu}), 
and provide supersymmetric generalizations to the original  Janus solution of \cite{Bak:2003jk}.
In M-theory, bubbling solutions were derived in \cite{Lin:2004nb}, while a wealth of 
half-BPS solutions with $AdS$ factor is available in 
\cite{Lunin:2007ab,D'Hoker:2008wc,D'Hoker:2008qm,D'Hoker:2009gg,D'Hoker:2009my}. 
 
\sm

Supersymmetry provides a powerful tool for the  construction of explicit solutions, as second order 
field equations may be replaced by first order BPS equations. 
In the case of defect solutions with 16 supersymmetries,
the BPS equations in Type IIB and M-theory suffice to solve the full Bianchi and  field equations.
The space-time structure of the resulting half-BPS solutions to Type IIB and M-theory is of the form, 
\bea
\label{1a1}
AdS_{p_0} \times S^{p_1} \times S^{p_2} \times \Sigma
\eea
where  $\Sigma$ is a Riemann surface with boundary over which the product
$AdS_{p_0} \times S^{p_1} \times S^{p_2}$  is warped, while the total dimension $p_0+p_1+p_2+2$ 
equals  10 for Type IIB and 11 for M-theory. All these solutions have co-homogeneity 2, following 
the definition of \cite{besse}. 
Remarkably, although various  integrable systems on $\Sigma$ arise during the course of the 
analysis of the BPS equations \cite{D'Hoker:2008ap}, the final solutions are invariably expressible 
in terms of holomorphic functions  on $\Sigma$, parametrized by strata of
finite-dimensional moduli spaces.

\sm

Supersymmetry can still be a powerful tool with fewer supersymmetries. 
One class of examples\footnote{Other examples are bubbling $AdS_3$  solutions, see e.g.  \cite{Martelli:2004xq,Liu:2004hy,Gauntlett:2006ns,Boni:2005sf}.} are  Janus-like 
solutions with 8 supersymmetries arise by considering Type IIB supergravity on  the following
space-times involving a $K3$ surface, 
\bea
\label{1a2}
AdS_2 \times S^2 \times K3 \times \Sigma
\eea
where the product $AdS_2 \times S^2 \times K3$ is warped over the Riemann surface $\Sigma$
with boundary. The solutions of \cite{Chiodaroli:2009yw,Chiodaroli:2009xh} (see also
\cite{Kumar:2002wc,Kumar:2003xi,Kumar:2004me,Lunin:2008tf,Chiodaroli:2010mv})
are locally asymptotic to the $AdS_3 \times S^3 \times K3$ vacuum solution, which 
has 16 supersymmetries. Although less supersymmetric than their counterparts of (\ref{1a1}),
these solutions are still expressible in terms of holomorphic 
functions  on $\Sigma$ and parametrized by interesting moduli spaces.

\sm

However, the solutions of \cite{Chiodaroli:2009yw,Chiodaroli:2009xh} did not turn on 
the moduli and fluxes of the $K3$ factor. In principle this can be done using a ten-dimensional Ansatz, but
it will be simpler to consider this problem directly within the framework of the 
six-dimensional supergravity obtained by the compactification of Type IIB on $K3$. 
The moduli of the $K3$ manifold arise as scalar fields which, together with the dilaton and axion, 
live in a coset manifold. 

\sm

Six-dimensional supergravity admits dyonic string solutions preserving four supersymmetries 
which are analogues to $(p,q)$ strings in ten dimensions. In \cite{Dasgupta:1997pu,Sen:1997xi} 
it was shown that in ten dimensions there exist
BPS configurations where three $(p,q)$ strings join at a point, and that such configurations can
be combined to construct networks of strings.

\sm

A similar analysis was carried out in six dimensions in \cite{Chiodaroli:2010mv}, 
finding that there exist supersymmetric junctions with an arbitrary number of dyonic strings, 
provided that the angles between the strings are adjusted according to their charges. 
Naturally, the solutions of \cite{Chiodaroli:2009yw,Chiodaroli:2009xh} were interpreted as
 the smooth near-horizon geometries produced by the back-reaction of these junctions of dyonic strings.

\sm

In this paper, we shall construct exactly the general  solution with 8 supersymmetries
to the six-dimensional chiral $(0,4)$ - or "Type 4b" - supergravity theory which has 16 supersymmetries.
The $(0,4)$ theory is in many respects the six-dimensional analogue of ten-dimensional Type IIB supergravity.
The $(0,4)$ theory was constructed by Romans \cite{Romans:1986er} 
(see also \cite{Deger:1998nm,Avramis:2006nb}), and
consists of a supergravity multiplet together with $m$ tensor supermultiplets.
Classically, all values of $m$ are allowed, but anomaly cancellation restricts $m$ to the values  5 or 21.
The scalars live in the  $SO(5,m)/(SO(5)\times SO(m))$ coset manifold. 
For $m=5$,  the $(0,4)$ theory may be obtained as a chiral truncation of the 
$(4,4)$ six-dimensional supergravity 
theory with 32 supersymmetries, which was constructed in \cite{Tanii:1984zk}.
The $m=21$ supergravity is given by Kaluza-Klein compactification of Type IIB on $K3$,
but it appears unknown, at present, whether this truncation is consistent or not.\footnote{We 
thank the referee for stressing this point.}

\sm

The $(0,4)$ supergravity admits a vacuum solution with space-time manifold $AdS_{3}\times S^{3}$
and the maximal number of 16 supersymmetries. Its isometries are the $SO(2,2)\times SO(4)$ 
maximal bosonic subalgebra of the full $PSU(1,1|2)^2$ symmetry superalgebra. The vacuum solution is 
analogous to the $AdS_5 \times S^5$ solution to Type IIB with 32 supersymmetries. The space-time 
manifold of six-dimensional half-BPS Janus solutions is of the form, 
\bea
\label{1a3}
AdS_2 \times S^2 \times \Sigma
\eea
where the product $AdS_2 \times S^2$ is warped over the Riemann surface $\Sigma$.
The solutions are invariant under the isometry  $SO(2,1)\times SO(3)$ which is the maximal
bosonic subalgebra of the full $PSU(1,1|2)$ symmetry superalgebra of the six-dimensional Janus solutions.
Contrarily to the half-BPS Janus slicing of $AdS_5 \times S^5$ in Type IIB, 
the warped space-time of (\ref{1a3}) 
provides an essentially unique $AdS$-slicing of the vacuum solution  $AdS_{3}\times S^{3}$.
Asymptotically $AdS_3 \times S^3$ bubbling solutions with geometry 
$S^1 \times S^1 \times {\rm \bf R} \times \Sigma_3$
where $\Sigma_3$ is a three-dimensional base, were obtained in \cite{Martelli:2004xq,Lunin:2008tf,Liu:2004hy}.  
Together with these bubbling solutions, the
$AdS_3 \times S^3$ Janus solutions account for the most general 
presently known class of half-BPS solutions with $AdS_3 \times S^3$ asymptotics in six-dimensional supergravity.

\subsection{Summary of results and organization}  
 
Since this paper is relatively long and technical, we shall present here a brief summary of our results. 
There are two parts to the solution; one local, the other global.

\sm

First,  we shall derive a complete local solution, characterized by one real harmonic function and $m+2$ 
holomorphic differential one-forms on the Riemann surface $\Sigma$.  Such a local solution solves 
the BPS conditions, the Bianchi identities, and the field equations point-wise on $\Sigma$. The local 
solution will be obtained in several steps. In Section \ref{sec3} the BPS equations will be reduced 
using the $SO(2,1) \times SO(3)$-invariant Ansatz for the metric, 
\bea
\label{1b1}
ds^{2}= f_{1}^{2 } ds^{2}_{AdS_{2}} + f^{2}_{2}ds^{2}_{S^{2}} +ds^2_{\Sigma}
\eea
as well as corresponding Ans\"atze for the scalar and anti-symmetric tensor  fields. 
Solving the BPS equations in Section \ref{sec4}, we shall find that the scalar fields are restricted to live 
on a $SO(2,m)/(SO(2)\times SO(m))$ sub-manifold of the full $SO(5,m)/(SO(5)\times SO(m))$ 
scalar coset, and that the anti-symmetric tensor fields are restricted accordingly. However,
the solutions to the BPS equations do not automatically solve the Bianchi and field equations.
In Section~\ref{sec5}, the local solutions to the BPS equations are combined with the Bianchi
identities, and used to produce solutions to the full set of Bianchi and field equations. These
solutions are parametrized by a single real  harmonic function $H$, and $m+2$ holomorphic one-forms
$\Lambda ^A$ with $A=1, \cdots, m+5$, and $\L^3=\L^4=\L^5=0$, subject to the relations,\footnote{Throughout,
we shall use the $SO(5,m)$-invariant metric $\eta = {\rm diag} (I_5, - I_m)$ to raise and lower indices
$K^A = \eta ^{AB} K_B$,  for $\bC^{5,m}$ vectors $K$, and to form  their inner product  defined by
$K \cdot L \equiv \eta  _{AB} K^A L^B$.}
\bea
\label{1b2}
\Lambda \cdot \Lambda  & = & 2 (\partial H)^2
\no \\
\Lambda \cdot \bar \Lambda & \geq & 2 |\p H|^2
\eea
where $ \eta$ is the $SO(5,m)$-invariant metric. The expressions for the metric factors
$f_1$, $f_2$, and $ds^2_\Sigma$, as well as for the scalars, anti-symmetric tensor fields
and associated charges of the solution may be found in Section \ref{sec54}.
  
\sm
  
Second, to obtain globally well-defined and regular solutions, which are locally asymptotically 
$AdS_3 \times S^3$, the harmonic function $H$ and the holomorphic one-forms $\Lambda^A$ 
must be subject to certain topology and regularity conditions, which will be spelled out in Section 
\ref{sec6}. A first topological condition requires the boundary of the Riemann surface $\Sigma$ 
to correspond to a vanishing volume for the $S^2$ fiber, so that $f_2=0$ on $\p \Sigma$, and $f_2 \not=0 $ in the 
interior of $\Sigma$. For simplicity in the present paper, we shall assume that $\p \Sigma$
consists of a single connected component, though we know from \cite{Chiodaroli:2009xh}
that this assumption may be easily relaxed, and that $\p \Sigma$ can support multiple connected 
boundary components. A second topological condition requires that the space-time manifold
will have $N$ of asymptotic regions where the geometry is locally asymptotically $AdS_{3}\times S^{3}$. 
Finally, the solution should otherwise be smooth.

\begin{figure}[ht]
\begin{centering}
\includegraphics[scale=0.35]{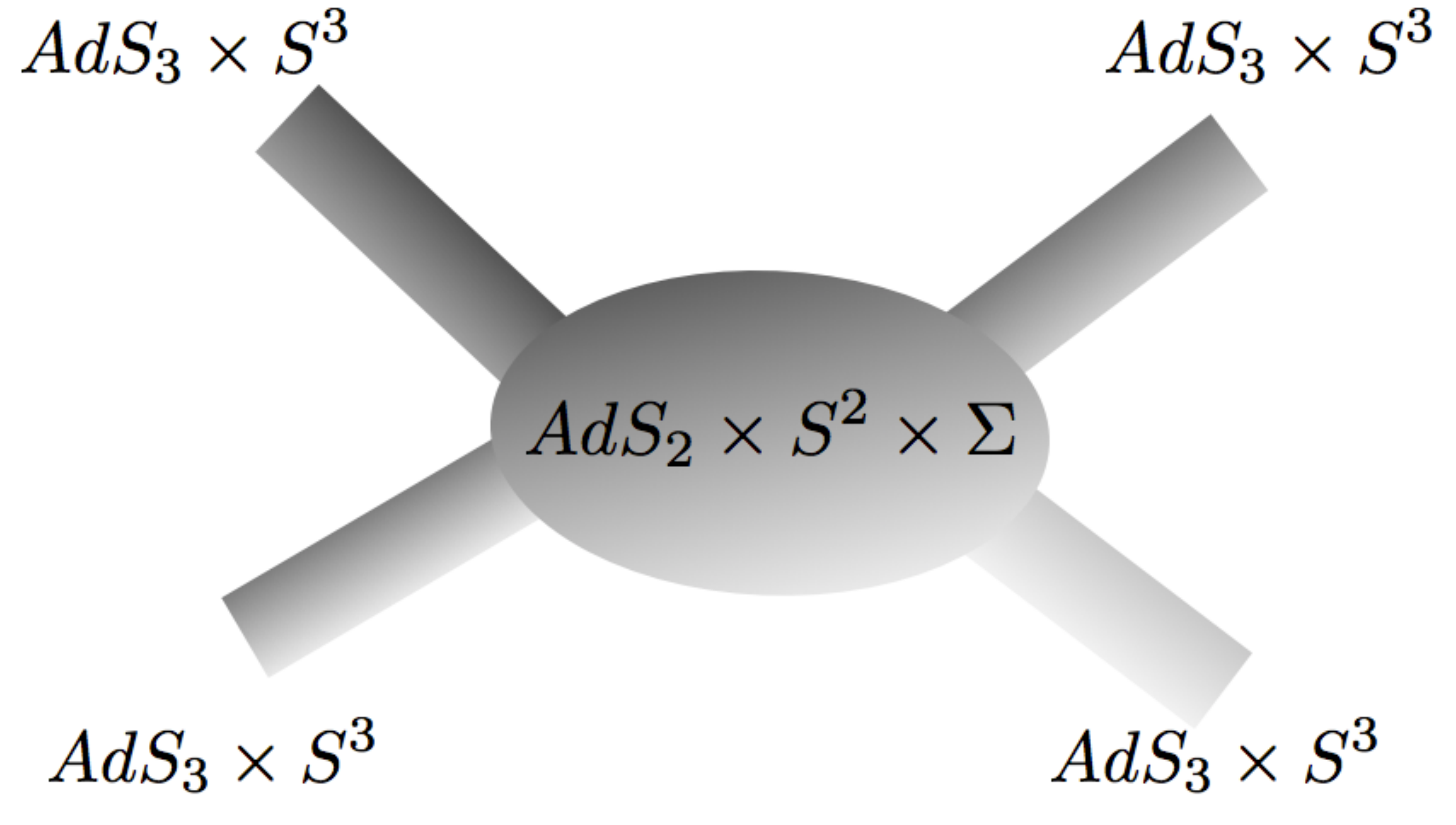}
\caption{Global structure of the $N=4$ solutions.}
\label{Fig1}
\end{centering}
\end{figure}

Parameterizing $\Sigma$ by the upper half-plane,  the above topological and regularity conditions 
are solved by a harmonic function $H$ and $m+2$ holomorphic  one-forms $\Lambda ^A$ given by,
\bea
\label{1b3}
H(w,\bar w) &=&  \sum _{n=1}^N  \left ({ i \, c_n \over w - x_n} - { i \, c_n \over \bar w - x_n} \right )
\no \\
\L ^A (w)  &=&  \sum _{n=1}^N \left ( { -i \kappa^A _n \over (w-x_n)^2} +  { -i \mu^A _n \over w-x_n} \right ) dw
\eea
where the positions of the poles $x_n$ and their residues $c_n, \kappa _n ^A, \mu _n ^A$ are 
real, with $c_n >0$, and restricted by the fact that $H$ and $\Lambda^A$ must obey both relations 
of (\ref{1b2}). To each pole $x_n$ corresponds a distinct asymptotic $AdS_{3}\times S^{3}$ region,
where  $\mu_n ^A$ gives the $m+2$-dimensional vector of three-form charges\footnote{This counting holds
when $\Sigma$ is simply connected; otherwise additional homology  three-cycles and charges will arise,
which must be included in the counting towards the total number of charges.} on the 
corresponding asymptotic  three-cycles $S^3$. There are also $2m$ scalar expectation values
associated with each asymptotic $AdS_{3}\times S^{3}$ region. The attractor mechanism fixes $m$ 
of these  in terms of  the charges $\mu_n^A$. The values of the  other $m$ scalars are related to
the residues $\kappa _n ^A/c_n$, and are (locally) arbitrary.  Thus, each asymptotic region has
$2m+2$ moduli. Charge conservation demands that the sum of  the charge vectors $\mu_n ^A$ vanishes.
Consequently, the total number of independent physical parameters  coming from all   asymptotic regions  is, 
\bea
\label{1b4}
\hbox{number of independent real moduli} ~ = ~ 2(m+1)N-(m+2)
\eea
It will be shown in Section \ref{sec7} that the number of physically independent moduli of (\ref{1b4}) 
precisely matches the number of free parameters in the solution (\ref{1b3})
subject to (\ref{1b2}).  

\sm

For general values of $m$ and $N$, the constraints imposed by (\ref{1b2}) on the parameters 
$x_n, c_n, \kappa ^A_n$, and $\mu^A_n$ consist of nested quadratic relations, and are highly 
non-trivial to solve analytically. Therefore, in Section \ref{sec8}, we shall begin by exhibiting 
solutions in a number of special cases, namely by solving (\ref{1b2}) numerically for $N=3,4,5$. 
These results demonstrate that regular solutions with asymptotic $AdS_3 \times S^3$ regions 
do exist, and may be obtained by straightforward numerical analysis.

\sm

Finally, in Section \ref{sec9}, we shall present a completely explicit analytical solution with the 
full number of free parameters given in (\ref{1b4}), for general $m$, and general number $N$ 
of asymptotic  $AdS_3 \times S^3$ regions. To do so, a light-cone like parametrization of the 
functions $\l^A = \L^A/\p H$ is used to solve the first constraint of (\ref{1b3}) in a rational manner.
The corresponding holomorphic light-cone functions may be parametrized explicitly by $2N-2$
{\sl auxiliary poles} and associated residues, in terms of which the second constraint of (\ref{1b3}) 
reduces to a set of Schwarz identities, which are found to be easily satisfied.
For $m \leq 2$, our solutions are found to reduce precisely to the solutions obtained 
in \cite{Chiodaroli:2009yw,Chiodaroli:2009xh}, as should have been expected.

\sm

The solutions constructed in this paper are to Type 4b six-dimensional supergravity
with $m$ tensor multiplets. Certain classes of these solutions are known to uplift to 
{\sl exact solutions} of the full ten-dimensional Type IIB supergravity. They include all 
solutions with $m=5$, since they can be lifted to Type IIB solutions via the consistent 
truncation of this theory on the zero modes of $T^4$. They also include all solutions for $m=21$ for which
the scalars take values in a $SO(2,2)/SO(2)\times SO(2)$ submanifold of the full
$SO(5,21)/SO(5)\times SO(21)$ coset,  where no moduli of the K3 are turned on and the 2-form fields restricted accordingly \cite{Chiodaroli:2009yw}.
Whether the general solutions, beyond these two special cases, also lift to {\sl exact solutions}
of Type IIB is unknown to us since it is, at present, unknown whether six dimensional Type 4b theory with $m=21$
arises as a {\sl consistent truncation} of Type IIB on $K3$, or not. All of our solutions are,
however, approximate solutions to Type IIB in the usual sense of Kaluza-Klein 
reduction, provided the scale of the solution is much larger than the scale of the $K3$
compactification manifold. Since each one of our solutions, for given asymptotic charges $\mu_n$
with $\mu _n \cdot \mu_n >0$, is completely regular as a 6-dimensional space-time, 
there will always be a sufficiently small scale of the $K3$ compactification (determined by the 
moduli of the solution) for which the solution is a reliable approximation to a full Type IIB solution
with the same charge assignments.

\sm

The holographic interpretation of our solutions is  given as the dual of interfaces between 
two CFTs, or more general as a junction of an arbitrary number of CFTs. The discussion   
parallels the  one given in \cite{Chiodaroli:2009yw} which we will briefly review in the following.
Each one of the $N$ asymptotic  $AdS_3\times S^3$ regions is dual to a 1+1-dimensional 
CFT defined on a half line ($\times$ time).  Each CFT 
is associated with the near-horizon limit of a half-BPS dyonic string in the six-dimensional supergravity 
carrying the same charges.  The CFTs are joined at a zero-dimensional defect which on the gravity side can be thought of as the boundary of the $AdS_2$ factor. Hence 
the solution is the holographic dual of a junction of $N$ CFTs  glued together at a one dimensional defect. The values of the un-attracted 
scalars in each asymptotic region correspond to exactly marginal deformations 
(i.e. moduli) of each CFT.   This construction generalizes the Janus solution which has  only two asymptotic AdS regions and corresponds to an interface of two CFTs where the moduli can jump when crossing the one dimensional interface.  

\sm

Since each asymptotic $AdS_3\times S^3$ region is associated with  half-BPS dyonic string the 1/2 BPS solutions presented in this paper can be viewed as near horizon limit of 1/4 BPS junctions of strings in 6 dimensional flat space.  The supersymmetry gets enhanced in the near horizon limit.   The existence of exact solutions of junctions of CFTs makes it possible to holographically calculate important observables such as  the entanglement entropy and reflection and transmission coefficients. It might be possible that our  solutions have some relevance for the description of junctions of quantum wires.

   \newpage

\section{Review of six-dimensional supergravities}
\label{sec2}
\setcounter{equation}{0}

In this section, we shall give a brief overview of   supergravity theories in 
six space-time dimensions, with various degrees of supersymmetry.

\sm

In six dimensions, a Weyl spinor is complex but self-conjugate. The Majorana condition 
cannot be imposed consistently on a single Weyl spinor; however, the symplectic 
Majorana condition may be imposed consistently on pairs of Weyl spinors. 

\sm

As a result, supergravity theories may be distinguished by a pair of integers $(n_+, n_-)$ 
where $n_\pm$ count the number of Weyl spinor supersymmetries of each chirality. 
The total number of (real) components of the supersymmetry generators equals $2n_+ + 2n_-$. 
If the symplectic-Majorana condition is imposed in each chirality sector, both integers 
$n_\pm$ must be even.
The corresponding R-symmetry is USp($n_+ ) \times$ USp($n_-$), under which the 
supersymmetries transform in the representation $(n_+,1) \oplus (1,n_-)$. The maximal 
number of (real)  components of the supersymmetry is 32, while the minimum number is 4.
Many, but not all, of the corresponding supergravity theories may be obtained by
dimensional reduction from eleven-dimensional M-theory, or from the ten-dimensional 
Type IIA, Type IIB, or $\cN=1$ (heterotic) supergravities. Below, we provide a list of 
supergravity  theories for which all supersymmetries are subject to the 
symplectic Majorana condition,

\begin{itemize}
\item 32 supersymmetries: The theory with $(n_+,n_-)=(4,4)$ is non-chiral; it may be obtained 
by compactification of Type II (A or B) supergravity on $T^{4}$.  Its field content consists of
the metric, 8 Rarita-Schwinger fields, 5 anti-symmetric tensor fields, 16 vector fields,
40 spinor fields, and 25 scalars parameterizing a $SO(5,5)/(SO(5)\times SO(5))$ coset.
Its Lagrangian and supersymmetry transformations were constructed in \cite{Tanii:1984zk}.
\item 
32 supersymmetries: The maximally chiral $(0,8)$ theory has been studied in \cite{Hull:2000zn},
where its field multiplet structure was determined (at the linearized level). To date, full field 
equations and Lagrangians for the $(0,8)$ and $(2,6)$ theories have not been written down.

\item 
16 supersymmetries: The Type 4a theory with  $(n_{+},n_{-})=(2,2)$ can be obtained 
by compactification of Type IIA theory on $K{3}$ \cite{Giani:1984dw}. Its field content consists of
the metric, 4 Rarita-Schwinger fields, vector fields, spinor fields, and scalars.
\item 
16 supersymmetries: The Type 4b theory with  $(n_{+},n_{-})=(0,4)$, can be obtained 
by compactification of Type IIB theory on $K{3}$  \cite{Romans:1986er} 
(see also \cite{Deger:1998nm,Avramis:2006nb}). This is the theory
of central interest to us, and its properties will be given in detail below.
\item 
8 supersymmetries: The $(0,2)$-theories have been 
discussed in detail in \cite{Avramis:2006nb}.
\end{itemize}

\subsection{The $(0,4)$ theory}

The six-dimensional Type 4b supergravity with $(n_+,n_-)=(0,4)$ is a close analogue of 
ten-dimensional Type IIB supergravity. The field content of the theory consists of a 
supergravity multiplet, and $m$ anti-symmetric tensor  
multiplets.  The supergravity multiplet contains the metric\footnote{
A detailed summary of notations and conventions for indices may be found in Appendix \ref{appA}.}
$ g_{\mu \nu}$, four negative chirality Rarita-Schwinger fields $\psi ^\alpha _\mu$,  
and five anti-symmetric tensor fields $B^I _{\mu \nu}$ with self-dual field strength.
The $m$ anti-symmetric tensor multiplets account for $m$ anti-symmetric tensor
fields $B^R_{\mu \nu}$ with anti-self-dual field strength, $4m$ Weyl spinors $\chi ^{r \alpha}$ of positive chirality,
and $5m$ scalars $V ^i{}_R$, which parametrize the coset $SO(5,m)/(SO(5)\times SO(m))$.
The multiplet structure is summarized in Table 1 below, 

\begin{table}[htdp]
\begin{center}
\begin{tabular}{|c||c|c|c|c||c|c|}
\hline
{\rm Field}  &  $g_{\mu\nu}$&$B^I_{\mu\nu}$&$B^R_{\mu\nu}$
&$ V^i{}_R$&   $\psi_{\mu}^{\alpha}$&$ \chi^{r \alpha}$  \\
\hline\hline
SO(5)  & {\bf 1}  & {\bf 5} & {\bf 1} & {\bf 5} & {\bf 4} & {\bf 4}  \\
\hline
  SO($m$) & {\bf 1} & {\bf 1} & {\bf m}  & {\bf m} & {\bf 1} & {\bf m} \\
  \hline
\end{tabular}
\caption{Field content of six-dimensional $(0,4)$ supergravity}
\end{center}
\end{table}

\subsubsection{Chiral truncation of the $(4,4)$ theory}
\label{sec22}

Consider now the following chiral truncation of the $(4,4)$ theory. We set to zero  
the sixteen vector fields, the four positive chirality components of the Rarita-Schwinger fields 
and of the supersymmetry parameters, as well as the twenty negative chirality components 
of the spinor fields. The field contents, Lagrangian, and supersymmetry transformations 
of this chiral restriction precisely coincide with those of the $(0,4)$ theory for 
the special choice of $m=5$. Note that the 5 anti-symmetric tensor fields of the $(4,4)$
theory split into self-dual and anti-self-dual, thereby providing the appropriate  fields required for both
the supergravity and anti-symmetric tensor multiplets of the $(0,4)$ theory.

\sm

It follows immediately that every classical solution to the field equations of the $(0,4)$ 
theory for $m \leq 5$ may be lifted to a solution of the $(4,4)$ theory with the same 
number of residual supersymmetries. On the other hand, cancellation of fermion anomalies
in the $(0,4)$ theory by itself requires $m=21$, in which case  the theory can be obtained by 
compactification of ten-dimensional Type IIB string theory on $K{3}$. In our search for 
classical (BPS) solutions in the $(0,4)$ theory, we shall keep $m$ arbitrary as long as possible.

\subsubsection{Possible truncations of the $(0,8)$ and $(2,6)$  theories ?}
\label{sec22a}

General considerations of supersymmetry representation theory indicate that field theories
should exist with $(0,8)$ and $(2,6)$ supersymmetry (totalling 32 supersymmetries). 
For the $(0,8)$ theory, the field multiplet  at the linearized level is known
\cite{Hull:2000zn}, but its extension to the fully interacting level is not known. 
Even at the linearized level, this theory is clearly out of the ordinary,
as the customary metric field for the graviton is replaced here by a rank-four tensor
with the symmetries of the Riemann tensor. Even less is known for the $(2,6)$ theory. 
Yet, both theories are expected to reduce to $N=8$ supergravity in five and in four dimensions,
which makes them interesting, at least in principle. Can one expect the $(0,4)$ theory to be a
truncation of the $(0,8)$ and $(2,6)$ ? If so, the solutions 
obtained for the $(0,4)$ theory in this paper would lift also to solutions
of the $(0,8)$ and $(2,6)$ theories.

\subsection{Scalar fields}

We begin by spelling out the coset parametrization of the scalar fields, and the duality 
restrictions on the anti-symmetric tensor fields for $(0,4)$  supergravity \cite{Romans:1986er} 
(see also \cite{Avramis:2006nb,Deger:1998nm}) in the subsequent section. Consistently with the discussions 
of the preceding section, we shall keep the number $m$ of anti-symmetric tensor multiplets arbitrary.

\sm

The scalar fields take values in the coset space $SO(5,m)/(SO(5) \times SO(m))$, and may 
be parametrized in terms of a {\sl frame field} $V $ which takes values in $SO(5,m)$.
A~convenient explicit representation may be obtained by block-decomposing the $SO(5,m)$-invariant 
metric $\eta$ and the frame $V$ according to the maximal compact subgroups $SO(5)\times SO(m)$,
\bea
\label{2a1}
\eta = \left ( \matrix{ I_5 & 0 \cr 0 & - I_m \cr } \right )
\hskip 1in 
V=V^{(i,r)}{}_A = \left ( \matrix{ V^i{}_I  & V^i {}_R \cr V^r{}_I & V^r{}_R \cr } \right )
\eea
The index $A=(I,R)$ will label the fundamental representation of $SO(5,m)$, while the indices 
$(i,r)$ will label the fundamental representation of $SO(5) \times SO(m)$, with $I,i=1,\cdots , 5$
and $R,r=6, \cdots, m+5$. Since the indices $i$ and $r$ label different objects, we shall refrain 
from assembling  them into a single index, following  \cite{Romans:1986er}. The $SO(5,m)$ index 
$A$ is raised, lowered, and contracted with the $SO(5,m)$-invariant 
metric $\eta$, while we will raise, lower, and contract indices $i$ and $r$ with 
the positive Euclidean metrics $\delta _{ij}$ and $\delta _{rs}$. 
Since $V \in SO(5,m)$, it obeys, 
\bea
\label{2a2}
\eta= V^t \eta V \hskip 1in  V^{-1} = \eta V^t \eta
\eea
The block decomposition of the right-invariant flat $SO(5,m)$-connection  is given by,
\bea
\label{2a3}
\p_\mu V \, V^{-1} = \left ( \matrix{ Q_\mu & \sqrt{2} P_\mu \cr \sqrt{2}  R_\mu  & S_\mu \cr } \right )
\eea
Since $\p_\mu V \, V^{-1}$ takes values in the Lie algebra of $SO(5,m)$, the product $ \p_\mu V \, V^{-1}\eta $
is antisymmetric, which implies the following relations between the connection components,
\bea
\label{2a4}
Q_\mu ^t = - Q_\mu \hskip 1in R_\mu = P_\mu ^t \hskip 1in S_\mu ^t = - S_\mu
\eea
Here, $Q_\mu$ and $S_\mu$ are canonical connections taking values in the Lie algebras of $SO(5)$
and $SO(m)$ respectively, while $P_\mu$ is a canonical frame on $SO(5,m)/(SO(5) \times SO(m))$. 

\sm

We represent the coset $SO(5,m)/(SO(5) \times SO(m))$ by a $5m$-dimensional slice in $SO(5,m)$.
We follow the conventions introduced in \cite{Romans:1986er}, which are the natural ones
associated with the right-invariant connection $\p_\mu V V^{-1}$.  A transformation
under the action of $\cM \in SO(5,m)$ of a point $V$ on the slice to another point $V'$ on the slice
is by right-multiplication, and must be compensated by a  gauge transformation 
$M \in SO(5)\times SO(m)$ acting on the left,
\bea
\label{2a4a}
V \cM = M^{-1} V' \hskip 1in M = \left ( \matrix{ M_5 & 0 \cr 0 & M_m \cr} \right )
\eea
The canonical frame and connections then transform as follows,
\bea
\label{2a4b}
\p_\mu V' (V')^{-1} = M \p_\mu V V^{-1} M^{-1} + \p_\mu M M^{-1}
\eea
The component formulas for the canonical frame and connection fields are, 
\bea
\label{2a5}
\sqrt{2} P^{ir}_\mu & = &    - (\p_\mu V^i{}_A) V^r{}_B \, \eta ^{AB}
\no \\
Q_\mu ^{ij} & = &  + (\p_\mu V^i{}_A) V^j{}_B \, \eta ^{AB}
\no \\
S_\mu ^{rs} & = &   - (\p_\mu V^r{}_A) V^s{}_B \, \eta ^{AB}
\eea
A good parametrization
of the coset $SO(5,m)/ (SO(5) \times SO(m))$ is provided by $V^i{}_R$.

\subsection{Anti-symmetric tensor fields}

The anti-symmetric rank-two tensor fields $B^A_{\mu \nu}$, 
respectively $B^I_{\mu \nu}$ of the supergravity multiplet,
and $B^R _{\mu \nu}$ of the anti-symmetric tensor multiplet, give rise to field strength 
three-forms,
\bea
\label{2b1}
G^A  & = & d B^A 
\eea
The associated $SO(5,m)$-covariant field strength three-forms $H^i $ and 
$H^r$, defined by,
\bea
\label{2b2}
H^i  & = & V^i{}_A \, G^A
\no \\
H^r  & = & V^r{}_A \, G^A 
\eea
are respectively self-dual and anti-self-dual three-forms, obeying, 
\bea
\label{2b3}
{}*H^i & = & + H^i  
\no \\
{}*H^r & = & - H^r 
\eea

\subsection{Bianchi identities}

As a result of their definition in (\ref{2a3}), the fields $P, Q$ and $S$ obey the following Bianchi
identities, expressed here in exterior form notation,
\bea
\label{2c1}
dP^{ir} - Q^{ij}  \wedge P^{jr} - S^{rs} \wedge P^{is}  & = & 0
\no \\
dQ^{ij} - Q^{ik} \wedge Q^{kj} - 2 P^{ir} \wedge P^{jr} & = & 0
\no \\
dS^{rs} - S^{rt} \wedge S^{ts} - 2 P^{ir} \wedge P^{is} & = & 0
\eea
where summation over repeated indices is implied.
Similarly, the three-forms $G$ obey simple Bianchi identities,
\bea
\label{2c2}
d G^A = 0
\eea
As a result of the relation (\ref{2b2}), the fields $H^i$ and $H^r$ then obey the Bianchi identities,
\bea
\label{2c3}
dH^i - Q^{ij} \wedge H^j - \sqrt{2} P^{ir} \wedge H^r & = & 0
\no \\
dH^r - S^{rs} \wedge H^s - \sqrt{2} P^{ir} \wedge H^i & = & 0
\eea

\subsection{Field equations for bosonic fields}

The Bianchi identities and field equations  for the bosonic fields are as follows.
For the field strengths $H^i$ and $H^r$, the Bianchi identities and field equations 
are equivalent to one another in view of their duality properties. Substituting (\ref{2b3}) into
(\ref{2c3}), we find the following field equations for $H^i, H^r$,
\bea
\label{2d1}
d*H^i - Q^{ij} \wedge *H^j + \sqrt{2} P^{ir} \wedge *H^r & = & 0
\no \\
d*H^r - S^{rs} \wedge *H^s + \sqrt{2} P^{ir} \wedge *H^i & = & 0
\eea
In components these equations read as follows,
\bea
\label{2d2}
\nabla ^{\mu} H^{i}_{\mu\nu\rho} - (Q^\mu)^{ij} H^j _{\mu \nu \rho} + \sqrt{2} (P^{\mu})^{ir} H^{r}_{\mu\nu\rho} 
& = & 0
\no\\
\nabla ^{\mu} H^{r}_{\mu\nu\rho} - (S^\mu) ^{rs} H^s _{\mu \nu \rho} +\sqrt{2} (P^{\mu})^{ir} H^{i}_{\mu\nu\rho}
& = & 0
\eea
where $\nabla ^\mu$ is the covariant derivative with respect to the affine connection.
The Einstein equations are given by,
\be
\label{2d3}
R_{\mu\nu}- H^{i}_{\mu\rho\sigma}H^{i\; \rho\sigma}_{\nu}
- H^{r}_{\mu\rho\sigma}H^{r\; \rho\sigma}_{\nu}-2 P_{\mu}^{ ir} P_{\nu}^{ir}=0
\ee
The field equation for the scalars is given by,
\be
\label{2d4}
\nabla^{\mu} P_{\mu}^{ir} - (Q^{\mu})^{ij}   P_{\mu}^{jr} - (S^{\mu})^{rs} P_{\mu}^{is}  
-{\sqrt{2}\over 3} H^{i\; \mu\nu\rho}H^{r}_{\mu\nu\rho}=0
\ee

\subsection{Supersymmetry transformation for fermionic fields}

The fermionic fields $\psi _\mu ^\a$ and $\chi ^{r\a}$ of the $(0,4)$ supergravity,
as well as the local supersymmetry spinor parameter $\ep^\a$, have definite chiralities,
and obey the symplectic Majorana conditions,  given by, 
\bea
\label{2e1}
\g^7 \psi _\mu ^\a =  - \psi ^\a _\mu \hskip 0.05in 
& \hskip 1in &
\hskip 0.05in \psi ^\a _\mu = \cB \, \cC^{\a}_{~\b} \, \big( \psi _\mu^{\b}\big)^*
\no \\
\g^7 \chi ^{r\a} = + \chi ^{r\a}
& \hskip 1in &
\chi ^{r\a} = \cB \, \cC^{\a}_{~\b} \, \big( \chi^{r \b} \big)^*
\no \\
\g ^7 \ep ^\a = - \ep ^\a \hskip 0.08in 
&&
\hskip 0.08in  \ep^{\a} = \cB \, \cC^{\a}_{~\b} \, \big( \ep^{\b}\big)^* 
\eea
Representations for Dirac matrices for $SO(1,5)$ and $SO(5)$ may be found in 
Appendix \ref{appA}, where a number of useful Dirac algebra relations have also
been collected. In particular,  $\cB$ is the $SO(1,5)$ complex conjugation 
matrix, while $\cC$ is the $SO(5)$ 
charge conjugation matrix, whose definitions and representations are given in 
(\ref{A10}) and (\ref{A16}) of Appendix \ref{appA}.

\sm

The supersymmetry transformations  on the fermionic fields are 
given by,\footnote{The Dirac matrices $\g^\mu$ with respect to coordinate indices are related to
the Dirac matrices $\g^M$ with respect to frame indices by  $\g^M = e^M {}_\mu \g^\mu$.}
\bea
\label{2e3}
\delta \psi_{\mu}^{\alpha}
&=&
D_{\mu}\ep^{\alpha}-{1\over 4} H^{i}_{\mu\nu\rho} \gamma^{\nu\rho} \;
(\Gamma^{i})^{\alpha}_{\;\; \beta}\ep^{\beta}
\no\\
\delta \chi^{r \alpha}
&=&
{1\over \sqrt{2}} \gamma^{\mu} P_{\mu}^{ir}
(\Gamma^{i})^{\alpha}_{\;\; \beta} \ep^{\beta}
+{1\over 12} \gamma^{\mu\nu\rho}H^{r}_{\mu\nu\rho}\ep^{\alpha}
\eea
The covariant derivative in (\ref{2e1})  is taken with respect to the $SO(1,5)$ spin
connection $\o_\mu $ and the $SO(5)$-connection $Q_\mu $, and given by,
\bea
\label{2e4}
D_{\mu}\ep^{\alpha} \equiv \partial_{\mu} \ep ^\a
+ {1\over 4} \omega_{\mu}^{MN}\gamma_{MN} \ep ^\a
-{1\over 4} Q_{\mu}^{ij} (\Gamma^{ij})^{\alpha} {} _\b\, \ep^\beta 
\eea
It is manifest that these relations are consistent with the chirality 
restrictions of (\ref{2e1}). They are also consistent with the 
symplectic Majorana condition, as may be established by 
analyzing the complex conjugate relations. Finally, using the chirality of $\ep^\a$, as well as
formula (\ref{A9}), we derive the relation,
$\g^{\mu \nu \rho} H^r _{\mu \nu \rho} \ep ^\a = - \g^{\mu \nu \rho} (*H^r) _{\mu \nu \rho} \ep ^\a$.
Therefore, the last term in $\delta \chi ^{r \a}$ is non-zero only if $*H^r = - H^r$,
which is indeed the case from (\ref{2b3}).

\sm

The BPS equations correspond to the vanishing of the supersymmetry transformations,
\bea
\label{2e5}
\delta \psi ^\a _\mu = \delta \chi ^{r \a} =0
\eea
for all fermionic fields, while setting the fermionic fields to zero. For non-trivial $\ep ^\a$,
the corresponding bosonic solutions are invariant under some degree of supersymmetry.

\newpage

\section{Ansatz and reduced BPS equations} 
\setcounter{equation}{0}
\label{sec3}

In this section, we construct the general Ansatz invariant under $SO(2,1) \times SO(3)$
for the bosonic fields and supersymmetry transformation parameters of (0,4) supergravity. 
The BPS equations of (\ref{2e5}) will then be reduced by restricting them to this Ansatz.

\subsection{The $SO(2,1) \times SO(3)$-invariant Ansatz}
\label{ansatzsol}

We seek the most general Ansatz in six-dimensional  (0,4) supergravity with
$SO(2,1) \times SO(3)$ symmetry. The $SO(2,1)$ requires the geometry to 
contain a factor of $AdS_2$, while the $SO(3)$ requires a factor of $S^2$,
both of which are {\sl warped} over a two-dimensional surface $\Sigma$, yielding a 
total warped space-time structure $AdS_2 \times S^2  \times \Sigma$.
The Ansatz for the metric is given by,
\bea
\label{3a2}
ds^2 = f_1^2 ds^2 _{AdS_2} + f_2 ^2 ds^2 _{S^2} +  ds^2 _\Sigma
\eea
where $f_1$ and $f_2$ are real functions on $\Sigma$. Orthonormal frames are defined by,
\bea
\label{3a3}
ds^2 _{AdS_2} = \eta _{mn}^{(2)} \, \hat e^m \otimes \hat e^n
& \hskip 0.8in &
e^m = f_1 \, \hat e^m \hskip 0.97in m=0,1
\no \\
ds^2 _{S^2}  = \delta _{pq} \, \hat e^{p} \otimes \hat e^{q} \hskip 0.1in 
&&
e^{p} \, = f_2 \, \hat e^{p} \hskip 1.05in p =2,3
\no \\
ds^2 _\Sigma   = \delta _{ab} \, e^{a} \otimes e^{b} \hskip 0.1in 
&&
e^{a} \, = \rho ~\hat e^{a} \hskip 1.1in a=4,5
\eea
The metrics $ds^2 _{AdS_2}$ and $ds^2 _{S^2}$, as well as the orthonormal frames 
$\hat e^m$ and $\hat e^{p}$, respectively refer to the spaces $AdS_2$ and $S^2$
with unit radius.
The Ansatz for the scalar field strength one-forms $P^{ir}$, 
as well as the composite  connection  one-forms $Q^{ij}$ and $S^{rs}$ is as follows,
\bea
\label{3a4}
P^{ir} & = & p^{ir}_{a} e^{a}
\no \\
Q^{ij} & = & q^{ij}_{a}e^{a}
\no \\
S^{rs} & = & s^{rs}_{a}e^{a}
\eea
Finally, the Ansatz for the three form field strength is,
\bea
\label{3a5}
H^{i}= g^{i}_{a} e^{01a}+ h^{i}_{a} e^{23a}
\no \\
H^{r}= \tilde g^{r}_{a} e^{01a}+ \tilde h^{r}_{a} e^{23a}
\eea
The functions $f_1, f_2, \rho$, $p^{ir}_{a}, q^{ij}_{a}$, 
$s^{rs}_{a}, g^{i}_{a}, h^{i}_{a}, \tilde g^{r}_{a}, \tilde h^{r}_{a}$ depend only on the coordinates on $\Sigma$.
The duality relations (\ref{2b3}), combined with $* e^{01a} =  \epsilon ^a{}_b e^{23b}$ and 
$* e^{23a} = - \epsilon ^a{}_b \, e^{01b}$, produce the following relations, 
\bea
\label{3a6}
h^i_a & = & - \epsilon_a{}^b g^i_b
\no \\
\tilde h^r_a & = & + \epsilon_a{}^b \tilde g^r_b
\eea

\subsection{Killing spinor basis}

A convenient basis of Killing spinors on $AdS_2 \times S^2 \times \Sigma$ is provided by the spinors 
$\chi ^{\eta_1, \eta_2, \eta_3}$ of $SO(2,1) \times SO(3)\times SO(2)$, which are defined by the 
following equations,
\bea
\label{3b1}
\left ( \hat \nabla _m  - \half \eta_1 \tilde\gamma_m \otimes I \otimes I \right )
\chi ^{\eta _1, \eta _2, \eta_3}
    & = & 0 \hskip 1in m=0,1
\no \\
\left ( \hat \nabla _{p} - {i \over 2} \eta_2 I  \otimes \tilde\gamma_{p} \otimes I \right )
\chi ^{\eta _1, \eta _2, \eta_3}
    & = & 0 \hskip 1in p=2,3
    \no \\
 I \otimes I \otimes \left ( I - \eta _3 \sigma ^3 \right ) \chi ^{\eta _1, \eta _2, \eta_3}
    & = & 0
\eea
Integrability  requires $\eta _1^2 = \eta _2 ^2= \eta _3^2 =1$.
Given $\eta$, there are 4 independent solutions $\chi ^{\eta _1, \eta _2, \eta_3}$ to (\ref{3b1}).
An explicit representation for the Dirac matrices $\tilde\g_m$ and $\tilde \g_p$ on $AdS_2$ and $S^2$
respectively is given in Appendix \ref{appA}.  
The chirality matrices on $AdS_2 \times S^2$ reverse the signs of 
 the $\eta_1$ and $\eta_2$, allowing us to make the following identifications,
\bea
\label{3b2}
(\s^3 \otimes I \otimes I ) \chi ^{\eta _1, \eta _2, \eta_3}
    & = & \chi ^{- \eta _1,  \eta _2, \eta_3}
\no \\
(I \otimes \s^3 \otimes I ) \chi ^{\eta _1, \eta _2, \eta_3}
    & = & \chi ^{\eta _1,  - \eta _2, \eta_3}
\eea
Under charge conjugation of $\chi$, we produce a spinor $ \chi^c$, defined by,
\bea
\label{3b3}
\left ( \chi ^c \right ) ^{\eta _1, \eta _2, \eta_3}  =\cB
\left ( \chi ^{\eta _1 , \eta _2, \eta_3} \right )^*
\eea
Since six-dimensional Minkowski space-time does not support Majorana spinors, $\chi^c$
can never be identified with $\chi$, and therefore produces an independent spinor.
It will be convenient to chose the space of $\chi$ and $\chi ^c$ spinors as a basis
for the full 8-components supersymmetry spinors $\ep ^\a$, which gives us the following decomposition, 
\bea
\label{3b4}
\ep^\a = \sum _{\eta _1, \eta _2,\eta_3 } \left (
\chi ^{\eta _1, \eta _2, \eta_3} \otimes \zeta^\a _{\eta _1, \eta _2, \eta_3}
+  (\chi^c) ^{\eta _1, \eta _2, \eta_3} \otimes  \hat \zeta^\a _{\eta _1, \eta _2, \eta_3} \right )
\eea
where the coefficients $\zeta^\a _{\eta _1, \eta _2, \eta_3}$ and $\hat \zeta^\a _{\eta _1, \eta _2, \eta_3}$
are spinors of $SO(5)$, as indicated by the index $\a$.
As explained in Appendix \ref{appA}, there is an additional two-fold degeneracy of $\chi ^{\eta _1 , \eta _2, \eta_3}$
which will double the number of supersymmetries.

\sm

The six-dimensional chirality matrix $\g^7$, which in our basis is given by 
$\g^{7} = \s^3 \otimes \s^3 \otimes \s^3$, relates different components of $\chi$ and $\chi^c$,
\bea
\label{3b5}
\g^7 \, \chi ^{\eta _1, \eta _2, \eta_3} & = & \eta _3 \, \chi ^{-\eta _1, -\eta _2, \eta_3}
\no \\
\g^7 \, (\chi ^c) ^{\eta _1, \eta _2, \eta_3} & = & \eta _3 \, (\chi ^c) ^{-\eta _1, -\eta _2, \eta_3}
\eea
As a result, the negative chirality condition $\g^{7} \ep^\a = - \ep^\a$ imposes the following relations between 
the different components of $\zeta$,  and between the different components of $\hat \zeta$,
\bea
\label{3b6}
\zeta^\a _{\eta _1, \eta _2, \eta_3} & = & - \eta_3 \, \zeta^\a _{-\eta _1, -\eta _2, \eta_3}
\no \\
\hat \zeta^\a _{\eta _1, \eta _2, \eta_3} & = & - \eta_3 \, \hat \zeta^\a _{-\eta _1, -\eta _2, \eta_3}
\eea
The symplectic Majorana condition $\ep^{\a} =\cB \, \cC^{\a}{}_\b  \, (\ep^{\b})^*$ imposes relations between 
$\zeta $ and $\hat \zeta$,
\bea
\label{3b7}
\zeta^\a _{\eta _1, \eta _2, \eta_3} & = &  - \cC^\a{}_\b \,  (\hat \zeta^{\b}_{\eta _1, \eta _2, \eta_3})^*
\no \\
\hat \zeta^\a _{\eta _1, \eta _2, \eta_3} & = & + \cC^\a {}_\b \, (\zeta^{\b}_{\eta _1, \eta _2, \eta_3})^*
\eea
where $\cC$  is the charge conjugation matrix of $SO(5)$, given in our basis by $\cC= \s^1 \otimes \s^2$.
As a result, $\hat \zeta$ may be eliminated in terms of $\zeta ^*$, yielding our final decomposition
formula for the supersymmetry spinor,
\bea
\label{3b8}
\ep^\a = \sum _{\eta _1, \eta _2,\eta_3 } \left (
\chi ^{\eta _1, \eta _2, \eta_3}
\otimes
\zeta^\a _{\eta _1, \eta _2, \eta_3}
+ (\chi^c) ^{\eta _1, \eta _2, \eta_3}
\otimes
\cC^{\a}_{~\b} \, \zeta^{\b*}_{\eta _1, \eta _2, \eta_3} \right )
\eea
Since the BPS equations are compatible with the symplectic Majorana condition, 
the reduced BPS equations for $\zeta$ and $\zeta ^*$ will automatically be complex
conjugates of one another.

\subsection{Derivation of the reduced BPS equations}

The detailed derivation of the reduced BPS equations will be presented in Appendix B.
Here, we shall provide only a summary of the results. We use the standard notation
for the Pauli matrices $\tau^a$, with $a=0,1,2,3$, acting on the indices $\eta$ of $\zeta$, 
\bea
\label{3c1}
\left ( \tau^{(abc)} \zeta^\a \right ) _{\eta _1, \eta _2, \eta_3}
\equiv
\sum _{\eta _1', \eta _2', \eta_3'}
(\tau ^a)_{\eta _1, \eta _1'} (\tau ^b )_{\eta _2, \eta _2'} (\tau ^c )_{\eta _3, \eta _3'} 
\zeta^\a  _{\eta _1', \eta _2', \eta_3'}
\eea
In this notation, the chirality condition becomes,
\bea
\tau^{(113)} \zeta = - \zeta 
\eea
The reduced BPS equations for the variation of the gravitini field are,
\bea
\label{3c2}
(m) \quad 0 &=&
 { 1\over  f_1} \t^{(300)} \zeta+ {D_{a} f_1 \over  f_1} \t^{(11a)} \zeta
+  g^i_a \t^{(01a)}\Gamma_{i}\zeta
\\
(p) \quad 0&=&
{ i\over  f_2} \t^{(130)}\zeta + {D_{a} f_2 \over  f_2}  \t^{(11a)} \zeta
-  i h ^i_a \t^{(10a)}\Gamma_{i}\zeta
\no \\
(a) \quad 0&=&
\left ( D_a + {i\over 2} \hat \omega_a \tau^{(003)} 
-{1\over 4} q_a^{ij} \Gamma_{ij} \right ) \zeta
+{1\over 2} g^i_a \t^{(100)} \Gamma_{i}\zeta
- {i\over 2}  h^i_a \t^{(010)} \Gamma_{i}\zeta
\no 
\eea 
while the reduced BPS equations for the fermions $\chi^{r \alpha}$ are given by, 
\bea
\label{3c3}
(c) \quad  0 & = &\sqrt{2} p_{a}^{ir} \t^{(11a)} \Gamma_{i} \zeta
- \tilde g_{a}^{r} \t^{(01a)} \zeta +i \tilde h ^r _a \tau ^{(10a)} \zeta 
\eea
Next, we perform the substitution $\zeta=- \tau^{(113)} \zeta$ in the second 
term of equation (m), in the fourth term of equation (a),
and in the second term of equation (c), and then multiply equations
(p) and (c) to the left by $\tau^{(100)}$.
We see that the two eigenmodes of $\tau^{(300)}$ decouple from one another and 
are mapped into one another by reversing the sign of $f_1$. Given one sign for the 
physical field $f_1$, one or the other eigenmode must vanish. We choose the 
non-vanishing mode to be the one corresponding to eigenvalue $+1$, and omit the
first index entry in the $\tau$-matrices, so that $\tau ^{(0bc)} \to \tau ^{(bc)}$.
In the same vein, we continue to use the notation $\zeta$ for the eigenmode of $\tau^{(300)}$
with eigenvalue $+1$.
Finally, we perform a cyclic permutation on the Pauli matrices,  
\bea
\label{3c5}
\t^{(10)} \rightarrow \t^{(20)},
\qquad
\t^{(20)} \rightarrow \t^{(30)},
\qquad
\t^{(30)} \rightarrow \t^{(10)}
\eea
After multiplying the resulting equation (p') on the left by $\tau^{(13)}$, 
and eliminating $h^i_a$ and $\tilde h^r _a$ in terms of respectively $g^i_a$ and $\tilde g^r _a$
using the duality equations (\ref{3a6}),  the equations become,
\bea
\label{3c6}
(m'') \quad 0 &=&
 { 1\over  f_1} \zeta - {D_{a} f_1 \over  f_1} \t^{(0a)}\t^{(03)} \zeta
+  g^i_a \t^{(2a)}\Gamma_{i}\zeta
\no \\
(p'') \quad 0&=&
{ 1\over f_2} \t^{(33)}\zeta - {D_{a} f_2 \over  f_2}  \t^{(0a)} \t^{(03)}\zeta
- g^i_a \t^{(2a)}\Gamma_{i}\zeta
\no \\
(a'') \quad 0&=&
(D_a + {i\over 2} \hat \omega_a \tau^{(03)})\zeta
-{1\over 4} q_a^{ij} \Gamma_{ij}\zeta
- {1\over 2} g^i_a \t^{(23)} \Gamma_{i}\zeta
+ {i\over 2} \epsilon_{ab} g^i_b \t^{(20)} \Gamma_{i}\zeta
\no \\
(c'') \quad  0 & = &{1\over \sqrt{2}} p_{a}^{ir} \t^{(0a)}\t^{(03)}\Gamma_{i} \zeta
+ \tilde g_{a}^{r} \t^{(2a)} \zeta
\eea

\subsection{Chiral form of the reduced BPS equations}

It will be convenient to further reduce the representation of the spinors, by reformulating 
the  frame $e^a$ with $a=4,5$ on the Riemann surface $\Sigma$ in terms of a 
complex basis, $e^a = (e^z, e^{\bar z})$, with metric and $\epsilon$-tensor
normalized by,
\bea
\label{3d1}
\delta _{z \bar z} = \delta _{\bar z z} = 2
\hskip 1in 
\epsilon _{z \bar z} = - \epsilon _{\bar z z} =2i
\eea
and as given by the following explicit formulas,
\bea
\label{3d2}
e^z = (e^4 + i e^5)/2
& \hskip 1in &
e_z = e^4 - i e^5
\no \\
e^{\bar z} = (e^4 - i e^5)/2
& \hskip 1in &
e_{\bar z} = e^4 + i e^5
\eea
Using these conventions the metric factor on $\Sigma$ is given by
\be
ds_\Sigma^2 = 4 \rho^2 |dw|^2
\ee
Corresponding relations hold for the the fields $p_a^{ir}, q_a^{ij}, g_a^i$,
so that, for example, $p^{ir}_z = p^{ir}_4 - i p^{ir}_5$, and $p^{ir}_{\bar z} = p^{ir}_4 + i p^{ir}_5$.
The Pauli matrices in this basis take the form,
\bea
\label{3d3}
\sigma ^z = \tau ^z= \left ( \matrix{ 0 & 1 \cr 0 & 0 \cr} \right )
\hskip 1in
\sigma ^{\bar z} = \tau ^{\bar z} = \left ( \matrix{ 0 & 0 \cr 1 & 0 \cr} \right )
\eea
We decompose $\zeta^\a$ into its two components of the $\eta_3$ index, leaving the $\eta_2$
index free,
\bea
\label{3d4}
\xi_{\eta_2} = \zeta_{\eta_2,+}
\qquad
\psi_{\eta_2} = \zeta_{\eta_2,-}
\eea
The BPS equations in the chiral form may be grouped into algebraic equations in $\psi, \xi$, 
\bea
\label{3d5}
(m_1) \qquad 0 &=&
 { 1\over f_1} \xi + {D_{z} f_1 \over f_1} \psi
+  g^i_z \t^{(2)}\Gamma_{i}\psi
\no \\
(m_2)\qquad 0 &=&
 { 1\over f_1} \psi - {D_{\bar z} f_1 \over f_1} \xi
+  g^i_{\bar z} \t^{(2)}\Gamma_{i}\xi
\no \\
(p_1) \qquad 0&=&
{ 1\over  f_2} \t^{(3)}\xi + {D_{z} f_2 \over  f_2}  \psi
- g^i_z \t^{(2)}\Gamma_{i}\psi
\no \\
(p_2) \qquad 0&=&
{ 1\over  f_2} \t^{(3)}\psi + {D_{\bar z} f_2 \over  f_2}  \xi
+  g^i_{\bar z} \t^{(2)}\Gamma_{i}\xi
\no \\
(c_1) \qquad 0 & = &-{1\over \sqrt{2}} p_{z}^{ir}\Gamma_{i} \psi
+ \tilde g_{z}^{r} \t^{(2)} \psi
\no \\
(c_2) \qquad  0 & = &{1\over \sqrt{2}} p_{\bar z}^{ir}\Gamma_{i} \xi
+ \tilde g_{\bar z}^{r} \t^{(2)} \xi
\eea
and differential equations in $\psi, \xi$, 
\bea
\label{3d5a}
(az_1) \qquad 0&=&
(D_z + {i\over 2} \hat \omega_z )\xi
-{1\over 4} q_z^{ij} \Gamma_{ij}\xi
- g^i_z \t^{(2)} \Gamma_{i}\xi
\no \\
(az_2) \qquad 0&=&
(D_z - {i\over 2} \hat \omega_z )\psi
-{1\over 4} q_z^{ij} \Gamma_{ij}\psi
\no \\
(a\bar z _1) \qquad 0&=&
(D_{\bar z} + {i\over 2} \hat \omega_{\bar z} )\xi
-{1\over 4} q_{\bar z}^{ij} \Gamma_{ij}\xi
\no \\
(a\bar z _2) \qquad 0&=&
(D_{\bar z} - {i\over 2} \hat \omega_{\bar z} )\psi
-{1\over 4} q_{\bar z}^{ij} \Gamma_{ij}\psi
+ g^i_{\bar z} \t^{(2)} \Gamma_{i}\psi
\eea
In a system of local complex coordinates $w, \bar w$ on $\Sigma$,  
we have,
\bea
\label{3d6}
e^z = \rho dw & \hskip 0.7in & D_z = \rho ^{-1} \p_w 
	\hskip 1in \hat \o_z = +i \rho ^{-2} \p_w \rho
\no \\
e^{\bar z} = \rho d \bar w && D_{\bar z} = \rho ^{-1} \p_{\bar w} 
	\hskip 1in \hat \o _{\bar z} = -i \rho ^{-2} \p_{\bar w} \rho
\eea

\newpage

\section{Solving the reduced BPS equations}
\setcounter{equation}{0}
\label{sec4}

In this section, we present a systematic solution for the BPS equations
with 8 supersymmetries. This section is somewhat technical, and the impatient reader
may wish to skip it and move directly to Section \ref{sec5} where a summary 
of the solution to the BPS equations is given.

\sm

Our starting point will be the set of reduced BPS equations in chiral form of (\ref{3d5}) and (\ref{3d5a}).
Viewed as equations on the spinors $\xi$ and $\psi$, equations  $(m_{1,2}), (p_{1,2})$,
and $(c_{1,2})$ are purely algebraic, while $(az_{1,2}), (a\bar z _{1,2})$ 
are partial differential equations. We shall now solve those in succession.

\subsection{Algebraic relation between $\xi$ and $\psi$}

The combinations $(m_1)+(p_1)$ and $(m_2)+(p_2)$ no longer involve $g_z^i$
and produce purely algebraic relations between the components of $\xi$ and $\psi$,
\bea
\label{4a1}
0 & = & \left ( f_2 \pm f_1 \right ) \xi _\pm + D_z (f_1f_2) \psi _\pm
\no \\
0 & = & \left ( f_2 \mp f_1 \right ) \psi _\pm - D_{\bar z} (f_1f_2) \xi _\pm
\eea
Defining the tensor $Y$ of type $(1,0)$ by,
\bea
\label{4a2}
Y \equiv   { D_z(f_1f_2) \over f_1+f_2}
\eea
the $f_2+f_1$ relations in (\ref{4a1})  may be recast in the following form, 
\bea
\label{4a2a}
\xi _+ = - Y \psi _+
\hskip 1in
\xi _-  =   (Y^*)^{-1}  \psi _-
\eea
while the $f_2-f_1$ relations take the form, 
\bea
\label{4a3}
{ f_1-f_2 \over f_1 + f_2} \xi _- = Y \psi _-
\hskip 1in
{ f_1-f_2 \over f_1 + f_2} \psi _+ = -Y^* \xi _+
\eea
Consistency of (\ref{4a2a}) and (\ref{4a3}) requires that
\bea
\label{4a4}
|Y |^2= { f_1-f_2 \over f_1 + f_2}
\eea
and 
\bea 
\label{4a5}
|D_z (f_1f_2) |^2 = f_1^2 - f_2^2
\eea
Condition of (\ref{4a5}) is equivalent to (\ref{4a4}) upon
use of (\ref{4a2}). When conditions (\ref{4a2}),  (\ref{4a4}), and (\ref{4a5}) are obeyed,
the relations of (\ref{4a3}) follow and may be consistently omitted. 
This provides the most general solution to equations $(m_1)+(p_1)$ and $(m_2)+(p_2)$.

\subsection{Identifying the harmonic function $H$}

Next, we shall solve combinations of the differential equations. We begin by eliminating $\xi_\pm$
from the differential equation $(az_1)$ in (\ref{3d5a}), in favor of $\psi _\pm$ 
and $Y$, using (\ref{4a2}). The resulting equations are,
\bea
\label{4d1}
0 & = &
(D_z + {i\over 2} \hat \omega_z + D_z \ln Y )\psi _+
-{1\over 4} q_z^{ij} \Gamma_{ij}\psi _+
-{ i \over |Y |^2} g^i_z \Gamma_{i}\psi _-
\no \\
0 & = &
(D_z + {i\over 2} \hat \omega_z - D_z \ln Y^* )\psi _-
-{1\over 4} q_z^{ij} \Gamma_{ij}\psi _-
+ i |Y |^2 g^i_z \Gamma_{i}\psi _+
\eea
Using equation $(az_2)$ of (\ref{3d5a}) to eliminate  $D_z \psi _\pm$ we
find the algebraic equations,
\bea
\label{4d2}
0 & = &
( i \hat \omega_z + D_z \ln Y )\psi _+
-{ i \over |Y |^2} g^i_z \Gamma_{i}\psi _-
\no \\
0 & = &
( i \hat \omega_z - D_z \ln Y^* )\psi _- + i |Y |^2 g^i_z \Gamma_{i}\psi _+
\eea
Eliminating $i g_z ^i \G^i \psi _\pm$ between these equations and the
corresponding ones in (\ref{4b2}) gives
\bea
\label{4d3}
0 & = & D_z (f_1-f_2) - (f_1-f_2) ( i \hat \o _z + D_z \ln Y)
\no \\
0 & = & D_z (f_1+f_2) - (f_1+f_2) ( i \hat \o _z - D_z \ln Y^*)
\eea
Using $\hat \o _z = i \rho ^{-2} \p_w \rho$ and $D_z = \rho ^{-1} \p_w$,
we identify holomorphic one-forms $c_\pm (w)$, given by, 
\bea
\label{4d4}
\rho (f_1-f_2) /Y ^* & = & c_-( w)
\no \\
\rho (f_1+f_2) Y & = & c_+( w)
\eea
Taking the ratio,  and using (\ref{4a4}), we see that $c_+(w)=c_-(w)=c(w)$.
Eliminating $\xi$ also from equation $(a\bar z_1)$ using (\ref{4a2}), and using
equation $(a\bar z _2)$ to eliminate $D_{\bar z} \psi$ reproduces the 
same algebraic equations  of (\ref{4d3}), with the same solutions.

\sm

Using  the defining equation for $Y$ in (\ref{4a2}), and 
eliminating the combination $\rho (f_1+f_2) Y$ in favor of $c_+(w)=c(w)$
using the second equation in (\ref{4d4}), we find, $\p_w (f_1f_2) = c(w)$.
Since $f_1f_2$ is real, this equation may be integrated in terms of a 
real harmonic function $H$,
\bea
\label{4d6}
f_1 f_2 & = & H \hskip 1in c(w) = \p_w H
\no \\
\rho^2 (f_1^2 - f_2^2) & = & |\p_w H|^2 
\eea
where the second equation of (\ref{4d6}) results from using the second equation in (\ref{4a4}).
These equations can be solved for $f_1$ and $f_2$ in terms of $\rho$ and $H$.

\subsection{Algebraic projector conditions}

Next, we consider the remaining  combinations,
$f_1 (m_1) - f_2 \tau^{(2)} (p_1)$ and $f_1 (m_2) + f_2 \tau^{(2)} (p_2)$,
and eliminate $\xi_\pm$ using (\ref{4a2}) and (\ref{4a4}), which gives, 
\bea
\label{4b2}
D_z(f_1-f_2) \psi _+  & = &  +i g_z ^i \G_i ( f_1 +  f_2) \psi _-
\no \\
D_z(f_1+f_2) \psi _- & = & -  i g_z ^i \G_i ( f_1 -  f_2 ) \psi _+
\no \\
D_{\bar z} (f_1+f_2)  \psi_+ & = &  +i g_{\bar z} ^i \G_i ( f_1 +  f_2 ) \psi _-
\no \\
D_{\bar z} (f_1-f_2)   \psi _- & = & - i g_{\bar z} ^i \G_i ( f_1 -  f_2 ) \psi _+
\eea
Any single one of these equations may be retained as giving $\psi _-$ in terms
of $\psi _+$ or vice-versa. The existence of non-vanishing solutions requires
these equations to be compatible with one another. Compatibility of the first
group of two equations requires,
\bea
\label{4b3}
(f_1^2 -f_2^2)  g_z ^i g_z^i = D_z(f_1-f_2) D_z(f_1+f_2) 
\eea
while compatibility of the second group  is given by the complex
conjugate of (\ref{4b3}). Compatibility of the first group with the second group 
may be expressed as projector equations,
\bea
\label{4b4}
g_z ^i g_{\bar z} ^j \G_i  \G_j  \psi_\pm & = & N_\mp^2 \psi_\pm 
\no \\
g_{\bar z} ^i g_{ z} ^j \G_i  \G_j  \psi_\pm & = & N_\pm ^2 \psi_\pm 
\eea
where we have defined $N_\pm^2$ by, 
\bea
\label{4b5}
(f_1^2 - f_2^2) N_\pm ^2 = |D_z (f_1 \pm f_2)|^2 
\eea
Since $g_z ^i g_{\bar z} ^j \G_i  \G_j = (g_z ^i \G_i) (g_z ^i \G_i) ^\dagger$ we have 
$N_\pm ^2 \geq 0$, and as a result $f_1^2 > f_2^2$.
Assuming (\ref{4b3}) the equations in (\ref{4b4}) are equivalent to one another.
Equations (\ref{4a2}), (\ref{4a4}), and (\ref{4b2}) are equivalent to the 
algebraic equations $(m_{1,2})$ and $(p_{1,2})$.

\subsection{Solving the projector conditions}

In view of the equivalence of the equations of (\ref{4b4}), we retain only the first,
recast it in terms of the real components of $g_a^i$ using $g_4^i-i g_5^i = g_z ^i$ and 
$g_4^i + i g_5^i = g_{\bar z} ^i$, and work out the $SO(5)$-gamma matrices as follows,
\bea
\label{4c1}
\left ( g^i _a g^i _a I  + i g_4 ^i g_5^j \G^{ij} \right ) \psi_\pm = N_\mp^2 \psi_\pm 
\eea
In view of the relation,
\bea
\label{4c2}
\left (  i g_4 ^i g_5^j \G^{ij} \right ) ^2 = \Delta ^2 I
\hskip 1in \Delta ^2 \equiv  g_4^i \left ( g_4^i g_5^j  - g_4^j g_5^i \right ) g_5^j
\eea
and hermiticity of $i g_4 ^i g_5^j \G^{ij}$, the real function $\Delta^2$ is non-negative. 
Since $ i g_4 ^i g_5^j \G^{ij}$ is traceless, its eigenvalues are $+\Delta$ and $- \Delta$,
both with multiplicity 2. The case $\Delta =0$, for which $g_4^i$ and $g_5^i$ are parallel, 
does not lead to interesting solutions. Thus,  we shall assume $\Delta \not=0$.
Choosing a basis of four eigenvectors $\phi ^{(\pm )} _\sigma$, with $\sigma =1,2$,  we obtain,
\bea
\label{4c3}
\left (  i g_4 ^i g_5^j \G^{ij} \right ) \phi ^{(\pm )} _\sigma = \pm \Delta \phi ^{(\pm )} _\sigma 
\eea
we may decompose $\psi _\pm$ as follows,
\bea
\label{4c4}
\psi _\pm = \sum _{\sigma =1,2} \, \sum _{\eta=\pm} \psi _{\pm, \sigma} ^{(\eta)} \, \phi_\sigma ^{(\eta)}
\eea
This results in the following conditions on the components,
\bea
\label{4c5}
\left ( g^i _a g^i _a + \eta \Delta - N_\mp^2 \right ) \psi ^{(\eta)} _{\pm , \sigma } =0
\eea
Using (\ref{4b3}), and the definition of $N_\pm^2$ in (\ref{4b4}), we readily derive the relation 
$|g_z ^i g_z^i |^2 = N_+^2 N_-^2$, which may also be recast as follows,
\bea
\label{4c6}
(g_a^i g_a^i + \Delta ) (g_b^j g_b^j - \Delta ) =  N_+^2 N_-^2
\eea
We are now in a position to analyze and solve equations (\ref{4c5}). 

\sm

Neither $\psi _+$ nor $\psi_-$ can vanish identically, since then both would have to vanish by 
(\ref{4b2}). Without loss of generality, we may assume that at least some component 
with $\eta =+$ is non-vanishing, since this condition may always be achieved possibly
upon reversing the sign of $\Delta$, which had not been fixed yet. Thus, we have 
$g^i _a g^i _a + \Delta - N_-^2 =0$. Using the relation (\ref{4c6}), we find also
$g^i _a g^i _a - \Delta - N_+^2 =0$, so that $N_-^2-N_+^2 = 2 \Delta \not= 0$. 
As a result, we have $g^i _a g^i _a \pm \Delta - N_\pm^2 \not=0$,
so that we must have for both $\sigma =1,2$, 
\bea
\label{4c7}
\psi ^{(+)}_{-,\sigma}= \psi ^{(-)} _{+,\sigma}=0 
\eea
This result is consistent with the fact that $g_z^i \G_i$ and $g_{\bar z}^i \G_i$ anti-commute 
with $i g_4 ^i g_5^j \G^{ij} $.

\subsection{Solving the remaining differential equations}

The remaining differential equations $(az_2)$ and $(a\bar z_1)$, with $\xi$ 
eliminated in terms of $\psi $  using (\ref{4a2}), are given as follows,
\bea
\label{4e1}
(az_2) \qquad 0&=&
D_z \psi_+ - {i\over 2} \hat \omega_z \psi_+ -{1\over 4} q_z^{ij} \Gamma_{ij}\psi _+
\no \\
0&=&
D_z \psi _- - {i\over 2} \hat \omega_z \psi_- -{1\over 4} q_z^{ij} \Gamma_{ij}\psi _-
\no \\
(a\bar z _1) \qquad 0&=&
 D_{\bar z} \psi_+ + {i\over 2} \hat \omega_{\bar z} \psi_+ +  (D_{\bar z} \ln Y)  \psi_+
-{1\over 4} q_{\bar z}^{ij} \Gamma_{ij}\psi _+
\no \\
0&=&
D_{\bar z} \psi_- + {i\over 2} \hat \omega_{\bar z} \psi_- -  (D_{\bar z} \ln Y^*) \psi_-
-{1\over 4} q_{\bar z}^{ij} \Gamma_{ij}\psi _-
\eea
We shall now express these equations in the basis provided by the decomposition 
of (\ref{4c4}) with (\ref{4c7}). 

\sm

To do so, it will be convenient to make a choice of $SO(5)$ gauge. This is permitted
since the consistent formulation of the scalar fields $V$ demands gauge covariance 
of the BPS equations under local $SO(5)$ gauge transformations. The functions $g^i_a$
transform homogeneously as a 5-vector of $SO(5)$. Thus, we may choose a gauge 
in which only the first component, $i=1$, of $g_4^i$ is non-vanishing, which leaves invariance
under a residual $SO(4)$. We may use this invariance in turn to choose a gauge in
which the first two components $i=1,2$ of $g_5^i$ are non-vanishing. Actually, it 
will turn out to be slightly more convenient to leave the first two components, $i=1,2$,
of $g^i_a$ non-zero, leaving over a residual $SO(2)\times SO(3)$ gauge freedom.
We note that, generally in a theory with holomorphic and harmonic functions, such 
gauge choices may upset holomorphicity and harmonicity, but here this will not be the case. 

\sm

The presence of the residual gauge symmetry $SO(2)\times SO(3)$ forces the 
reduced BPS equations in this gauge to have this symmetry manifest. Thus, 
in this gauge, the residual BPS equations must be invariant under the $SO(2)$ 
generator $\G^{12}$. As a result of this invariance, we find that
the $SO(5)$-connection is consistently reduced to an $SO(2)\times SO(3)$-connection,
\bea
\label{4e2}
q_z^{13}=q_z ^{14}=q_z^{15}= q_z^{23}=q_z ^{24}=q_z^{25}= 0
\eea
Following the conventions of Appendix A, the $SO(2)\times SO(3)$ generators 
take the form, 
\bea
\label{4e3}
\G^{12} = i \sigma _3 \otimes I
& \hskip 1in &
\G^{34} = I \otimes i \sigma _3
\no \\ &&
\G^{45} = I \otimes i \sigma _1
\no \\ &&
\G^{53} = I \otimes i \sigma _2
\eea
With this gauge choice, and the sign choice $\Delta = -g_4 ^1 g_5^2 + g_4^2 g_5^1$
in (\ref{4c3}), we find that the basis of $\phi ^{(\eta)} _{\pm , \sigma}$, namely,
\bea
\label{4e4}
(\sigma _3 \otimes I ) \phi ^{(\eta)} _\sigma = \eta \phi ^{(\eta )} _\sigma
\eea
so that we have $\psi _{+,\a}^{(-)} = \psi ^{(+)}_{-,\a}  = 0$.
In this basis, the remaining equations in (\ref{4e1}) decompose as follows,
\bea
\label{4e5}
(az_2) \qquad 0&=&
\left ( D_z - {i\over 2} \hat \omega_z - {i \over 2} q_z^{12} -  \cA_z \right )\psi_+^{(+)} 
\no \\
0&=&
\left ( D_z - {i\over 2} \hat \omega_z +  {i \over 2} q_z^{12} - \cA_z \right ) \psi_-^{(-)} 
\no \\
(a\bar z _1) \qquad 0&=&
\left ( D_{\bar z} + {i\over 2} \hat \omega_{\bar z} 
 -{i \over 2} q_{\bar z}^{12} - \cA_{\bar z} + D_{\bar z} \ln Y \right ) \psi_+^{(+)}
\no \\
 0&=&
\left ( D_{\bar z} + {i\over 2} \hat \omega_{\bar z} 
+ {i \over 2} q_{\bar z}^{12} - \cA_{\bar z} - D_{\bar z} \ln Y^*  \right ) \psi_-^{(-)}
\eea
where we have denoted the remaining $SO(3)$-connection by
\bea
\label{4e6}
\cA_a = { i \over 2} \left ( q_a^{34} \s^3 + q_a^{45} \s^1 + q_a^{53} \s^2 \right )
\eea
From equation (\ref{4b2}), we have a relation between $\psi_+^{(+)}$ and $\psi_-^{(-)}$ given by,
\bea
\label{4e7}
 \psi _-^{(-)} = X \psi _+^{(+)}
 \hskip 1in
 X \equiv -  i (g^1_z + i g^2_z) { f_1 -  f_2 \over  D_z(f_1+f_2)}
\eea
Eliminating $\psi_-^{(-)}$ in equations (\ref{4e5}) in favor of $\psi _+^{(+)}$ and $X$, 
and taking the sum or difference between similar equations, we find the differential equations,
\bea
\label{4e8}
0 &=&
\left ( D_z - {i\over 2} \hat \omega_z -{i \over 2} q_z^{12} - \cA_z \right ) \psi_+^{(+)}  
\no \\
0 &=&
\left ( D_{\bar z} + {i\over 2} \hat \omega_{\bar z} 
-{i \over 2} q_{\bar z}^{12} - \cA_{\bar z}+ D_{\bar z} \ln Y \right ) \psi_+^{(+)} 
\eea
as well as the equations,
\bea
\label{4e9}
q_z^{12}  &=& + i \, D_z \ln X
\no \\
q_{\bar z} ^{12} &=& -i \, D_{\bar z} \ln \bar X 
\eea
In order to solve these equations, it is useful to rewrite the $D_{\bar z} Y$ 
term in equation (\ref{4e8}) using the relation (\ref{4a4}). We obtain
\bea
\label{4e10}
D_{\bar z} \ln Y =  
{1 \over 2} D_{\bar z} \ln{f_1-f_2 \over f_1 + f_2} + {1 \over 2} D_{\bar z} \ln {c( w ) \over \bar c(\bar w)} 
\eea
At this point, it is helpful to introduce the notation
\bea 
\label{4e11}
\Psi_+^{(+)} = 
\left\{ { i \rho X \over  c(\bar w)}  \right\}^{1 \over 2} \psi^{(+)}_{+}
\eea
Equations (\ref{4e8})  now assume a particularly simple form,
\bea 
\label{4e12}
(\partial_w -   \cA_w) \Psi^{(+)}_+ &=& 0 
\no \\
(\partial_{\bar w} -  \cA_{ \bar w}) \Psi^{(+)}_+ &=& 0  
\eea
where we have set $\cA_w=\rho \cA_z$ and $\cA_{\bar w} = \rho \cA_{\bar z}$.
These equations pose no further restriction on the number of independent 
components of $ \Psi^{(+)}_+$ provided that the integrability condition
\bea 
\label{4e13}
\partial_{w} \cA_{\bar w} - \partial_{\bar w} \cA_w 
-   \cA_w \cA_{\bar w} + \cA_{\bar w} \cA_w    = 0  
\eea
is obeyed. Hence, $\mathcal{A}$ needs to be pure gauge and there must exist an $SO(3)$
gauge transformation such that $\mathcal{A}= 0$ or, equivalently,
\be 
q^{34}_z= q^{45}_z=q^{53}_z=0 
\ee
In other words, barring global issues, it is possible to find a gauge in which 
all the components of $q^{ij}_z$ are identically zero, with the exception of
$q^{12}_z$. In this gauge, $\Psi^{(+)}_+$ is a constant spinor.

\subsection{Solving for $p^{ir}_a$ and $\tilde g^r_a$}

 The  equations  $(c_1)$  and $(c_2)$ in (\ref{3d5}) can be simplified by eliminating $\xi$ 
 in favor of $\psi$ of $Y$, using equations (\ref{4a2a}), and we find, 
\bea
\label{4f1}
(c_1) \qquad 0 & = &{1\over \sqrt{2}} p_{z}^{ir}\Gamma_{i} \psi_+  + i  \tilde g_{z}^{r} \psi_-
\no \\
0 & = &{1\over \sqrt{2}} p_{z}^{ir}\Gamma_{i} \psi_-  - i  \tilde g_{z}^{r} \psi_+
\no \\
(c_2) \qquad  0 & = &{|Y|^2\over \sqrt{2}} p_{\bar z}^{ir}\Gamma_{i} \psi _+
+ i \tilde g_{\bar z}^{r}  \psi _-
\no \\
0 & = &{1\over \sqrt{2}} p_{\bar z}^{ir}\Gamma_{i} \psi _-
- i |Y|^2 \tilde g_{\bar z}^{r}  \psi _+
\eea
Next, we use the decomposition of (\ref{4c4}) with the projector conditions of (\ref{4c7}),
and the explicit characterization (\ref{4e4}) of the basis spinors $\phi ^{(\pm)} _\sigma$,
to analyze (\ref{4f1}). Clearly, the equations decompose onto $\phi ^{(\pm)} _\sigma$ as follows,
\bea
\label{4f2}
\phi ^{(+)} _\sigma & \hskip 0.8in & \psi _+, ~ \G^3 \psi _+,  ~ \G^4 \psi _+,  ~ \G^5 \psi _+,  ~ \G^1 \psi _-, ~ \G^2 \psi _-
\no \\
\phi ^{(-)} _\sigma &  \hskip 0.8in & \psi _-, ~ \G^3 \psi _-,  ~ \G^4 \psi _-,  ~ \G^5 \psi _-,  ~ \G^1 \psi _+, ~ \G^2 \psi _+
\eea
As a result, we immediately have,
\bea
\label{4f3}
0 & = & ( p_{z}^{3r}\s_{1} + p_{z}^{4r}\s_{2} + p_{z}^{5r}\s_{3}) \psi_+^{(+)}
\no \\
0 & = & ( p_{\bar z }^{3r}\s_{1} + p_{\bar z}^{4r}\s_{2} + p_{\bar z}^{5r}\s_{3}) \psi_+^{(+)}
\eea
Since $\psi ^{(+)}_+ \not= 0$, it follows that 
\bea
\label{4f4}
 p_{z}^{3r} =  p_{z}^{4r} = p_{z}^{5r} = 0
\eea
In the remaining equations, we eliminate $\psi_-$ in favor of  $\psi_+$ and $X$, using (\ref{4e7}),
resulting in a set of four equations which are all proportional to $\psi ^{(+)}_+$. Again, since this
quantity is non-vanishing, it may be omitted, and we obtain the following equations on the coefficients,
\bea
\label{4f5}
(c_1) \qquad 0 & = & {1\over \sqrt 2}(p_{z}^{1r} + i p_{z}^{2r})  + i \tilde g^r_z X 
\no \\
0 & = & {1\over \sqrt 2}(p_{z}^{1r} - i p_{z}^{2r}) X  - i \tilde g^r_z  
\no \\
(c_2) \qquad 0 & = & {|Y|^2\over \sqrt 2}(p_{\bar z}^{1r} + i p_{\bar z}^{2r})  + i \tilde g^r_{\bar z} X 
\no \\
0 & = & {1\over \sqrt 2}(p_{\bar z}^{1r} - i p_{\bar z}^{2r}) X  - i |Y|^2 \tilde g^r_{\bar z}  
\eea
Compatibility of the first and the fourth equations of (\ref{4f5}) requires
\bea
\label{4f6}
|X|^2 = |Y|^2
\eea
which in turn implies the consistency of the second and third equations.
The remaining equations give $p_z ^{1r}$ and $p_z ^{2r}$ in terms of $\tilde g^r _z$,
and may be summarized as follows,
\bea
\label{4f7}
{1 \over \sqrt{2} } \left ( p^{1r} _z + i \alpha  p^{2r} _z \right ) = - i \a X^\a \tilde g^r _z \hskip 1in \a = +,-
\eea
which concludes the complete parametrization of the solutions to the BPS equations,
in terms of a single complex function $X$. A summary of the solution will be the 
starting point of the subsequent section.

\newpage

\section{Constructing the local solution}
\setcounter{equation}{0}
\label{sec5}

We begin by summarizing the complete solution to the BPS equations, obtained in the previous section.
Combining these results with the Bianchi identities, and the general structure of the scalar field equations, 
we produce the complete solution to all the BPS, Bianchi and field equations. We will work for a general 
number of tensor multiplets $m$.

\subsection{Summary of the solutions to the BPS equations}

Given the solutions to the BPS equations with the gauge choice $g^i_a=0$ for $i=3,4,5$, 
and the resulting solutions
$p^{ir}_a=0$ for $i=3,4,5$, the basis of frames takes the simplified form,
\bea
\label{7a1}
D_a V \, V^{-1} = \left ( \matrix{
q_a^{ij} & 0 &  \sqrt{2} p_a^{is}  \cr
0 & 0 &  0  \cr
\sqrt{2} p_a^{jr} & 0 &  s_a^{rs}  \cr}\right )
\eea
where $i,j=1,2$ and $r,s=6,\cdots, m+5$. The block $q^{ij}_a$ generates
an $SO(2)$-connection, namely $SO(2)_q$, which we shall
abbreviate as follows,
\bea
\label{7a2}
q_z = q^{12} _z = + i D_z \ln X
\no \\
q_{\bar z} = q_{\bar z} ^{12} = - i D_{\bar z} \ln \bar X
\eea
As a result, the frame $V$ itself takes the reduced form, 
\bea
\label{7a3}
V = \left ( \matrix{ 
V^i {}_I & 0 & V^i {}_R \cr
0 & I_3 & 0 \cr
V^r {}_I & 0 & V^r {}_R \cr } \right )
\eea
where $i,I=1,2$, and $r,R=6, \cdots, m+5$. The block $I_3$ represents the identity in the 
indices $i,I=3,4,5$. The defining property for $SO(5,m)$ is $V^{-1} = \eta V^t \eta$
where $\eta = {\rm diag} (I_5 , -I_m)$. It will be useful to write in components both 
equations $VV^{-1} = V^{-1} V=I$,
\bea
\label{7a5}
V^i{}_I \, V^j{}_I - V^i {}_R \, V^j{}_R = \delta ^{ij} & \hskip 1in & V^i{}_I \, V^i{}_J - V^r{}_I \, V^r{}_J = \delta _{IJ}
\no \\
V^i{}_R \, V^r{}_R - V^i{}_I \, V^r{}_I =0 ~ ~& & V^i{}_I \, V^i{}_R - V^r{}_I \,  V^r{}_R =0
\no \\
V^r {}_R \, V^s{}_R - V^r{}_I \, V^s{}_R = \delta ^{rs} && V^r{}_R \, V^r{}_S - V^i{}_R \,  V^i {}_S = \delta _{RS}
\eea
Summation over repeated indices is assumed here; specifically over the indices $I=1,2$ and $R=6, \cdots, m+5$ 
in the left column, and over the indices $i=1,2$ and $r=6, \cdots, m+5$ in the right column.
Using the same conventions, the expressions for $q,p,s$ are found to be, 
\bea
\label{7a6}
\sqrt{2} p_a^{ir}  & = &  (D_a V^i{}_R) V^r{}_R - (D_a V^i{}_I ) V^r{}_I 
\no \\
q_a ^{ij} & = & (D_a V^i{}_I ) V^j {}_I  - (D_a V^i {}_R ) V^j {}_R
\no \\
s_a^{rs} & = & (D_a V^r{}_R) V^s{}_R - (D_a V^r{}_I ) V^s{}_I
\eea
It will be natural to decompose all fields in a basis in which the $SO(2)_q$ action
is diagonal. To this end, we introduce with $\alpha = +,-$, 
\bea
\label{7a7}
g _z ^\a = { 1 \over \sqrt{2} } \left ( g^1 _z + i \a g_z^2 \right ) ~~
& \hskip 1in &
g _{\bar z} ^\a = { 1 \over \sqrt{2} } \left ( g^1 _{\bar z} + i \a g_{\bar z}^2 \right )
\\
\label{7a8}
p_z ^{\a r}  =  {1\over \sqrt{2}} \left ( p_z ^{1r} + i \a p_z ^{2r}  \right )
& \hskip 1in &
p_{\bar z} ^{\a r}  =  {1\over \sqrt{2}} \left ( p_{\bar z} ^{1r} + i \a p_{\bar z} ^{2r} \right )
\eea
As a result of the BPS equations, the fluxes $g_a^\a$ may then be parametrized by the field $X$,
and the radii $f_1$ and $f_2$, as follows,
\bea
\label{7a9}
g_z^+ =  i { X \over \sqrt{2} } \, { D_z (f_1+f_2) \over f_1-f_2}
& \hskip 1in & 
g_z^- = - i { \bar X \over \sqrt{2} } \, { D_z (f_1-f_2) \over f_1-f_2}
\no \\
g_{\bar z}^+ =  i  { X \over \sqrt{2} } \, { D_{\bar z} (f_1-f_2) \over f_1-f_2}
&&
g_{\bar z}^- = - i { \bar X \over \sqrt{2} } \, { D_{\bar z} (f_1+f_2) \over f_1-f_2}
\eea
As a result of (\ref{4a4}), (\ref{4a5}), and (\ref{4f6}), the radii $f_1$, $f_2$, the metric $\rho^2$, 
and the field $X$ are related as follows,
\bea
\label{7a10}
f_1 f_2 =H 
\hskip 0.8in 
|X|^2 = { f_1 - f_2 \over f_1 + f_2}  
\hskip 0.8in 
\rho ^2 (f_1^2-f_2^2) = |\p_w H|^2
\eea
By the solution to the BPS equations of (\ref{4f7}), the functions $p_a ^{ir}$ and $\tilde g _a ^r$ are 
related as follows,
\bea
\label{7a11}
p_z ^{\a r} & = & - i \a X^\a \ti g^r _z
\no \\
p_{\bar z} ^{\a r} & = & - i \a \bar X ^{-\a} \ti g^r _{\bar z} \hskip 1in \a  = +,-
\eea

\subsection{Solving for $V$ and $|X|$}

We begin by eliminating $\tilde g^{r}_a$ between $p^{\pm r}_a$, and find,
\bea
\label{7b1}
p_z^{+r} = - X^2 p_z ^{-r}
\eea
together with its complex conjugate. Next, we write out this equation in terms of
the frame variables, and we find, 
\bea
\label{7b2}
- (D_z V^+{}_I) V^r{}_I + (D_z V^+{}_R) V^r{}_R
=  X^2 \bigg (  (D_z V^-{}_I) V^r{}_I - (D_z V^-{}_R) V^r{}_R \bigg )
\eea

\subsubsection{Solving for $V^i{}_I$}

Contracting (\ref{7b2})  by $V^r{}_J$, and using the top two relations on the right of  
(\ref{7a5}),  we find, 
\bea
\label{7b4}
&&
D_z V^+ {}_I - \left [ (D_z V^+{}_J) V^-{}_J - (D_z V^+{}_R) V^-{}_R \right ] V^+{}_I
\no \\ && \hskip 0.5in
= - X^2 \left  ( D_z V^-{}_I  - \left [ (D_z V^-{}_J ) V^+{}_J - (D_z V^-{}_R) V^+{}_R \right ] V^-{}_I \right )
\eea
Recognizing the following combinations, 
\bea
\label{7b5}
q^{+-} _z & = & (D_z V^+{}_J) V^-{}_J - (D_z V^+{}_R) V^-{}_R
\no \\
q^{-+} _z & = & (D_z V^-{}_J) V^+{}_J - (D_z V^-{}_R) V^+{}_R
\no \\
q_z ^{+-} & = & - q^{-+}_z = - i q_z ^{12} = D_z \ln X
\eea
equation (\ref{7b4}) takes the following form,
\bea
\label{7b6}
D_z V^+{}_I - (D_z \ln X) V^+{}_I
= - X^2 \left  ( D_z V^-{}_I + (D_z \ln X ) V^-{}_I \right )
\eea
which may also be rewritten (in components) as, 
\bea
\label{7b7}
D_z \left ( { V^+{}_I \over X} + X V^-{}_I \right ) =0
\eea
These are Cauchy-Riemann equations, and are solved by a pair of arbitrary holomorphic 
functions $\lambda _I$, with $I=1,2$. In local complex coordinates $w, \bar w$ on $\Sigma$,
the Cauchy-Riemann equations simply read $\p_{\bar w} \lambda _I=0$.  As  a result, we have,
\bea
\label{7b8}
{ V^+{}_I \over X} + X V^-{}_I  & = & \bar \lambda _I
\no \\
{ V^-{}_I \over \bar X} + \bar X V^+{}_I  & = & \lambda _I
\eea
They are solved by,
\bea
\label{7b9}
\left ( 1 - |X|^4 \right ) V^+{}_I & = & X \bar \lambda _I - X |X|^2 \lambda _I
\no \\
\left ( 1 - |X|^4 \right ) V^-{}_I & = & \bar X \lambda _I - \bar X |X|^2 \bar \lambda _I
\eea

\subsubsection{Solving for $V^i{}_R$}

Next, we contract (\ref{7b2})  by $V^r{}_S$, use the bottom two relations
in the right column of (\ref{7a5}), and employ
the combinations of (\ref{7b4}) to recast the equations in the following form,
\bea
\label{7b10}
D_z \left ( { V^+{}_R \over X} + X V^-{}_R \right ) =0
\eea
which are solved by  further  arbitrary holomorphic functions $\lambda _R$, with $R=6, \cdots, m+5$,
\bea
\label{7b11}
{ V^+{}_R \over X} + X V^-{}_R = \bar \lambda _R
\eea
giving rise to the following explicit solutions, 
\bea
\label{7b12}
\left ( 1 - |X|^4 \right ) V^+{} _R & = & X \bar \lambda _R - X|X|^2 \lambda _R
\no \\
\left ( 1 - |X|^4 \right ) V^-{}_R & = & \bar X \lambda _R - \bar X |X|^2 \bar \lambda _R
\eea

\subsubsection{Solving for $|X|$}

Having now obtained explicit solutions for the fields $V^\pm {}_I$ and $V^\pm {}_R$
in terms of holomorphic functions $\l_I, \l_R$, their complex conjugates, and the 
function $X$, we may use the top left relation of (\ref{7a5}) to determine $|X|$.
In $\pm$ components, this equation amounts to one complex and one real equation, 
respectively given by,
\bea
\label{7b13}
V^+{}_I V^+ {}_I - V^+{}_R V^+{}_R & = & 0
\no \\
V^+{}_I V^- {}_I - V^+{}_R V^-{}_R & = & 1
\eea
Using the explicit solutions of (\ref{7b9}) and (\ref{7b12}), the first equation becomes,
\bea
\label{7b14}
\bar \l  \cdot  \bar \l + |X|^4 \l \cdot  \l - 2 |X|^2 \l \cdot \bar \l  =0
\eea
where we have omitted overall factors of powers of $(1-|X|^4)$, since they are non-vanishing,
and the functions $\l^A, \bar \l^A$ are contracted with the $SO(2,m)$-invariant metric $\eta$.
Vanishing of the imaginary part of (\ref{7b14}) implies, $\Im ( \l \cdot \l) =0$.
Since $\l \cdot \l$ is holomorphic, this can only be realized by having
this quantity be a real constant $b$, so that $\l \cdot \l = b$.
The real part of (\ref{7b14}) and the second equation of (\ref{7b13}) then become respectively,
\bea
\label{7b18}
b (1 +|X|^4) - 2 |X|^2 \l \cdot \bar \l & = & 0
\no \\
(1 +|X|^4) |X|^2 \l \cdot \bar \l - 2 b |X|^4 & = & (1 -|X|^4)^2
\eea
Eliminating $\l \cdot \bar \l$ between the equations of (\ref{7b18}) gives a relation involving 
only $|X|$, which can hold for non-trivial $|X|$ if and only if $b=2$. As a result, 
$|X|$ is determined in terms of $m+2$ holomorphic functions $\l_A$ as follows,
\bea
\label{7b21}
\l \cdot \l  & = & 2
\no \\
\l \cdot \bar \l  & = & |X|^2 + |X|^{-2} 
\eea
By construction of the local solution, $|X|$ is real and positive and hence $\l \cdot \bar \l \geq 2$.

\subsubsection{Solving for  $V^r{}_I$ and $V^r{}_R$}

The relations  (\ref{7a5}) between the scalar fields allow us to solve for 
$V^r{}_A$ in terms of $V^i{}_A$ via smooth relations. By first diagonalizing the real
symmetric matrix $V^r{}_R V^r {}_S=\delta _{RS} + V^i{}_R V^i{}_S$ by an 
orthogonal matrix $M_1$ and a real diagonal matrix $D$, we find
\bea
\label{7f18}
V^r{}_R & = & (M_2)^r{}_s D^s{}_S (M_1)^S{}_R
\no \\
V^r{}_I & = & (M_2)^r{}_s (D^{-1})^s {}_R (M_1)^R{}_S V^i {}_T V^j{}_I \delta ^{ST}\delta _{ij}
\eea  
where $M_2$ is a second orthogonal matrix, and where summation is implied for any pair
of repeated raised and lowered indices.

\subsection{Calculating the flux potentials}

Conserved charges arise from integrating the closed three-form fields $G^A$ of (\ref{2b1})
over closed compact three-cycles, such as the asymptotic $S^3$ components
of the boundary. The field $G^A$  may be obtained by inverting relations of (\ref{2b2}),
and expressing $H^i$ and $H^r$ in terms of (\ref{3a5}), 
\bea
\label{7d1}
\eta _{AB} G^B = 
+ \left ( V^i {}_A g^i _a - V^r{}_A \tilde g^r _a \right ) f_1 ^2 \hat e^{01} e^a
+ \left ( V^i {}_A h^i _a - V^r{}_A \tilde h^r _a \right ) f_2 ^2 \hat e^{23} e^a
\eea
The forms $\hat e^{01}$ and $\hat e^{23}$ are the unit volume forms respectively 
on $AdS_2$ and $S^2$. They are real and closed by construction, so that 
$G^A$ may be recast in terms of derivatives of real-valued flux potentials $\Phi ^A$ 
and $ \Psi ^A$, which are functions of $\Sigma $ only, 
\bea
\label{7d2}
G^A =  d  \Psi ^A \wedge \hat e^{01} + d \Phi ^A \wedge \hat e^{23} 
\eea
The conserved charges of interest here arise from the contributions above 
which are proportional to the volume form $\hat e^{23}$ on $S^2$. Since 
the volume form $\hat e^{01}$ assigns infinite volume to $AdS_2$, its
contributions does not lead to finite conserved charges. Thus, we focus
on the behavior of $\Phi ^A$, which may be determined from the differential 
equations,  
\bea
\label{7d4}
\eta _{AB} D_z \Phi ^B = -i f_2^2 \left ( V^i{}_A \, g^i _z + V^r {}_A \, \tilde g^r_z \right ) 
\eea
where we have used  (\ref{3a6}) to re-express $h^i_a$ and $\tilde h^r _a$ 
in terms of $g^i_a$ and $\tilde g^r _a$ respectively. 

\sm

We now seek to solve for $\Phi ^A$ in terms of the harmonic function $H$, the 
holomorphic functions $\l_A$ and their complex conjugates. To do so, 
we express the components of $g^i_z$ given in (\ref{7a9}) in terms of $H$ and $|X|$ 
using the following formulas,
\bea
\label{7d5}
(f_1+f_2)^2 & = & { 4 H \over 1 - |X|^4} 
\no \\
(f_1-f_2)^2 & = & { 4 H |X|^4 \over 1 - |X|^4} 
\eea
As a result, we have 
\bea
\label{7d6}
g^+ _z & = & + {i X \over 2 \sqrt{2} |X|^2 } D_z \ln \left ( { 4 H \over 1 - |X|^4}  \right )
\no \\
g^- _z & = & - {i \bar X \over 2 \sqrt{2} } D_z \ln \left ( { 4 H |X|^4 \over 1 - |X|^4}  \right )
\eea
To compute the flux potentials, only the general relations obeyed by
$V^r{}_A$ will be needed, but not their explicit forms of (\ref{7f18}).
We compute  $V^r {}_A \, \tilde g^r_z$ using (\ref{7a11}) and (\ref{7a6}),
\bea
\label{7d7}
V^r {}_A \, \tilde g^r_z & = & {i \over X} V^r{}_A \, p^{+r}_z=
{ i \over \sqrt{2} X} \left ( - D_z V^+{}_J \, V^r{}_A  V^r{}_J + D_z V^+{}_R \, V^r{}_A V^r{}_R \right )
\eea
We now make use of the $SO(2,m)$ relations in the right column of (\ref{7a5}).
Recognizing again the combinations of (\ref{7b5}), we obtain, 
\bea
\label{7d9}
V^r{}_A \, \tilde g^r _z = { i \over \sqrt{2}} D_z \left ( { V^+{}_A \over X} \right )
\eea
Using  (\ref{7d4}), (\ref{7d6}), (\ref{7d9}), and the expressions for $V^i_I$ and $V^i_R$, 
(\ref{7b9})  and (\ref{7b12}),  we obtain a formula for the derivative of the potential 
$\Phi^A= \eta ^{AB} \Phi _B$ in terms of $\l_A$, $H$ and $|X|$, 
\bea
\label{7d10}
D_z \Phi _A  =
D_z \left (  - {1 \over \sqrt{2} } \, {H |X|^2 (\l_A + \bar \l_A) \over (1+|X|^2)^2}  \right ) 
+ { 1 \over 2 \sqrt{2} } \l_A D_z H
\eea
Next, we use the second  formula of (\ref{7b21}) to express $|X|$ in terms of  
$\l \cdot \bar \l $, and to obtain an  expression for $D_z \Phi^A$ in terms of only 
$\l_A, \bar \l_A$ and $H$.  Integrating this equation, we find the 
following final form for the flux potentials, 
\bea
\label{7d12}
\Phi _A = \tilde \Phi _A  -  \sqrt{2} \, { H \, \Re (\l_A) \over \l \cdot \bar \l +2} 
\hskip 1in 
\tilde \Phi_A = { 1 \over 2 \sqrt{2} } \int  \l_A \p_w H   +{\rm c.c.}
\eea
Note that the contributions $\tilde \Phi _A$ are real harmonic functions on $\Sigma$.
An integration constant may be added to each $\tilde \Phi _A$, 
which amounts to a gauge transformation on the two-form field $B^A$, and does not 
affect the field strengths $G^A$, or $H^i, H^r$.

\subsection{Summary of the local solution}
\label{sec54}

The data of the general local solution are the real harmonic function $H$, and 
an $SO(2,m)$ vector of (locally) holomorphic functions $\l_A$ which must satisfy the 
following conditions,
\bea
\label{7c0}
\l \cdot \l = 2 \hskip 1in  \l \cdot \bar \l  \geq  2
\eea
Combining the formulas for the radii $f_1, f_2$ and worldsheet metric $\rho^2$ of (\ref{7a10})
with the expression for $|X|$ in terms the harmonic function $H$ and the holomorphic 
functions $\l_A$, we find the following explicit expressions for the radii and metric, 
\bea
\label{7c1}
f_1 ^4 & = & H^2 ~ { \l \cdot \bar \l +2 \over \l \cdot \bar \l -2} 
\no \\
f_2 ^4 & = & H^2 ~ { \l \cdot \bar \l -2 \over \l \cdot \bar \l +2} 
\no \\
\rho^4 & = & {|\p_w H|^4 \over 16 H^2} \Big ( \l \cdot \bar \l -2 \Big ) \Big ( \l \cdot \bar \l +2 \Big )
\eea
Note that the quantity $\l \cdot \bar \l$ has indefinite metric.

\sm

The solution for the scalar fields may be derived from eliminating $|X|$
in favor of $\l \cdot \bar \l = |X|^2 + |X|^{-2}$ from (\ref{7b9}) and (\ref{7b12}).
Overall phases remain in view of the $SO(2)$ gauge non-invariance of the
scalar fields. Thus we have, 
\bea
\label{7c5}
V^+{}_A = X \left ( \bar \lambda _A - |X|^2 \lambda _A \right ) / \left ( 1 - |X|^4 \right ) 
\eea
together with its complex conjugate. Expressions for $V^r{}_A$ may be derived using (\ref{7f18}). 

\sm

The solutions for the three-form field $G^A=dB^A$ may be expressed in terms of real-valued 
flux potential functions $\Phi ^A$ and $\Psi ^A$ by (\ref{7d2}), via the relations,
\bea
\label{7c2}
B^A = \Psi ^A \hat e^{01} + \Phi ^A \hat e^{23}
\eea
where the flux potential functions for the solution are given by,
\bea
\label{7c3}
\Phi ^A =  -  \sqrt{2} \, { H \, \Re (\l^A) \over \l \cdot \bar \l +2} + \tilde \Phi ^A 
& \hskip 0.6in &
\tilde \Phi_A = { 1 \over 2 \sqrt{2} } \int  \l_A \p_w H   +{\rm c.c.}
\no \\
\Psi ^A =  -  \sqrt{2} \, { H \, \Im (\l^A) \over \l \cdot \bar \l - 2} + \tilde \Psi ^A 
& \hskip 0.6in &
\tilde \Psi_A = { i \over 2 \sqrt{2} } \int  \l_A \p_w H   +{\rm c.c.}
\eea
If a closed homology three-cycle $\cS$ in the solution manifold (typically a sphere $S^3$) is parametrized 
as a two-sphere $S^2$ warped over an interval $\cI$ in $\Sigma$, then the conserved Page
charge (see for example \cite{Marolf:2000cb}) vector $Q_\cS^A$ through the cycle $\cS$ is given by,
\bea
\label{7c4}
Q_\cS^A  = \oint _\cS G^A =  \sqrt{2} \pi \int  _{\cI} \l^A \p_w H   +{\rm c.c.}
\eea
where we have used  $\oint _{S^2} \hat e^{23} = 4 \pi$.
Generally, the interval $\cI$ need  not be closed on $\Sigma$.

\sm

Finally, we have verified that the local solution obtained here solves the  Bianchi identities 
which have not been solved explicitly in this section,  as well as the field equations. 
The detailed calculations may be found in \cite{guothesis}.

\newpage

\section{Topology and regularity assumptions}
\setcounter{equation}{0}
\label{sec6}

In the preceding section, we have obtained the complete and most general local half-BPS
solution to six-dimensional $(0,4)$ supergravity with $SO(2,1) \times SO(3)$ symmetry.
The solutions are parametrized by a real harmonic function $H$, and $m+2$ 
holomorphic functions $\lambda _I, \l_R$ on a Riemann surface $\Sigma $ with boundary,
subject to the conditions,
\bea
\label{6a1}
\l \cdot \l = \delta _{IJ} \l^I \l^J - \delta _{RS} \l^R \l^S & = & 2 \hskip 1in I,J=1,2
\no \\
\l \cdot \bar \l = \delta _{IJ} \l ^I \bar \l^J - \delta _{RS} \l^R \bar \l^S & \geq  & 2 \hskip 1in R,S=6, \cdots, m+5
\eea
The expressions for the metric factors $f_1, f_2$, $\rho$, the scalars, and for the flux potentials 
$\Phi^A, \Psi ^A$ were given in terms of these functions in (\ref{7c1}), (\ref{7c2}), and (\ref{7c3}).
In this section, we spell out the topology and regularity conditions which we shall
impose to obtain regular global solutions with prescribed asymptotic behavior.

\sm

Motivated by holographic applications in the study of junctions of two-dimensional CFTs, 
we restrict the topology of solutions to having $N$ 
distinct asymptotic regions each of which is homeomorphic to $AdS_3 \times S^3$ (namely 
a maximally symmetric asymptotic solution supported by non-vanishing three-form flux). 
This condition is the direct six-dimensional analogue of the restriction to 
locally asymptotically $AdS_5 \times S^5$ for ten-dimensional Type IIB.
It is further motivated by the study in \cite{Chiodaroli:2009yw, Chiodaroli:2009xh, Chiodaroli:2010mv}
of Type IIB on locally asymptotically  $AdS_3 \times S^3 \times K3$.

\sm

However, we know from \cite{Chiodaroli:2009xh} that there exist regular
solutions with different asymptotics. At the same time, it may
also be possible to find a physical interpretation for some of the
singular solutions. The study of  solutions with different topologies
and asymptotic regions remains an open and interesting problem, which we
defer to future work.

\subsection{Basic regularity assumptions}
\label{sec61}

Motivated by the locally asymptotically $AdS_3 \times S^3$ structure of the solutions,
we impose the following basic regularity assumptions on the physical fields needed to achieve 
regularity under the assumption of this topological structure. 

\begin{description}
\item{(1)}
The physical data of the solution, namely the metric factors $f_1, f_2, \rho$, 
the scalars $V^i{}_R$, and the reduced field strengths $d\Phi^A, d\Psi ^A$ 
are regular and single-valued inside $\Sigma$. 
\item{(2)} 
These physical data  extend  smoothly to the boundary $\p \Sigma$, except 
at isolated points where they may diverge to produce a locally asymptotically 
$AdS_3 \times S^3$ region.
\item{(3)} 
The condition $f_1>0$ holds throughout  $\Sigma$ and on the boundary 
$\p \Sigma$. 
\item{(4)} 
The condition $f_2>0$ holds throughout  the interior of $\Sigma$. \\
The condition $f_2=0$ defines the position of the boundary  $\p \Sigma$.
\end{description}

For simplicity, we shall assume here that the boundary $\p \Sigma$ 
consists of a single  connected component, though we know that this restriction 
may easily be lifted, as was done in \cite{Chiodaroli:2009xh}.

\subsection{Implications on the (locally) holomorphic data}
\label{sec62}

The above assumptions on topology and regularity on the physical fields have 
simple implications on the (locally) holomorphic and harmonic data in terms of 
which the complete local solutions of the preceding section have been expressed. 
We now spell out these implications.  
\begin{description}
\item{(a)} 
The real harmonic function $H$ is smooth, single-valued, and satisfies $H>0$ inside $\Sigma$.
\item{(b)}  The condition $H=0$ gives an equivalent definition of the position of the boundary $\p \Sigma$. 
\item{(c)} The condition $\l \cdot \bar \l > 2$ holds inside $\Sigma$.
\item{(d)}  The relation $\Im (\l_A)=0$ holds on $\p \Sigma$, except at isolated points.
\item{(e)} The one-forms $\Lambda _A \equiv \l_A \p H$ are holomorphic
and single-valued inside $\Sigma$. They extend smoothly to $\p \Sigma$,
except at isolated points.
\end{description}

Points (a) and (b) follow from the relation $f_1 f_2 =H$ and points (1), (2), (3) and (4) of 
Section \ref{sec61}, while point (c) results upon the use of equation (\ref{7c1}).
Point (d) follows from the regularity of $f_1$, which via the first
relation of (5.37) requires that $\lambda \cdot \bar \lambda -2$
must vanish at the same rate as $H^2$ does at the boundary. Regularity 
of the flux potential $\Psi ^A$ in (5.40) then requires that $\Im \lambda ^A=0$
at the boundary. Point (e) follows from point (1) of Section \ref{sec61} for
the flux field strengths, and their resulting equations (\ref{7c3}) from which
it is manifest that the form $\l _A \p_w H$ must be smooth inside $\Sigma$.

\sm 

Concretely, in view of the restriction to a boundary with a single connected component
made in this paper for simplicity's sake, we shall parametrize $\Sigma$ by the upper half 
complex plane, whose boundary $\p \Sigma$ is the real line. Equivalently, we may
conformally map the upper half-plane to the unit-disk, whose boundary is the unit circle.

\newpage

\section{Regular locally asymptotically $AdS_3 \times S^3$ solutions}
\setcounter{equation}{0}
\label{sec7}

In this section, we shall produce the general form of the harmonic function $H$ and 
the holomorphic one-forms $\Lambda^A$ which will solve the topology and regularity 
conditions of Section~\ref{sec6} required for regular 
solutions with $N$ distinct  $AdS_3 \times S^3$ asymptotic regions. We shall 
parametrize $\Sigma$ by the upper half complex plane, and its boundary $\p \Sigma$
by the real line.

\subsection{Solving the regularity conditions on $H$}

Positivity of the harmonic function $H$ inside $\Sigma$ by point (a) of Section \ref{sec62}
prohibits any poles in the interior of $\Sigma$. Since $H$ is real, all poles of $H$ must lie 
on the boundary $\p \Sigma = \bR$. We shall restrict attention to simple poles,
as higher order poles may be reached by coalescing simple poles.
Thus, we parametrize $H$ by $N$ simple poles $x_n$ on the real axis,
\bea
\label{7f1}
H =  \sum _{n=1}^N { i  c_n \over w - x_n} + {\rm c.c.}
\eea
The vanishing of $H$ on $\p \Sigma$ by point (b) of Section \ref{sec62} requires the residues 
$c_n$ to be  real, while the positivity of $H$  inside $\Sigma$ by point (a) 
requires $c_n >0$. 
An alternative expression for $H$ is then given by,
\bea
\label{7f2}
H = \Im (w) \sum _{n=1}^N { 2 c_n \over |w -x_n|^2} 
\eea
making the above enforced properties of $H$ manifest. In view of the relation $f_1f_2=H$,
it is clear that one or both of the radius functions $f_1,f_2$ must diverge at the poles $x_n$,
while the radii are finite elsewhere. We shall establish in detail later that each pole indeed
corresponds to a distinct asymptotically $AdS_3 \times S^3$ region.

\subsection{Solving the regularity conditions on $\Lambda ^A$}

By point (e) of Section \ref{sec62}, the one-forms $\Lambda^A = \l^A \p_w H$ 
must be holomorphic inside $\Sigma $, but may have poles on $\p \Sigma$. 
Recasting the constraint relations (\ref{6a1}) in terms of $\L^A$,
\bea
\label{7f4}
\L \cdot \L  & = & 2 \left ( \p_w H \right )^2
\no \\
 \L \cdot \bar \L  & \geq &  2 \, |\p_w H|^2
\eea
it is clear that $\L^A$ must have double poles at each one of the poles $x_n$ of $H$.
For simplicity, we shall assume here that $\L^A$ has no other poles.
Thus, we have the following form for $\Lambda ^A$, 
\bea
\label{7f5}
\L ^A = \sum _{n=1}^N \left ( { -i \kappa^A _n \over (w-x_n)^2} + { -i \mu^A _n \over w-x_n} \right )
\eea
The residues $\kappa ^A_n, \mu ^A_n$ must be real. To establish this, we use (\ref{7f1}) to
see that $\p_wH$ is purely imaginary for $w$ real, which with the help of point 
(d) of Section \ref{sec62} implies,
\bea
\label{7f6}
\Im (\l^A)=\Re (\Lambda ^A)=0  \hskip 1in  \hbox{on} ~ \p \Sigma
\eea
The relation $\Re(\L^A)=0$ readily implies that the residues
$\kappa ^A_n$ and $ \mu ^A_n$ must be real, as announced.

\subsection{Regularity at the poles of $\l^A$}

Unlike the one-forms $\Lambda ^A$ which are holomorphic inside $\Sigma$, the scalars 
$\lambda ^A$ are allowed to have poles in the interior of $\Sigma$, namely at all the points
where $\p _w H$ vanishes. These points come in complex conjugate pairs, and will
be denoted by $w_q, \bar w_q$ with $\Im (w_q) >0$ and $q=1, \cdots, N-1$. 
Near a pole $w_q$ in $\Sigma$, the functions behave as follows, 
\bea
\label{7f10}
\l^A (w) \sim { P^A_q \over w-w_q}
\hskip 1in P^A_q = { \Lambda ^A (w_q) \over \p_w^2 H(w_q)}
\eea
where $P^A_q$ are vectors of complex residues. Conditions (\ref{6a1})  imply 
$P_q \cdot P_q=0$ and $P_q \cdot \bar P_q > 0$. In the limit $w \to w_q$
we have $\l \cdot \bar \l \to + \infty$. In view of (\ref{7b21}), we find two solutions
for $|X|^2$, namely $|X|^{\pm 2}  \sim (P_q \cdot \bar P_q) / |w-w_q|^2$. In either case, 
the metric factors $f_1, f_2, \rho$ of (\ref{7c1}), 
the scalars $V^+{}_A$ of (\ref{7c5}), and the flux potentials of (\ref{7c3}), are 
regular near $w_q$. The common phase to all $V^+{}_A$  is a gauge artifact,
and remains undetermined. We close by noting that at the poles $w_q$, the radius factors 
become equal, $f_1=f_2$, and multiply the Bertotti-Robinson metric on the space $AdS_2 \times S^2$.

\subsection{Solving the constraints}

The harmonic function $H$ and the holomorphic one-forms $\L^A$, provided by the 
parameterization of (\ref{7f1}) and (\ref{7f5}) respectively, must obey the constraints
of (\ref{7f4}) pointwise in order to produce genuine local supergravity solutions. In this section,
we shall derive the relations between the positions of the poles $x_n$ and the residues
$c_n$, $\kappa ^A_n$, and $\mu^A_n$ implied by the first equation of (\ref{7f4}).
The explicit solution to these equations, as well as to the inequality relation 
of (\ref{7f4}) will be constructed in Section \ref{sec8}.

\sm

The first relation of (\ref{7f4}) produces an equation with poles at $x_n$ of order up to four,
upon substituting  $H$ of (\ref{7f1}) and  $\Lambda ^A$ of (\ref{7f5}).
Matching the residues of the poles of orders 4, 3, 2, and 1 requires the following relations, 
\bea
\label{7h2}
\kappa _n \cdot \kappa _n  = 2 (c_n)^2
\hskip 1in 
\kappa _n \cdot \mu _n  = \cE_n^{(2)}=\cE_n ^{(1)} = 0
\eea
for all $n=1,\cdots, N$,  where we have defined the following combinations, 
\bea
\label{7h4}
\cE_n ^{(2)} 
& = & 
\mu _n \cdot \mu _n  +
2 \sum _{n'\not= n} {\kappa _n \cdot \mu  _{n'} \over x_n - x_{n'}} 
- 2 \sum _{n'\not=n} { 2c_n c_{n'} - \kappa  _n \cdot \kappa  _{n'} \over (x_n-x_{n'})^2} 
\no \\
\cE_n ^{(1)} 
& = &  
 \sum _{n' \not= n} { \mu _n \cdot   \mu _{n'} \over x_n - x_{n'}}
- \sum _{n'\not= n} { \kappa _n \cdot \mu  _{n'} - \mu _n \cdot \kappa _{n'}  \over (x_n - x_{n'})^2} 
+ 2 \sum _{n'\not=n} { 2c_n c_{n'} -  \kappa  _n \cdot \kappa _{n'} \over (x_n-x_{n'})^3} 
\eea
While the relations $\cE_n^{(2)}=0$ are all independent from one another, one has the 
following constraints between the relations $\cE_n ^{(1)}=0$,
\bea
\label{7h5}
\sum _n (x_n)^k \cE_n ^{(1)} \sim k \sum _n  (x_n)^{k-1} \cE_n ^{(2)} 
\hskip 1in k=0,1,2
\eea
Thus there are $N$ independent relations $\cE_n^{(2)}=0$, and
$N-3$  independent relations $\cE_n ^{(1)}=0$.

\sm

The second relation of (\ref{7f4}) limits the allowed ranges of the parameters 
$x_n, c_n, \kappa ^A_n, \mu ^A_n$ of the solutions. It may be simplified by 
adding to it the real part of the first equation of (\ref{7f4}), 
resulting in the following equivalent inequality,
\bea
\label{7n2}
\Re (\Lambda) \cdot \Re (\Lambda ) \geq 2  \Big ( \Re ( \p_w H) \Big )^2 
\eea
This inequality reduces to an equality on $\p \Sigma$, since
$\Re (\Lambda)= \Re ( \p_w H)=0$ for real $w$, but (\ref{7n2}) 
must be obeyed strictly inside $\Sigma$. More properly, (\ref{7n2}) should
be viewed as a {\sl family of inequalities}, parametrized by $w$. The fact 
that we shall be able to solve completely in Section \ref{sec8} this intricate
system is a highly non-trivial bonus of the BPS problem.

\subsection{Asymptotic behavior of the metric in $AdS_3 \times S^3$ regions}

The asymptotic behavior of the metric factors $f_1, f_2, \rho$ near the poles $x_n$
is determined by the data $x_n, c_n, \kappa ^A_n, \mu ^A_n$, 
and will now be derived. Expanding $H$ near the poles $x_n$, 
\bea
\label{8c1}
H^2 \sim   {  4 c_n^2 \Im (w)^2  \over |w-x_n|^4} \hskip 1in 
|\partial_w H| ^2 \sim  {c_n^2 \over |w-x_n|^4} 
\eea
we calculate the following asymptotic behaviors 
using (\ref{7c1}) and we find, 
\bea
\label{8c2}
\rho^4  \sim { \mu_n \cdot \mu_n \over 8 |w-x_n|^4}
& \hskip 1in &
f_1^4  \sim  {8 c_n^4 \over \mu_n \cdot \mu_n } {1\over |w-x_n|^4} 
\no \\ && 
f_2^4  \sim  { 2 \mu_n \cdot \mu _n  } { \Im (w)^4 \over |w-x_n|^4} 
\eea
Note that positivity of these metric factors requires 
\bea
\label{7n5}
\mu _n \cdot \mu_n >0
\eea
a property that can also be extracted directly by enforcing (\ref{7n2}) in the vicinity of $w\sim x_n$.
In terms of polar coordinates near the pole defined by
$w= x_n + r e^{i \theta}$, the full six-dimensional metric of the solution becomes,
\bea
\label{8c3}
ds^2 \sim  \sqrt{ 2 \mu_n \cdot \mu _n } \left (  {dr^2 \over r^2} 
 + {2 c_n^2 \over \mu_n \cdot \mu_n } {1\over r^2} ds_{AdS_2}^2+ d\theta^2 + \sin^2\theta \, ds_{S^2}^2 
 \right )
\eea
We recognize the metric of $AdS_3\times S^3$ which confirms that the vicinity of each pole $x_n$ indeed
produces an asymptotically $AdS_3\times S^3$ region.

\subsection{Calculation of the  charges and conservation}

The Page charges in each asymptotic region were defined in (\ref{7c4}),
and may be obtained  by taking the integral of $\L^A$ along a
small contour $\cI_n$ which starts and ends on the boundary and encloses the pole at $w=x_n$,
\bea
\label{7g1}
Q^A _n  
= i \sqrt{2} \pi   \int _{\cI_n}  \L^A + {\rm c.c.}
=  2 \sqrt{2} \pi^2   \mu ^A _n
\eea
Regularity of $\Lambda ^A$ as differential one-forms when $w \to \infty$ requires,
\bea
\label{7f21}
\sum _n \mu _n ^A=0
\eea
which corresponds to the condition of vanishing total Page charge of the solution.

\subsection{Attractor behavior  of scalars fields in $AdS_3 \times S^3$ regions}

The asymptotic behavior of the scalars $V^{(i,r)} {}_A$ near the poles is also 
determined by the data $x_n, c_n, \kappa ^A_n, \mu ^A_n$, up to $SO(5)\times SO(m)$ gauge
transformations. As was shown in (\ref{7f18}), the components $V^r{}_R$ are determined,
up to gauge transformations, by the components $V^\pm{}_A$, according to formulas  given in  
(\ref{7c5}). To compute their limits as $w \to x_n$, we first evaluate the limits of $\lambda ^A$ 
and $X$. Since $|X| \to 1$ as $w\to x_n$, we will need to keep higher order terms, and 
assume that the phase of $X$ has a finite limit $X(x_n)$  as $ w \to x_n$. 
To carry out these calculations, it will be useful to decompose the scalar fields as follows,
\bea
\label{8a2}
V^+{}_A (x_n) = X(x_n) \left ( \half  \, \Re \l_A (x_n) + i \lim _{w \to x_n} { \Im \, \l_A (w) \over |X(w)|^2-1} \right )
\eea
To compute the term whose limit is indefinite, we  determine the behavior
of $|X|$ and $\Im \l_A$ as $w \to x_n$. We decompose the relevant functions as follows,
\bea
\label{8a3}
\p_w H (w) & = & { -i \over (w-x_n)^2} \left ( c_n + (w-x_n)^2 \p H_n (w) \right )
\no \\
\L^A (w) & = & { -i \over (w-x_n)^2} \left ( \kappa ^A _n + (w-x_n) \mu _n ^A + (w-x_n)^2 \L^A _n (w) \right )
\eea
where $\p H_n(w)$ and $\L^A _n (w)$ admit finite limits as $w \to x_n$. 
The resulting values $\p H_n (x_n)$ and $\L^A _n (x_n)$ are real.
The equality $2 (\p_w H)^2 = \L\cdot \L$, expanded to second order in $w-x_n$, 
imposes the familiar conditions for poles of orders 4, 3, and 2 of (\ref{7h2}). 
Computing the combination $\l \cdot \bar \l$  to second  order in $w-x_n$ we find,
\bea
\label{8a6}
\l ^A(w) 
\sim { \kappa _n ^A \over c_n}  + (w-x_n) { \mu _n ^A \over c_n}
+ (w-x_n)^2 {\L^A _n (x_n) \over c_n}  - (w-x_n)^2  \p H_n (x_n) {\kappa _n ^A \over c_n^2}  
\eea
Using the equations of (\ref{7h2}) to simplify the terms of order 0 and 1, 
and to cancel the terms proportional to $(\Re(w) -x_n)^2 $, we find to leading order,
\bea
\label{8a9}
\l \cdot \bar \l & \sim & 2 + 2 \Im (w)^2  {\mu _n \cdot \mu _n \over c_n^2}
\no \\
|X|^2  & \sim &  1 \pm { \Im (w)  \over c_n}  \sqrt{ 2 \mu _n \cdot \mu _n  }
\eea
As a result, we find the following limiting values for the scalar fields,
\bea
\label{8a13}
V^+{}_A (x_n) = X(x_n) \left ( {\kappa _{nA} \over 2 c_n} - i {\mu _{nA} \over  \sqrt{ 2 \mu _n \cdot \mu _n  } } \right )
\eea
Since the values of $\mu_n ^A$ give the charges of the solution, 
we see that (ignoring the phase artifact of $X(x_n)$), the imaginary parts of the scalars are 
fully attracted, and their values are uniquely determined in terms of the charges.
The real parts depends on $N$ vectors $\kappa _n/c_n$, whose square equals 2,
and which are orthogonal to $\mu_n$ by (\ref{7h2}).

\subsection{SL(2,R) transformations on equations and solutions}
\label{sec77}

Under $SL(2,\bR)$ automorphisms of $\Sigma$, which leave $\p \Sigma$ invariant, 
the coordinate $w$ and the positions of the poles $x_n$ transform as follows,
\bea
\label{8c4}
w \to w'={ \a w + \b \over \g w + \delta }
\hskip 1in 
x_n \to x_n'= { \a x_n + \b \over \g x_n + \delta }
\eea
where $\a, \b, \g, \delta \in \bR$ and $\a \delta - \b \g =1$. Transforming also the residues
at the poles of $\p_w H$ and $\Lambda ^A$ by $\mu _n  \to \mu_n' = \mu _n $ and, 
\bea
\label{8c5}
c_n \to c_n' = { c_n \over (\g x_n + \delta )^2}  \hskip 1in  
\kappa _n \to \kappa _n ' = { \kappa _n \over (\g x_n + \delta )^2}  
\eea
we obtain the following transformations rules,
\bea
\label{8c6}
H & \to & H'(w',\bar w')= H(w,\bar w)
\no \\
\p_w H & \to & \p_{w'} H'(w')= (\g w+ \delta )^2 \p_w H (w) 
\no \\
\Lambda  & \to & \Lambda ' (w') = (\g w + \delta )^2 \Lambda (w)
\no \\
\lambda  & \to & \lambda ' (w') =  \lambda (w)
\eea
where the functions $H', \L', \l'$ are evaluated with the poles $x_n'$ and the residues
$c'_n, \kappa '_n, \mu '_n$.  As a result, the physical data of the solution, assembled
in Section \ref{sec54}, have simple transformation properties. The functions $f_1, f_2, \rho^2 |dw|^2$
of (\ref{7c1}), the scalars fields $V$ of (\ref{7c2}) and (\ref{7f18}), and the flux potentials 
$\Phi ^A, \Psi ^A$ transform as scalars, possibly up to gauge transformations. Also, 
the fundamental relations of (\ref{7f4}) defining the local solutions are invariant under $SL(2,\bR)$. 
Thus, the physical data of the solutions are invariant under $SL(2,\bR)$.

\subsection{Counting moduli of the solutions}

The enumeration of the basic parameters  of the solutions as they enter  $H$ and $\Lambda $,  
is as follows,
\bea
\label{7j1}
x_n , \, c_n & \hskip 1in & 2N
\no \\
\kappa _n, \, \mu _n & & 2(m+2)N
\eea
giving a total of $(2m+6)N$ real parameters.  The number of relations is as follows. 
The regularity conditions (\ref{7f21}) provide $m+2$ relations; matching of the poles or orders
4, 3, and 2 each provide $N$ relations; matching of the first order poles only provides $N-3$
relations in view of the dependencies of (\ref{7h5}), yielding a total of 
$4N+m-1$ relations. Thus, the number of independent parameters
which enter a solution is given by $(2m+6)N- (4N+m-1)= (2m+2)N-m +1$. The {\sl physical data} 
are invariant under the $SL(2,\bR)$ transformations of Section 
\ref{sec77}, bringing the total number of physical parameters of the solution to, 
\bea
\label{7j3}
(2m+2)N-m -2
\eea
We shall now show that this counting precisely matches the counting of physical parameters.

\subsection{Counting physical parameters of the solutions}

On physical grounds, the solutions are expected to be determined by the charges of the three-form fields 
$G^A$ on the 3-spheres in the asymptotic $AdS_3\times S^3$ regions, as well
as by the values of the scalar fields there. The counting proceeds as follows. There are $m+2$
independent charge parameters in each asymptotic region, and one overall charge 
conservation relation, thus yielding a total of $(m+2)(N-1)$ charge parameters.
There are $2m$ scalar entries  in each asymptotic region, but as is clear from 
(\ref{8a13}), half of those are subject to the attractor mechanism and already fixed by the 
charges $\mu_n$. Thus, only the  $m$  ``un-attracted" scalars must be counted 
as independent parameters per asymptotic region, giving a total of $mN$ scalar parameters. 
This gels nicely with the fact that in each asymptotic region there are $m+2$ entries 
$\kappa^A_n/c_n$ subject to the two constraints of (\ref{7h2}). 
In conclusion, the number of physical parameters of the solution is given by
$(m+2)(N-1) + mN = (2m+2)N -m-2$, 
which matches perfectly with the counting of (\ref{7j3}).

\newpage

\section{Numerical solutions,  $N=2,3,4,5$}
\setcounter{equation}{0}
\label{sec8}

In this section, we shall spell out the conditions on our solutions for the cases $N=2,3$, which
includes the vacuum solution $AdS_3 \times S^3$ and the Janus solutions found in \cite{Chiodaroli:2009xh}. 
Furthermore we will derive numerically examples of new solutions with $N=3,4,5$.

\subsection{Solutions with $N=2$}
\label{sec81}

For $N=2$ only two asymptotic $AdS_3 \times S^3$ regions appear, 
with opposite charges, $\mu_1+\mu_2=0$, and the following general relations, 
\bea
\label{9a1}
\kappa _1 \cdot \kappa _1 = 2 c_1 ^2 
& \hskip 1in & \kappa _1 \cdot \mu_1=0
\no \\
\kappa _2 \cdot \kappa _2 = 2 c_2 ^2 
& \hskip 1in & \kappa _2 \cdot \mu_2=0
\eea
Using $SL(2,\bR)$ invariance to fix the poles at $x_1=0$, $x_2=1$, 
the relations of (\ref{7h2}) reduce to, 
\bea
\label{9a2}
\mu_1 ^2 = \mu_2 ^2 = 4 c_1 c_2 - 2 \kappa _1 \cdot \kappa _2
\eea
while we find that the second condition of (\ref{7f4}) is automatic.
For $\kappa _2 = -  \kappa _1$, we have the vacuum solution $AdS_3 \times S^3$,
while for $\kappa _2 \not= - \kappa _1$, we have the basic Janus solution  
of the $(0,4)$ six-dimensional supergravity \cite{Chiodaroli:2009yw}.

\subsection{Solutions with $N=3$}
\label{sec82}

For $N=3$, there are three asymptotic $AdS_3 \times S^3$ regions, with $\mu _1 + \mu _2 + \mu _3 =0$, 
and, 
\bea
\label{9b1}
\kappa _n  \cdot \kappa _n & = & 2 c_n  ^2 
\hskip 1in n=1,2,3
\no \\
\kappa _n \cdot \mu_n & = &0
\eea 
Using $SL(2,\bR)$ symmetry, we fix the poles at $x_1=-1$, $x_2=0$, $x_3=1$.
The relations $\cE_n ^{(1)}=0$ of (\ref{7h2}) follow from the relations $\cE_n^{(2)}=0$,
which reduce to the following equations, 
\bea
\label{9b2}
0 & = & \mu_1 ^2 - \half \kappa _1 \cdot (\mu_2 - \mu_3) + 2 C_{12} + \half C_{13}
\no \\
0 & = & \mu_2 ^2 + 2 \kappa _2 \cdot (\mu_1 - \mu_3) + 2 C_{12} + 2 C_{23}
\no \\
0 & = & \mu_3 ^2 - \half \kappa _3 \cdot (\mu_1 - \mu_2) + \half  C_{13} + 2 C_{23}
\eea

\begin{figure}[ht]
\begin{centering}
\includegraphics[scale=0.75]{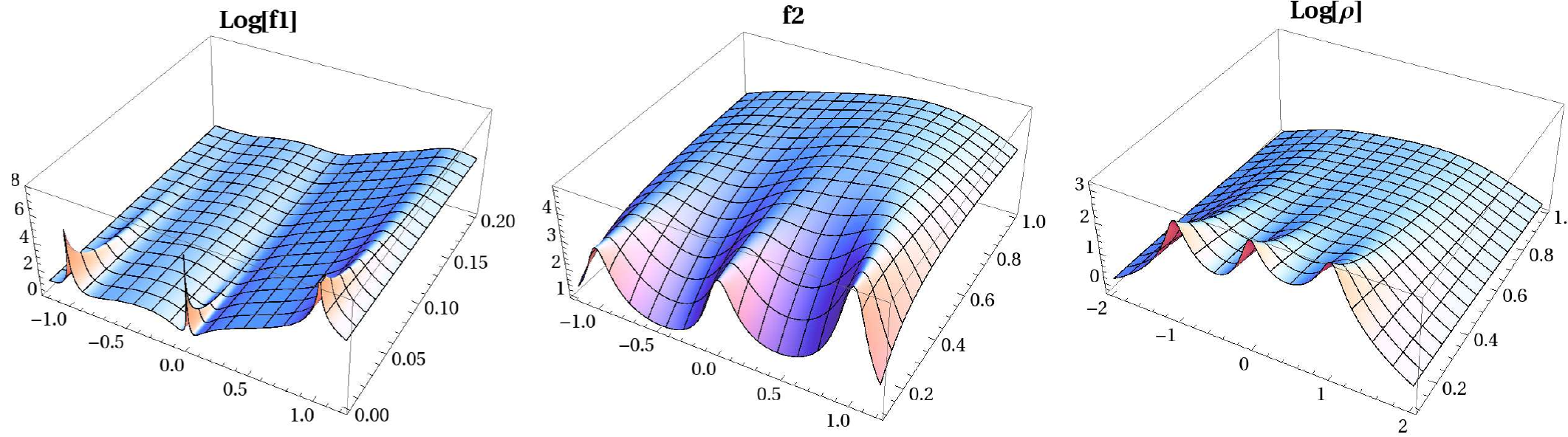}
\caption{Metric factors for the upper half-plane parametrization of $\Sigma$}
\label{Fig2}
\end{centering}
\end{figure}
\begin{figure}[ht]
\begin{centering}
\includegraphics[scale=0.2]{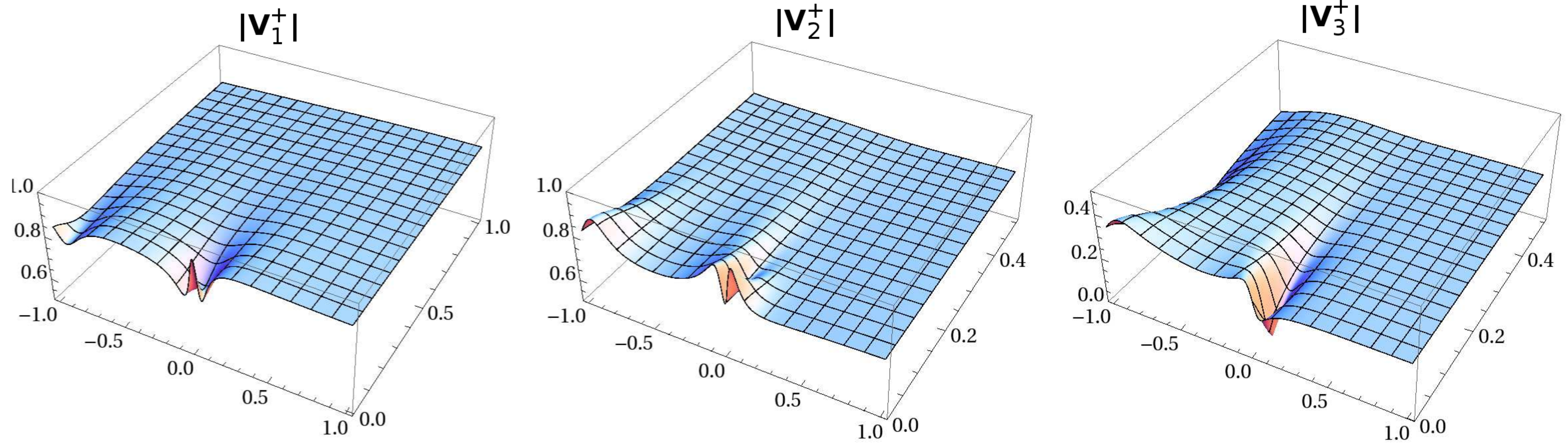}
\caption{Scalars for the upper half-plane parametrization of $\Sigma$}
\label{Fig3}
\end{centering}
\end{figure}
\noindent
where we have used the abbreviation $C_{mn} = \kappa _m \cdot \kappa _n - 2 c_m c_n$.
Obeying equations (\ref{9b1}) and (\ref{9b2}) guarantees that we satisfy the first relation
of (\ref{7f4}). To satisfy the second relation of (\ref{7f4}) imposes further non-trivial  
conditions, which may be readily solved  by numerical analysis,
thus proving that, as soon as $N\geq 3$, solutions exist which are inequivalent to the 
solutions of \cite{Chiodaroli:2009yw} in that the scalar fields take values in a coset larger than
the $SO(2,2)/(SO(2)\times SO(2))$ coset of \cite{Chiodaroli:2009yw}.
Here we study numerically the case $N=3$ and with the scalars taking values
in a $SO(2,4)/(SO(2)\times SO(4))$ coset space.
In Figures \ref{Fig2} and \ref{Fig3}, we present the numerical solution with poles
brought back to the values $x_1=-1$, $x_2=0$, and $x_3=1$ by $SL(2,\bR)$ symmetry, 
and for the following  data,
\bea
\kappa _1 & = & (-0.5621, -0.4099, -0.2461, - 0.0706, -0.08979, -0.3586)
\no \\
\kappa _2 & = & (-0.2835, -0.05616, -0.07683, -0.0174, -0.1350, 0.1157)
\no \\
\kappa _3 & = & (87.29, 30.68, -23.11, 40.00, -3.648, 33.22)
\no \\
\mu_1 & = & (3.681, -6.058, -0.8065, -0.2336, -0.5414, -0.4192)
\no \\
\mu_2 & = & (1.557, -9.885, 0.8671, -0.7100, -0.8268, 0.4859)
\eea
with $\mu_3=-\mu_1-\mu_2$, and the derived data $c_1=0.3753$, $c_2=0.1511$,
and $c_3=51.52$.

\subsection{Solutions with $Z_N$  symmetry}
\label{sec95}

An interesting subclass consists of solutions with cyclic $Z_N$ symmetry. To define 
these solutions, we map the upper half-plane coordinates $w, \bar w$ with $\Im (w) >0$  
to the unit-disk coordinates $z, \bar z$ with $|z|=1$, by the following relations,
\bea
\label{9c1}
z & = & { 1 +i w \over 1-iw} 
\no \\
z_n & = & { 1 +i x_n \over 1-i x_n} 
\eea 
Functions and forms transform as 
\bea
\label{9c2}
\hat H (z, \bar z) & = & H(w, \bar w)
\no \\
\hat \L^A (z) dz & = & \L^A (w) dw
\eea
In the new coordinates, the functions $\hat H$ and $\hat \L^A$ are found as follows,
\bea
\label{9c3}
\hat H (z, \bar z) & = & \hat H_0 - {1 \over 2} \sum _{n=0}^{N-1} { (z_n+1)^2 \ c_n  \over z-z_n} + {\rm c.c.}
\no \\
\hat \L^A (z) & = & \sum _{n=0}^{N-1} \left (- {1 \over 2} { (z_n+1)^2 \kappa ^A_n  \over (z-z_n)^2} - { i \mu_n^A \over z-z_n} \right )
\eea
where the values  $\mu_n^A$ are unchanged from the upper half-plane, 
and $\hat H_0$ is a constant such that $\hat H (z,\bar z)=0$ for $|z|=1$.

\sm

The $Z_N$ transformations on the unit-disk are generated by,
\bea
\label{9c6}
Z_N ~ : \quad z \to \ep z \hskip 1in \ep = e^{2 \pi i /N}
\eea
Invariance under $Z_N$ of the set of poles $z_n$ requires their values to be given by,
\bea
\label{9c7}
z_n = \ep ^n 
\eea
up to an immaterial overall phase factor. The harmonic function $\hat H$  transforms as a scalar under $Z_N$, while 
the holomorphic form $\hat \Lambda^A$ transforms as a vector of $SO(2,m)$,
\bea
\label{9c8}
\hat H (\ep z , \bar \ep \bar z) & = & \hat H(z,\bar z)
\no \\
\hat \Lambda ^A (\ep z , \bar \ep \bar z) d(\ep z) & = & E^A{}_B \hat \L^B (z) dz
\eea
where the matrix $E$  provides an $SO(2,m)$-valued representation of the $Z_N$ 
transformation of (\ref{9c6}), and thus obeys $E^t \eta E= \eta$ and $ E^N=I$.
Substituting the unit-disk expressions of (\ref{9c3}) into the $Z_N$ covariance relations
of (\ref{9c2}), and identifying residues of like poles, gives the following relations,
\bea
c_n & = & {4 \ep^n c_0 \over (\ep^n+1)^2 } 
\no \\ 
\kappa_n & = & {4 \ep^n E^n \kappa_0 \over (\ep^n+1)^2 } 
\no \\
\mu_n & = & \, E^n \mu_0
\eea
Note that the $Z_N$ symmetry relation on $c_n$ is consistent with $c_n >0$.
A convenient explicit representation for the matrix $E$ may be obtained,
using an $SO(2,m)$ rotation, by changing to a basis in which $E$ assumes the form, 
\bea
\label{9c11}
E = \left ( \matrix{
d_0 & 0 & 0 & \cdots & 0  \cr
0 & d_1 & 0 & \cdots & 0  \cr
0 & 0 & d_2 & \cdots & 0  \cr
& & & \cdots & \cr
0 & 0 & 0 & \cdots & d_{[m/2]}  \cr
} \right )
\eea
This form holds for $m$ even; for $m$ odd, an extra column and row need to be added 
whose only non-zero entry is 1 in the bottom right corner. 
The $2 \times 2 $ blocks $d_i$ are given as follows,
\bea
d _i =  \left ( \matrix{ \cos {2 \pi \nu_i \over N} & \sin { 2\pi \nu _i   \over N} \cr & \cr 
-\sin  {2 \pi \nu_i  \over N} & \cos { 2 \pi \nu _i \over N} \cr } \right )
\eea
The integers $\nu_i$ are defined mod $N,$ and characterize the representation 
of $Z_N$ into $SO(2,m)$.

\newpage

\subsection{Numerical solutions with $Z_4$ and $Z_5$  symmetry}
\label{sec94}

\begin{figure}[ht]
\begin{centering}
\includegraphics[scale=0.78]{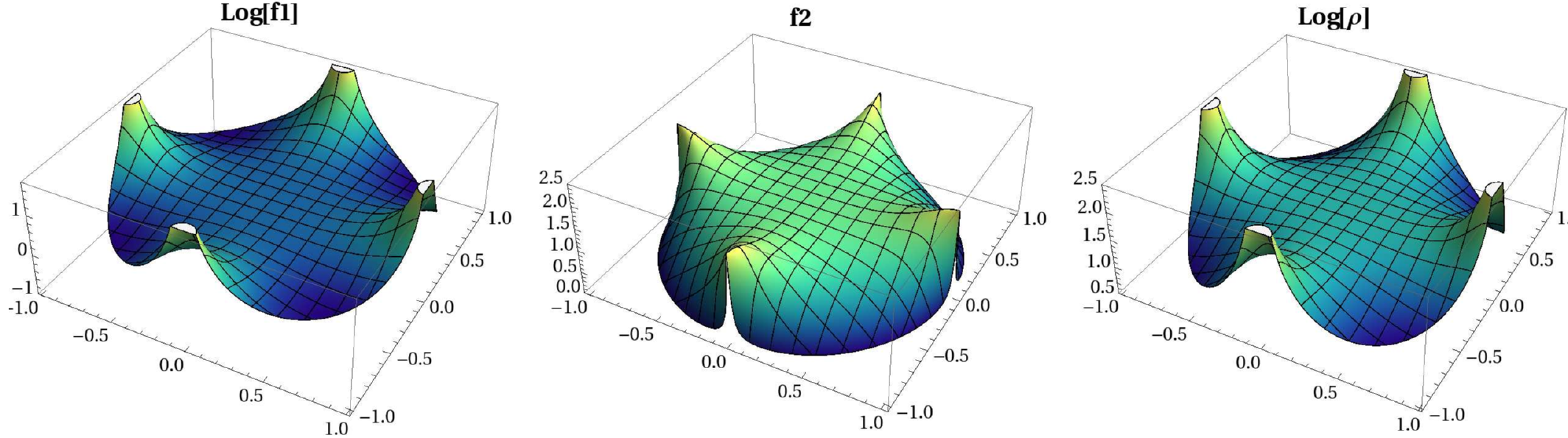}
\caption{Metric factors for the $Z_4$-symmetric solution  with data  (\ref{data4}) and (\ref{data4a}).}
\label{Fig4}
\end{centering}
\end{figure}

\begin{figure}[ht]
\begin{centering}
\includegraphics[scale=0.18]{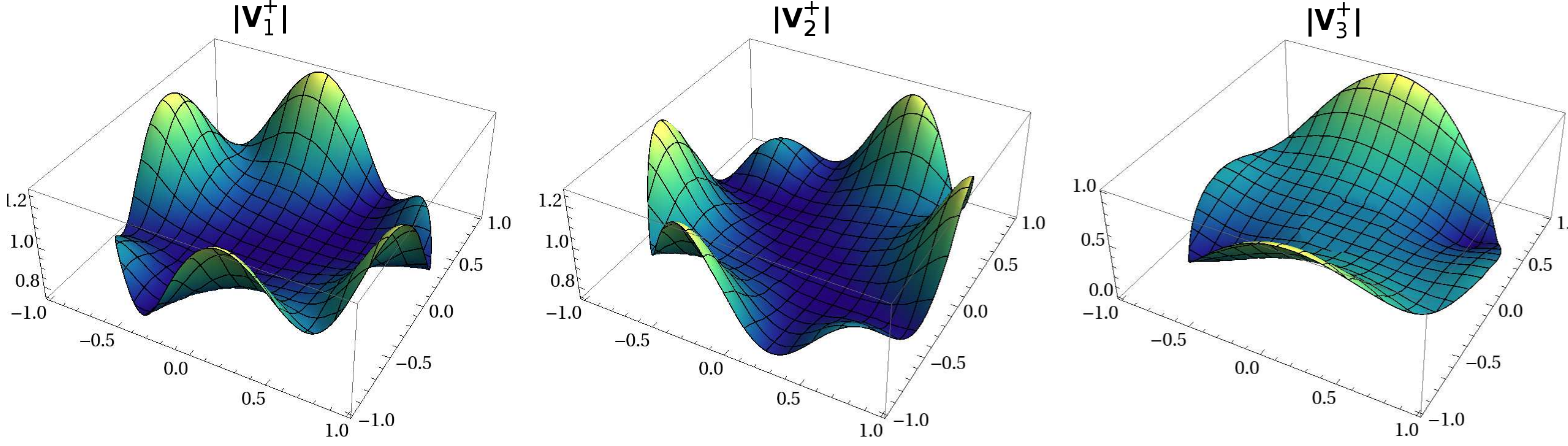}
\caption{Scalar components for the $Z_4$-symmetric solution  with data (\ref{data4}) and (\ref{data4a}).}
\label{Fig5}
\end{centering}
\end{figure}

In Figures \ref{Fig4} and \ref{Fig5}, we present a numerical solution with $N=4$, 
effective scalar coset space $SO(2,4)/(SO(2)\times SO(4))$, and  the values 
\bea
\label{data4}
\nu_1=1 \hskip 1in \nu_2=1 \hskip 1in   \nu_3 = 2
\eea
The remaining data have been chosen randomly subject
to the constraints (\ref{7h2}-\ref{7h4}), 
\bea
\label{data4a}
\mu _0 & = & (-2.631, 1.686, 0.4054, -0.5013, -0.7915, -0.6094)
\no \\
\kappa _0 & = & (0.3870, 0.5423, -0.1557, -0.4702, 0.2524, 0.1260)
\no \\
c_0 & = & 0.2438
\eea

\newpage

\begin{figure}[ht]
\begin{centering}
\includegraphics[scale=0.8]{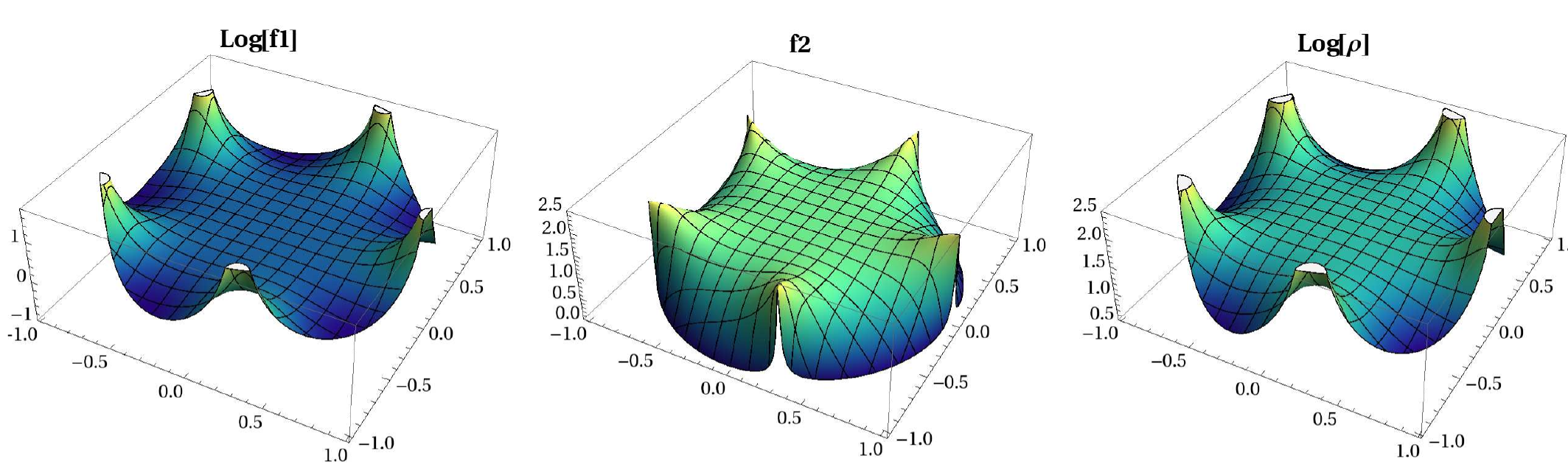}
\caption{Metric factors for the $Z_5$-symmetric solution with data (\ref{data5}) (\ref{data5a}).}
\label{Fig6}
\end{centering}
\end{figure}

\begin{figure}[ht]
\begin{centering}
\includegraphics[scale=0.19]{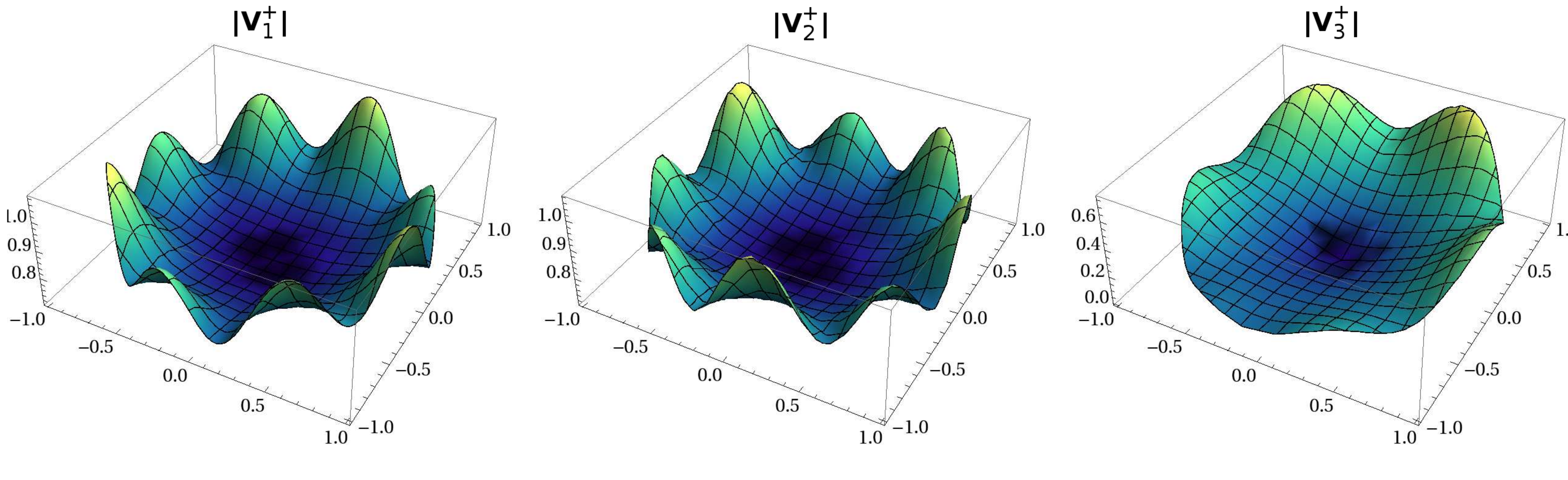}
\caption{Scalar components for the $Z_5$-symmetric solution  with data (\ref{data5}) and (\ref{data5a}).}
\label{Fig7}
\end{centering}
\end{figure}

In Figures \ref{Fig6} and \ref{Fig7}, we present a numerical solution with $N=5$,
scalar coset space $SO(2,4)/(SO(2)\times SO(4))$ and the values 
\bea
\label{data5}
\nu_1=1 \hskip1 in \nu_2=3 \hskip 1in  \nu_3 = 2
\eea
The remaining data have been chosen randomly subject to the constraints (\ref{7h2}-\ref{7h4}), 
\bea
\label{data5a}
\mu_0 &=& (-2.481, 2.273, -0.2239, -0.9870, 0.9926, 0.04811) \no \\
\kappa_0 &=& (0.3134, 0.3310, 0.2456, 0.1550, 0.1822, 0.03752) \no \\
c_0 &=& 0.2107 \eea


\newpage

\section{Maximal families of regular solutions for all $m,N$}
\setcounter{equation}{0}
\label{sec9}

In this section, we shall derive a general family of solutions
with the maximal number $(2m+2)N - m -2$ of free modular parameters.
The corresponding half-BPS solutions, and their moduli space, may be 
parametrized explicitly and completely in terms data satisfying only simple linear inequalities.
These data include a set of $2N-2$ {\sl auxiliary poles}, and associated residues,
along with the mild technical assumption on these poles.
The resulting families of regular  solutions provide a generalization of the solutions of 
\cite{Chiodaroli:2009yw} to the case of general $m $, including new solutions for $m>2$.

\subsection{Light-cone like solution of the first constraint}

We begin by spelling out a light-cone like decomposition which will be at the core
of the general and explicit parametrization of the families of solutions that we derive.
The constraints, formulated in terms of $\l^A$, are as follows,
\bea
\label{20a1}
2 & = & \l \cdot \l = \delta _{IJ} \l^I \l^J  - \delta _{RS} \l^R \l^S 
\no \\
2 & < & \bar \l \cdot \l = \delta _{IJ} \bar \l^I \l^J  - \delta _{RS} \bar \l ^R \l^S 
\eea
where $I,J=1,2$,  $R,S=6, \cdots, m+5$. We choose one index $I$ of positive signature, and one index $R$
of negative signature of the metric $\eta$. By $SO(2) \times SO(m)$ covariance of the 
solution, these may be chose to be the indices 2 and 6, without loss of generality. 
Accordingly, we define the inverse of the combination $\l^2-\l^6$, 
\bea
\label{20a2}
L^6 ={ 1 \over \l^2 - \l^6} 
\eea
and rescale the remaining fields as follows, 
\bea
\label{20a2a}
L^A = { L^6 \over \sqrt{2}} \, \l^A \hskip 0.7in A=1,7,8,\cdots , m+5
\eea
In terms of the functions $L^A$, the first constraint of (\ref{20a1}) may be solved rationally, 
so that all $\lambda ^A$ may be expressed in terms of the functions $L^1, L^6, L^7 \cdots , L^{m+5}$. 
The counting works out, and the explicit formulas are given as follows,
\bea
\label{20d2}
\l^A & = & { \sqrt{2} L^A \over L^6}  \hskip 2in A=1,7,8,\cdots, m+5
\no \\
\l^2 & = & {1 \over L^6} \left ( + \half - L^1 L^1 + \delta _{RS} L^R L^S \right )
\no \\
\l^6 & = & {1 \over L^6} \left ( - \half - L^1 L^1 + \delta _{RS} L^R L^S \right )
\eea

\subsection{Solving the second constraint}

In terms of the functions $L^A$, the second constraint of (\ref{20a1}) simplifies considerably. To
see this, we arrange the constraint in the following form, 
\bea
\label{20b1}
2 <   \bar \l^1 \l ^1 + \half ( \bar \l ^2 + \bar \l^6) (\l^2 - \l^6) + \half ( \bar \l ^2 - \bar \l^6) (\l^2 + \l^6)
 - \sum _{A=7}^{m+5} \bar \l^A \l^A 
\eea
Eliminating $\l^A$ in favor of $L^A$ using (\ref{20d2}), we obtain, 
\bea
\label{20b3}
0 < - (L^1 - \bar L^1)^2 + \delta _{RS} (L^R - \bar L^R) (L^S - \bar L^S)
\eea
where the summation over repeated indices now runs over $R,S=6,7, \cdots, m+5$.
Introducing the following harmonic functions 
\bea
\label{20b4}
h^A = \Im (L^A) \hskip 0.7in A=1,6,7, \cdots, m+5
\eea
equation (\ref{20b3}), and thus the second constraint of (\ref{20a1}), may  be recast in the following form,
\bea
\label{20b5}
 \delta _{RS} h^R h^S < h^1 h^1 
\eea
From the requirement that this inequality be obeyed strictly inside $\Sigma$,
we deduce that $h_1$ must maintain the same sign inside $\Sigma$, which
we shall take to be negative, $h_1<0$ inside $\Sigma$.

\subsection{Parametrization of solutions by ``auxiliary poles"}

From the general expressions for $\p_w H$ and $\Lambda ^A$, given  in (\ref{7f1}) and (\ref{7f5}),
we see that each meromorphic function $\l^A$ is a ratio of two polynomials of 
degree $2N-2$. Simple poles are allowed in  $\l^A$ at the $2N-2$ complex zeros $w_q, \bar w_q$
of $\p_w H $, with $q=1,\cdots, N-1$. Each $\l^A$ has $2N-2$ zeros. 
We shall denote the zeros of $\l^2-\l^6$ by $y_p$ where $p=1,\cdots, 2N-2$.
Hence, $L^1$ and $L^R$ have poles at the points $y_p$ and may be decomposed
as follows,
\bea
\label{20c2}
L^A (w) = \ell ^A_\infty  + \sum _{p=1}^{2N-2} { \ell ^A _p \over w-y_p}
\hskip 1in A=1,R
\eea
For simplicity, we shall restrict here to the case where the poles $y_p$ are all simple.

\sm

Next, we show that the poles $y_p$ and residues $\ell ^A_p$ have to be real.
Since $h^1$ is a real harmonic function, it will assume both positive and negative 
values in any open neighborhood of each one of its poles. Thus, the condition $h^1<0$ inside
$\Sigma$ prohibits any poles of $h^1$ from being complex, thereby forcing all poles $y_p$
to be real. Since $\l^A(w)$ must be real for real $w$, it follows that $L^6(w)$,
and hence all the $L^A (w)$, must be real for $w \in \bR$. In view of the decomposition
of (\ref{20c2}), we see that real $L^A(w)$ for real $w$ requires $\ell _\infty^A$ and $\ell^A _p$
to be real. As a result, the harmonic functions take on simplified forms, 
\bea
h^A = - \Im (w) \sum _{p=1}^{2N-2} { \ell ^A _p \over |w-y_p |^2}
\hskip 1in A=1,R
\eea
The absence of poles  at the points $y_p$ in $\l^2$ and $ \l^6$ (which were expressed in terms
of $L^A$ in (\ref{20d2}))  requires the following relation between the residues, 
\bea
\label{20c3}
\left ( \ell _p ^1 \right )^2 = \delta _{RS} \, \ell _p ^R \ell _p^S 
\eea
Consistently with the result $h^1<0$ derived earlier, we choose the positive roots,
\bea
\label{20c6}
\ell ^1 _p =  \sqrt{ \delta _{RS} \ell ^R _p \ell ^S_p}
\eea
Condition (\ref{20b5}) now becomes,
\bea
\label{20c4}
0 < \sum _{p, p' =1}^{2N-2} { \ell_p^1 \ell _{p'} ^1 - \delta _{RS} \ell _p ^R \ell _{p'} ^S \over 
\left | w-y_p \right |^2 \left | w-y_{p'} \right |^2}
\eea
In view of (\ref{20c6}) and the Schwarz inequality, 
\bea
\label{20c5}
0 \leq \ell_p^1 \ell _{p'} ^1 - \delta _{RS} \ell _p ^R \ell _{p'} ^S 
\eea
for all $p,p'=1,\cdots, 2N-2$, the inequality $\leq $ in (\ref{20c4}) follows automatically,
and for all values of $\ell _p ^R$, and all values of $y_p$. 

\sm

For the strict inequality $<$ to hold in (\ref{20c4}), as is required to have regular solutions,
it suffices to have $0 < \ell_p^1 \ell _{p'} ^1 - \delta _{RS} \ell _p ^R \ell _{p'} ^S $ for a
single pair of $p,p'$. If the contrary were true, then the vectors $\ell ^R_p$ would have 
to be parallel for all values of $p$ which leads to a trivial solution.

\subsection{Relating covariant data to auxiliary pole data}

Given a solution parametrized by the harmonic function $H$ and the auxiliary
pole data $y_p$ and $\ell _p ^A$, we seek to obtain formulas for the data $\kappa ^A_n$
and $\mu^A_n$, which include the three-form Page charges. First, we relate the data to the 
values of $\l^A$ and its first derivative at $x_n$,
\bea
\label{20d1}
\kappa _n ^A & = & c_n \l^A(x_n)
\no \\
\mu _n ^A & = & c_n \p \l^A (x_n)
\eea
Note that, in view of the relation $\l (w) \cdot \l(w) =2$, and its first derivative, 
the results of (\ref{20d1}) automatically satisfy $\kappa _n \cdot \kappa _n = 2 c_n^2$
as well as $\kappa _n \cdot \mu_n=0$. The values of $\l^A$ and its derivative may
in turn be obtained from the relations (\ref{20d2}).

\subsection{Counting data and explicit parametrization of solutions}

The family of solutions may be parametrized -- {\sl without leaving any unsolved equalities or inequalities} --
by the following data. We prescribe the positions of both sets of poles, $x_n$ and $y_p$
for $n=1,\cdots, N$ and $p=1,\cdots, 2N-2$; give the residues $c_n$ for $n=1,\cdots, N$, 
and $\ell ^R_p$ for $R=7,8,\cdots, m+5$; and prescribe the values at $w=\infty$, namely
$\ell ^A_\infty$ for $A=1, 7, 8, \cdots, m+5$. Finally, there is an overall real normalization 
factor $\ell ^6 _\infty$ for the function $L^6$. We will prove below that this set of data 
completely specifies the functions $H$ and $\l^A$, up to an overall redundancy 
by $SL(2,\bR)$ automorphisms of the upper half-plane.
In this parametrization, we have the following counting of free parameters,
\bea
\label{20e1}
x_n & \hskip 1in & N
\\
c_n & \hskip 1in & N
\no \\
y_p && 2N-2
\no \\
\ell ^R_p && (m-1)(2N-2)  \hskip 0.75in R=7,8,\cdots, m+5
\no \\
\ell^A_\infty && m+1 \hskip 1.5in A=1,6,7,8,\cdots, m+5
\no
\eea
The total number of data, including $-3$ degrees of freedom due to the action of $SL(2,\bR)$,
is given by $ (2m+2)N -m -2$ in agreement with the results of (\ref{7j3}),
which means that our family of solutions has maximal dimension.

\sm

To prove that the solutions are fixed completely by the parameters listed in (\ref{20e1}),
and that this parametrization solves all constraint equations, we proceed as follows.
The data of (\ref{20e1}) manifestly  fix the functions $H$ and 
$L^R$ for $R=7,8,\cdots, m+5$. This leaves the functions $L^1$ and $L^6$
to be determined.

\sm

First, we determine $L^6$ by realizing that its poles are the zeros of $\l^2 - \l^6$,
which by definition are the points $y_p$ for $p=1, \cdots, 2N-2$, while its zeros
are the poles of $\l^2 - \l^6$, which in turn are the zeros of $\p_wH$. Thus, 
we can have the following formula for $L^6$,
\bea
L^6 = i \ell _\infty ^6 \, { \prod _{n=1} ^N (w-x_n)^2 \over \prod _{p=1} ^{2N-2} (w-y_p)} \, \p_w H
\eea
where $\ell ^6 _\infty$ is a real overall normalization, which is being supplied by the 
parametrization of (\ref{20e1}), and is equivalent to giving the value of $L^6$
at $w=\infty$. The pre-factor $i$ is required because $L^6$ is real-valued for $w$ real, 
while $\p_wH$ is purely imaginary for $w$ real. From this formula, we extract
\bea
\ell ^6 _p = i \ell _\infty ^6 \, 
{ \prod _{n=1} ^N (y_p-x_n)^2 \over \prod _{p' \not =p} ^{2N-2} (y_p-y_{p'})} \, \p_w H (y_p)
\eea
for $p=1, \cdots, 2N-2$. Second, using the relation (\ref{20c6}), and the datum $\ell ^1 _\infty$ 
from the parametrization of (\ref{20e1}), it is clear that also $L^1$ is known explicitly. 
Third, having now at our disposal the functions $H$ and $L^A$ with $A=1,6,7,\cdots, m+5$,
we use (\ref{20d2}) to derive all the $\l^A$ functions. Thus we have shown by explicit 
construction that, starting from the data of (\ref{20e1}), the full solution is determined uniquely,
up to inequalities on the parameters.

\sm

Finally, there is only a single group of inequalities to be satisfied,
\bea
c_n  >  0 \hskip 1in n=1, \cdots, N
\eea
All others, such as the fact that $L^6$ has only complex zeros, and that
$h^1<0$ are automatically realized by our construction.

\subsection{Relation to the solutions of \cite{Chiodaroli:2009yw}}
\label{appC}

The solutions to ten-dimensional Type IIB supergravity compactified on $K3$ correspond 
to the case $m=2$ in our present work, since only dilaton, axion, four-form potential and $K3$ metric 
factor produce scalars
in the six-dimensional theory. The general solutions to the $K3$ compactification of
ten-dimensional Type IIB supergravity were obtained in \cite{Chiodaroli:2009yw} 
for all values of $N$, and provide solutions to the six-dimensional supergravity
for general $N$. In this section, we spell out the correspondence. 

\sm

To translate the solutions of \cite{Chiodaroli:2009yw} to the formulation used here, 
we may compare, for example, the expression for the metric $\rho ^2$ on $\Sigma$, given 
by
\bea
\label{8z1}
\rho^4 = {|\p_w H |^4 \over H^2 }
\left ( { (A+\bar A)K -B^2 - \bar B^2 \over B \bar B} -2 \right ) 
\left ( { (A+\bar A)K -B^2 - \bar B^2 \over B \bar B} + 2 \right ) 
\eea
The harmonic function $H$ is the same in both formulations. Comparing with (\ref{7c1}), we are led to making the following identification, 
\bea
\label{8z2}
\l \cdot \bar \l = {(A+\bar A) K  -B^2 - \bar B^2 \over B \bar B}
\eea
The combination $\lambda \cdot \bar \lambda$ of (\ref{8z2}) obeys   
the decomposition properties  into a sum of holomorphic times 
anti-holomorphic functions, together with the purely holomorphic relation, 
\bea
\label{8z3}
\l \cdot \bar \l & = &  \l_1 \bar \lambda _1 + \l_2 \bar \lambda _2 
- \l_6 \bar \lambda _6 - \l_7 \bar \lambda _7  
\no \\
2 & = & \l_1 ^2  + \l_2 ^2   - \l_6 ^2 - \l_7 ^2
\eea
where the functions $\l_A$ are given by, 
\bea
\label{8z4}
\l_1 = i \,   {A \cK  - B^2 +1 \over \sqrt{2} \, B} 
& \hskip 1in & 
\l_6 = i \,   {A \cK - B^2 -1 \over \sqrt{2} \, B} 
\no \\
\l_2 =   {A + \cK \over \sqrt{2} \, B} \hskip 0.5in
& \hskip 1in & 
\l_7 =  {A - \cK \over \sqrt{2} \, B} 
\eea
where $\cK$ is the holomorphic part of the harmonic function $K$ so that $K=\cK + \bar \cK$.
Using the above relations together with (\ref{20d2}), we can express the harmonic functions in \cite{Chiodaroli:2009yw} as
\bea B  &=& i \sqrt{2} L^6  \no \\
A + \bar A &=& - 2\sqrt{2} \big( h^1 + h^7 \big) \no \\
K &=& - 2\sqrt{2} \big( h^1 - h^7 \big) \eea
The reality and regularity conditions in terms of the functions $A,B,K,H$ were spelled out in 
\cite{Chiodaroli:2009yw,Chiodaroli:2009xh}, and are given as follows,
\begin{enumerate}
\item The functions $H,A,B,K$ must be regular in the interior of $\Sigma$;
\item The functions $H, \Re(A),K$ cannot vanish in the interior of $\Sigma$;
\item All the zeros of $B$ and $\p_w H$ must be common;
\item All poles of $A,B,\cK$ must be simple, common, and located on $\p \Sigma$, with related residues;
\item The inequality $\l \cdot \bar \l >2$ must hold in the interior of $\Sigma$;
\item The harmonic functions $H, \Re(A),\Re(B), K$ must obey vanishing Dirichlet 
boundary conditions on $\p \Sigma$, while the harmonic functions $\Im (A)$ and $\Im (B)$
must satisfy Neumann boundary conditions on $\p \Sigma$.
\end{enumerate}
The five topology and regularity conditions of Section \ref{sec62} on the locally holomorphic 
data for the solutions of this paper are seen to result from the above regularity assumptions 
on the solutions from Type IIB on $K3$. Specifically, 
condition (a) of Section \ref{sec62} follows from conditions 1 and 2 above; 
condition (b) follows from 6; 
condition (c) is identical to 5; 
condition (d) follows from the explicit form (\ref{8z4}) and the boundary conditions 
of 6, which may also be read as $A,B,K$ are purely imaginary on $\p \Sigma$; 
and condition (e) follows from the explicit form (\ref{8z4}), the relation 
$\Lambda ^A = \lambda ^A \p_w H$,  and condition 3. Condition 4 above is 
equivalent to our earlier assumptions in (\ref{7f1}) and (\ref{7f5}) on the functional form of 
$H$ and $\Lambda ^A$. 

\sm

Finally, the requirements that $\Re(A)$ and $K$ cannot vanish inside $\Sigma$ 
are consequences of the relation (\ref{8z2}) and condition 5. Indeed, at any point where 
$(A+\bar A) K =0$, we have $\l \cdot \bar \l = 2-(B+\bar B)^2/(B\bar B) \leq 2$
which violates condition 5. 

\newpage

\section{Discussion}
\setcounter{equation}{0}
\label{discussion}

In this paper, we have explicitly constructed a general maximal family of  global regular
half-BPS solutions to $(0,4)$ six-dimensional supergravity with $m$ tensor multiplets. 
The space-time of each one of these solutions consists of an 
$AdS_2\times S^2$ fibration over a Riemann surface $\Sigma$, and has $N$
distinct asymptotic $AdS_3\times S^3$ regions. In our construction, we have made the mild 
technical assumption that the one-forms $\Lambda^A$ have no poles on the real axis other than the poles of $H$.
It would be interesting to see whether and how this restriction might be lifted, 
and if solutions with some singularities can still have a physical interpretation.
Beyond this technical question, there are several directions in which the results 
presented here can be generalized and explored further.

\medskip

We have focused on solutions of  a particular six-dimensional supergravity, 
namely the $(n_+,n_-)=(0,4)$ theory. As discussed in Section \ref{sec22} it is 
possible to embed  our solutions into the $(n_+,n_-)=(4,4)$ theory. It is an open 
problem to apply the techniques of the present paper to six-dimensional 
supergravities with less supersymmetry and construct half-BPS string-junction solutions.

\medskip

In future work, we plan to investigate the different degenerations of our family of solutions. 
In particular, a natural limit which should be analyzed in detail is the one where two or more poles of 
$H$ coalesce. This limit should lead to the factorization of the solution into two solutions 
with a lower number of asymptotic regions. In turn, a careful study of the factorization properties 
of the solutions may help relate some solutions to the near-horizon limit of 
dyonic string \emph{networks} in six dimensions. Some of the bubbling solutions of Type IIB and 
M-theory have been related to brane webs in \cite{Lunin:2008tf}. 
A related question is whether some limit of the solutions constructed in this paper can describe 
configurations of intersecting branes similar to the ones discussed in 
\cite{Gaiotto:2008ak,Gaiotto:2008sa,Gaiotto:2008sd}
in a ten-dimensional setting.

\medskip

The  construction of globally regular solutions in Sections \ref{sec7}, \ref{sec8}, 
and \ref{sec9}  assumed that the Riemann surface $\Sigma$  has a single boundary 
component.  This analysis can 
be generalized to Riemann surfaces with two or more boundary components  along 
the lines of \cite{Chiodaroli:2009xh}.  Such solutions should have novel and interesting 
features. First,  additional homology three-spheres  which are not associated 
with poles of the harmonic function $H$ will appear. These cycles  can carry non-trivial anti-symmetric 
tensor flux. Second, in addition to degeneration limits where poles of $H$ coalesce, 
there are new degeneration limits where a boundary shrinks to zero size. 
In \cite{Chiodaroli:2009xh}, this degeneration limit was found to produce new asymptotic
regions of the form $AdS_2 \times S^2 \times S^1 \times {\rm \bf R}$.

\medskip

The holographic interpretation of all Janus-type solutions is given 
by an interface CFT  where different CFTs living on half-spaces are linked by a 
common interface world-line  \cite{Clark:2004sb}. A proposal 
for the realization of a holographic dual to a {\sl boundary} CFT as a limit of the 
Type IIB  Janus solutions  \cite{D'Hoker:2007xy,D'Hoker:2007xz} was recently set 
forth in \cite{Aharony:2011yc,Assel:2011xz}. 
It would be interesting to investigate whether a similar limit is possible for the solutions 
presented in this paper. A holographic realization of  boundary
CFT by introducing a brane which cuts off the AdS bulk space and removes a second AdS half space was originally proposed in \cite{Karch:2000gx} and has recently been discussed in more detail in \cite{Takayanagi:2011zk}  . 

\medskip

In a recent paper \cite{Chiodaroli:2010ur} two of the authors calculated the 
boundary entropy of an interface conformal field theory  using the holographic 
prescription developed in \cite{Ryu:2006bv} and the solutions constructed in \cite{Chiodaroli:2009yw}.
The solutions in the present 
paper are more general, and their parametrization via the harmonic function $H$ and the one-forms $\Lambda^A$ 
is more directly related 
to the physical charges and values of the scalars.
Hence, to calculate the holographic 
boundary entropy of these new solutions along the lines of \cite{Chiodaroli:2010ur} and \cite{Ryu:2006bv}
 seems a natural and promising enterprise. 
Finally, there are other observables, such as boundary correlation functions, which can be 
calculated holographically and compared to conformal field theory  results. 
Such calculations would be extremely valuable in determining the exact map between 
holographic BPS Janus solutions and interface CFTs.

\bigskip

\bigskip

\bigskip

\noindent{\Large \bf Acknowledgements}

\bigskip

The work of ED, MG and YG was supported in part by NSF grant PHY-07-57702. 
The work of MC  was supported in part by NSF grant PHY-08-55356.
MC would like to thank Murat G\"{u}naydin, Radu Roiban, and Nicholas Warner for useful conversations and 
the Kavli Institute for Theoretical Physics for hospitality during
the course of part of this work. MG would like to thank the Centro de ciencias de Benasque Pedro Pascual for hospitality while this paper was finalized.

\newpage

\appendix

\section{Notations and Conventions}
\setcounter{equation}{0}
\label{appA}

In this appendix, we spell out our core notations and conventions for the space-time
and internal degrees of freedom of six-dimensional supergravity.

\subsection{Six-dimensional Minkowski space-time}

Throughout, we shall use the following index notations for six-dimensional space-time,
\bea
\label{A1}
\mu, \nu = 0,\cdots,5 &\qquad& \textrm{coordinate indices}
\no \\
M,N = 0,\cdots,5 &\qquad& \textrm{frame indices}
\no \\
m,n = 0,1 &\qquad& AdS_2  ~\textrm{frame indices}
\no \\
p,q = 2,3 &\qquad& S^2  ~\textrm{frame indices}
\no \\
a,b = 4,5 &\qquad& \Sigma  ~\textrm{frame indices}
\eea
The six-dimensional space-time metric has Minkowski signature. The Minkowski frame 
metric is chosen to be $\eta_6={\rm diag} (-\, +++++)$ and will be
used to raise and lower the frame indices $M,N, \cdots$. The completely anti-symmetric
tensors $\epsilon$ acting on frame indices, in six and in two-dimensions,  are normalized by,
\bea
\label{A2}
\epsilon_{012345} = \epsilon_{45} =1
\eea
Our conventions for differential 
forms are  the standard ones, namely the $p$-form $T^{(p)}$ associated with a totally
anti-symmetric tensor $T^{(p)}_{\mu_1 \cdots \mu_p}$ is normalized by 
\bea
\label{A3}
T^{(p)} = {1 \over p!} 
T^{(p)}_{\mu_1 \cdots \mu_p} dx^{\mu_1} \wedge \cdots \wedge dx^{\mu_p}
\eea
The frame one-forms are defined by $e^M = e^M{}_\mu dx^\mu$.
Throughout, we shall use the notation 
$e^{M_1 \cdots M_p} = e^{M_1} \wedge \cdots \wedge e^{M_p}$
so that the symbol $e^{012345}$ stands for the space-time volume form. 
The $p$-form $T^{(p)}$ may be expressed alternatively in frame indices,
\bea
\label{A4}
T^{(p)} = {1 \over p!} 
T^{(p)}_{M_1 \cdots M_p} e^{M_1 \cdots M_p} 
\eea
The Poincar\'e dual of a $p$-form $T^{(p)}$ may be defined equivalently by the following relations, 
\bea
\label{A5}
(*T^{(p)})^{M_1 \cdots M_{6-p}} 
& = & {1 \over p!} \epsilon^ {M_1 \cdots M_{6-p} N_1 \cdots N_p}  T^{(p)}_{ N_1 \cdots N_p}
\no \\
S^{(p)} \wedge *T^{(p)} & = & 
{(-1)^{p(6-p)} \over p!} S^{(p) \, N_1 \cdots N_p} T^{(p)}_{ N_1 \cdots N_p} \, e^{012345}
\eea
where the second line is to hold for all $p$-forms $S^{(p)}$.

\subsection{$SO(5,m)/(SO(5)\times SO(m))$ coset space notations}

We use the following index notations for the coset space $SO(5,m)/SO(5)\times SO(m)$ variables,
\bea
\label{A6}
A,B= 1,\cdots m+5 &\qquad& SO(5,m)~ \textrm{defining representation indices}
\no \\
&& ~~ A=(I,R) \hskip 0.5in  I,J=1,\cdots, 5
\no \\
&& ~~ B=(J,S)  \hskip 0.5in R,S=6, \cdots , m+5
\no \\
i,j = 1,\cdots,5 &\qquad& SO(5)~ \textrm{defining representation indices}
\no \\
r,s = 1,\cdots,n &\qquad& SO(m)~ \textrm{defining representation indices}
\no \\
\a,\b = 1,\cdots,4 &\qquad& SO(5)~ \textrm{spinor representation indices}
\eea

\subsection{$SO(1,5)$-gamma matrices}

The $\g$-matrices in six dimensions are defined by
$ \{ \g^M, \g^N \} = 2 \eta ^{MN}_6$, and we choose the basis,
\bea
\label{A7}
i\gamma^{0} =  \sigma_{2}\otimes I_2\otimes I_2 
& \hskip 1in & 
\gamma^3 = \sigma_{3}\otimes \sigma_{2}\otimes I_2
\no\\
\gamma^{1} = \sigma_{1}\otimes I_2\otimes I_2 
&&
\gamma^{4} = \sigma_{3}\otimes \sigma_{3}\otimes \sigma_{1}
\no\\
\gamma^{2} = \sigma_{3}\otimes \sigma_{1}\otimes I_2
&&
\gamma^{5} = \sigma_{3}\otimes \sigma_{3}\otimes \sigma_{2}
\eea
Throughout, $SO(1,5)$ spinor indices will be suppressed.
The chirality matrix $\g^7$, defined by, 
\bea
\label{A8}
\gamma^{7} =  \gamma^{012345} = \sigma _3 \otimes \sigma _3 \otimes \sigma _3 
\eea
obeys $(\g^7)^2 = I $, as well as the following relations, 
\bea
\label{A9}
\g^{MNPQRS} & = & - \g^7 \epsilon ^{MNPQRS}
\no \\
\g^{MNP} \g^7 & = & - {1 \over 3!} \epsilon ^{MNPQRS} \g_{QRS}
\eea 
The complex conjugation matrix $\cB$ is defined to obey,
\bea
\label{A10}
(\gamma^M)^*  =   \cB \gamma^M \cB^{-1}
\eea
and is given in the representation (\ref{A7}) by, 
\bea
\label{A11}
\cB  =  I_2 \otimes \sigma _1 \otimes \sigma _2
\eea
The matrix $\cB$ obeys  $\cB \cB^* = -I_8$; it commutes with $\g^7$.  

\sm

Representing the six-dimensional space-time by the warped product $AdS_2 \times S^2 \times \Sigma$,
it is natural to restrict the Dirac matrices to these two-dimensional factors.
Denoting the corresponding reduced matrices by $\tilde \g ^M$, and their 
associated reduced chirality matrices respectively by $\tilde \g _{(1)}$, $\tilde \g _{(2)}$, 
and $\tilde \g _{(3)}$, we have the following induced representations, 
\bea
\label{A12}
 \tilde \gamma^{1} = \tilde \gamma^{2} = \tilde \gamma^{4} &=&  \sigma_{1}
\no \\
i \tilde \gamma^{0} = \tilde \gamma^{3} = \tilde \gamma^{5} &=&  \sigma_{2}
\no \\
\tilde \g_{(1)} = \tilde \g_{(2)} =\tilde \g_{(3)} & = & \sigma _3
\eea
In this representation, the chirality matrix $\g^7$ and the complex conjugation matrix $\cB$ become, 
\bea
\label{A13}
\gamma^{7} & = & \tilde \gamma_{(1)}  \otimes \tilde \gamma_{(2)} \otimes \tilde \gamma_{(3)}  
\no \\
\cB & = &  B_{(1)} \otimes  B_{(2)} \otimes B_{(3)} 
\eea
The complex conjugation matrices in each subspace may be chosen as follows,
\bea
\label{A14}
\left ( \tilde \g^m \right ) ^* =  + B_{(1)} \tilde \g ^m B_{(1)} ^{-1}
& \hskip .5in &   (B_{(1)})^* B_{(1)} = +I_2 \hskip .6in B_{(1)} = I_2
\no \\
\left ( \tilde \g^{p} \right ) ^* =  + B_{(2)} \tilde \g ^{p} B_{(2)} ^{-1}
& &   (B_{(2)})^* B_{(2)} =  +I_2 \hskip .6in B_{(2)} = \s_1
\no \\
\left ( \tilde \g^{a} \right ) ^* =  - B_{(3)} \tilde \g ^{a} B_{(3)} ^{-1}
& &   (B_{(3)})^* B_{(3)} =  -I_2 \hskip .6in B_{(3)} = \s_2
\eea
An alternative representation is obtained by reversing the roles of $B_{(2)}$ and $B_{(3)}$.

\subsection{$SO(5)$-gamma matrices}
\label{sofivedef}

The internal $SO(5)=USp(4)$ structure of six-dimensional supergravity involves
$\Gamma$-matrices, defined by $\{ \G^i, \G^j\}= 2 \delta ^{ij}$, where
$\delta ^{ij}$ in the $SO(5)$ frame metric. We use the following basis,
\bea
\label{A15}
\Gamma^1 = \sigma_{1}\otimes I_2 
& \hskip 1in & 
\Gamma^3 = \sigma_{3}\otimes \sigma_1 
\no\\
\Gamma ^2 = \sigma_{2}\otimes I_2 
&&
\Gamma^4 = \sigma_{3}\otimes \sigma_{2}
\no\\ &&
\Gamma^5 = \sigma_{3}\otimes \sigma_{3}
\eea
The charge conjugation matrix $\cC$ is defined by $(\G^i)^t = (\G^i)^* = \cC \G^i \cC^{-1}$.
 In the basis of (\ref{A15}), the matrix $\cC$ is given by,
\be
\label{A16}
\cC=\sigma_{1}\otimes \sigma_{2}
\ee
The following basis-independent properties of $\cC$ hold,
\bea
\label{A17}
\cC+ \cC^t = \cC\G^i + \left (\cC \G^i \right )^t = \cC \G^{ij} - \left ( \cC \G^{ij} \right )^t =0
\eea
The anti-symmetry property will be crucial for the reduction of the BPS equations.

\subsection{Choice of a spinor basis in $AdS_2 \times S^2\times \Sigma$}

In this section we briefly review the definitions of the Killing spinors on $AdS_2$ and 
$S^2$, an explicit construction of these spinors can be found for example in 
\cite{D'Hoker:2007fq,D'Hoker:2007xy}.
We choose the Killing spinor equation for $AdS_{2}$  which gives a real equation
in the basis of $\g$-matrices that we have chosen in (\ref{A12}).
This Killing spinor equation is given by
\be
\label{A18}
\nabla_m \chi^{\eta}_{(1)} - {\eta\over 2} \tilde \gamma_{m}\chi_{(1)}^{\eta} =0
\ee
Integrability demands that $\eta=\pm 1$ and there are two linearly independent
solutions for each choice of $\eta$.
The Killing spinor equation on $S^2$  is chosen as 
 \be
 \label{A19}
\nabla_p \chi^{\eta}_{(2),a} - {i\eta\over 2} \tilde \gamma_{p}\chi_{(2),a}^{\eta} =0
\ee
where $\eta=\pm 1$ from the integrability condition.   For each choice of $\eta$ there are two linear independent solutions denoted by the subscript $a$. These two solutions  are related by complex conjugation.
A suitable basis of invariant spinors on the full six-dimensional space-time
$AdS_2 \times S^2 \times \Sigma$ is then obtained as follows,
\bea
\label{A20}
\chi _a ^{\eta _1, \eta _2, \eta _3} \equiv \chi _{(1)} ^{\eta _1}
\otimes \chi _{(2)a} ^{\eta _2} \otimes \chi _{(3)} ^{\eta _3}
\hskip 1in a=1,2
\eea
The action of chirality is as follows,
\bea
\label{A21}
\g_{(1)} \chi ^{\eta _1, \eta _2, \eta _3} & = & \chi ^{-\eta_1, \eta _2, \eta _3}
\no \\
\g_{(2)} \chi ^{\eta _1, \eta _2, \eta _3} & = & \chi ^{\eta_1, -\eta _2, \eta _3}
\no \\
\g_{(3)} \chi ^{\eta _1, \eta _2, \eta _3} & = & \eta _3 \chi ^{\eta_1, \eta _2, \eta _3}
\eea
The action of complex conjugation is also well-defined in this basis,
\bea
\label{A22}
\cB^{-1} \left ( \chi ^{\eta _1, \eta _2, \eta _3} _1 \right )^*
& = &
- i \eta _2 \eta _3  \chi ^{\eta _1, -\eta _2, - \eta _3} _2
\no \\
\cB^{-1} \left ( \chi ^{\eta _1, \eta _2, \eta _3} _2 \right )^*
& = &
+ i \eta _2 \eta _3  \chi ^{\eta _1, -\eta _2, - \eta _3} _1
\eea
Finally, we recall the action of the covariant derivatives,
\bea
\label{A23}
\left ( \hat \nabla _m  - \half \eta_1 \tilde\gamma_m \otimes I \otimes I \right )
\chi ^{\eta _1, \eta _2, \eta_3}_a
    & = & 0 \hskip 1in m=0,1
\no \\
\left ( \hat \nabla _{p} - {i \over 2} \eta_2 I  \otimes \tilde\gamma_{p} \otimes I \right )
\chi ^{\eta _1, \eta _2, \eta_3}_a
    & = & 0 \hskip 1in p=2,3
\eea
A general spinor $\ep^\a$ (with $SO(5)$-spinor index $\a$, and 
$SO(1,5)$ index suppressed),
may now be decomposed onto the basis vectors $\chi$ as follows,
\bea
\label{A24}
\ep^\a = \sum _{\eta _1, \eta _2, \eta _3, a} \chi^{\eta _1, \eta _2, \eta _3}_a
\otimes \zeta _{\eta _1 , \eta _2, \eta _3,a}^\a
\eea
It is clear from the complex conjugation relation, however, that
the spinor $\chi_2$ is proportional to $\cB^{-1} (\chi _1)^*$. Thus, we
may equivalently use the following decomposition,
\bea
\label{A25}
\ep^\a = \sum _{\eta _1, \eta _2, \eta _3} \left ( \chi^{\eta _1, \eta _2, \eta _3}_1
\otimes \zeta _{\eta _1 , \eta _2, \eta _3}^\a +
\cB^{-1} \left (\chi^{\eta _1, \eta _2, \eta _3}_1 \right )^*
\otimes \hat \zeta _{\eta _1 , \eta _2, \eta _3}^\a \right )
\eea
Note that the $AdS_2$ spinor is doubly degenerate, so that each solution
of the type given above generates two supersymmetries.

\section{Reducing the BPS equations}
\setcounter{equation}{0}
\label{appB}

In this appendix, we present the details of the reduction  of the BPS equations. 
We first derive some results that will be used later.
Notice that the $\eta_3$ index is virtually the $\s^3$ basis for the space $\Sigma$, namely,
\bea
\label{B1}
\chi^{\eta_1,\eta_2, +} = \chi^{\eta_1,\eta_2} \otimes {1 \choose 0}
\hskip 1in 
\chi^{\eta_1,\eta_2, -} = \chi^{\eta_1,\eta_2} \otimes {0 \choose 1}
\eea
we have the following relation,
\bea
\label{B2}
\gamma^a \chi^{\eta_1 ,\eta_2,\eta_3} = \s^3 \otimes \s^3 \otimes \s^a \chi^{\eta_1 ,\eta_2,\eta_3} = \chi^{-\eta_1 ,-\eta_2,\eta_3'} (\s^a)_{\eta'_3 \eta_3}
\eea
and we can translate the action of $\s^a$ on $\chi$ to the action on $\zeta$ in the following way,
\bea
\label{B3}
\s^3 \otimes \s^3 \otimes \s^a (\chi^{\eta_1 ,\eta_2,\eta_3} \zeta_{\eta_1 ,\eta_2,\eta_3})
& = & \chi^{-\eta_1 ,-\eta_2,\eta_3'} (\s^a)_{\eta'_3 \eta_3} \zeta_{\eta_1 ,\eta_2,\eta_3}
\no \\
& = & \chi^{\eta_1 ,\eta_2,\eta_3} (\tau^{(11a)} \zeta)_{\eta_1 ,\eta_2,\eta_3}
\eea
where we make use of the $\tau$-matrix notation.
Similarly for the charge conjugate basis,
\bea
\label{B4}
\s^3 \otimes \s^3 \otimes \s^a \left (
(\chi^c)^{\eta_1 ,\eta_2,\eta_3} \cC^{\a}_{~\b} \zeta^{\b*}_{\eta_1 ,\eta_2,\eta_3} \right )
& = &  
\cB ~ (\s^3 \otimes \s^3 \otimes \s^{a}\chi^{\eta_1 ,\eta_2,\eta_3})^* \cC^{\a}_{~\b} \zeta^{\b*}_{\eta_1 ,\eta_2,\eta_3}
\no \\
& = & (\chi^c)^{\eta_1 ,\eta_2,\eta_3} \cC^{\a}_{~\b} (\tau^{(11a)} \zeta^\b)^*_{\eta_1 ,\eta_2,\eta_3}
\eea
As argued above in the previous section, the $(\chi^c) ^{\eta _1, \eta _2, \eta_3}$ term in $\ep^\a$ spinor will just give the complex conjugate of
the $\chi ^{\eta _1, \eta _2, \eta_3}$ term equations, hence below we shall focus on the $\chi ^{\eta _1, \eta _2, \eta_3}$ term, and omit its conjugate terms.

\subsection{Reduction of the gravitino BPS equation}

The gravitino equation may be expressed as follows,
\bea
\label{B5}
 d \ep + \omega \ep + \phi \ep + \theta \ep =0 
\eea
where the ingredients are defined as follows, 
\bea
\label{B6}
(\omega \ep)^\alpha & = & {1 \over 4} \omega_{MN} \gamma^{MN} \ep ^\alpha
\no \\ 
(\phi \ep)^\alpha  & = & -{1\over 4} Q^{ij} (\Gamma_{ij})^{\alpha}_{\;\beta}\ep^{\beta}
\no \\ 
(\theta \ep)^\alpha  & = & -{1\over 4} H^{i}_{MNP} e^M \gamma^{NP} \; (\Gamma^{i})^{\alpha}_{\;\; \beta}\ep^{\beta} 
\eea
To calculate the contribution $d \ep + \omega \ep$, we evaluate the connection.
The non-vanishing spin connection components are $\omega^{a} {}_{b}$ and 
\bea
\label{B7}
\omega^m {}_n = \hat \omega^m {}_n
&\qquad&
\omega^m {}_{a} = e^m {D_{a} f_1 \over f_1}
\no\\
\omega^{p} {}_{q} = \hat \omega^{p} {}_{q} ~
&\qquad&
\omega^{p} {}_{a} = e^{p} {D_{a} f_2 \over f_2}
\eea
The hats refer to the canonical connections on $AdS_2$ and $S^2$.
In components along respectively $e^m$, $e^p$, and $e^a$, the  contribution  
$d\ep + \omega \ep$ projects as follows,
\bea
\label{B8}
(m) &\qquad& \hat \nabla_m \ep^\a+ \half {D_{a} f_1 \over f_1} \g_{m} \g^{a} \ep^\a
\no\\
(p) &\qquad& \hat \nabla_{i} \ep^\a + \half {D_{a} f_2 \over f_2} \g_{i} \g^{a} \ep^\a
\no\\
(a) &\qquad& \nabla_{a} \ep^\a
\eea
where the hats on the covariant derivatives indicates that only the connection
along respectively $AdS_2$, and $S^2$ have been  included.
Using the Killing spinor equations (\ref{A18}) and (\ref{A19}) to eliminate the hatted covariant derivatives,
and combining the terms, we obtain, 
\bea
\label{B9}
(m) &\qquad&  \g_{m}
\sum_{\eta_1 ,\eta_2, \eta_3}  \chi^{\eta_1, \eta_2, \eta_3} \otimes
\left[ { 1\over 2 f_1} \tau^{(300)}\zeta^\a_{\eta_1, \eta_2, \eta_3} + {D_{a} f_1 \over 2 f_1} (\tau^{(11a)} \zeta^\a)_{\eta_1, \eta_2, \eta_3} \right ]
\no\\
(p) &\qquad&   \g_{p}
\sum_{\eta_1 ,\eta_2, \eta_3}  \chi^{\eta_1, \eta_2, \eta_3} \otimes
\left[ { i \over 2 f_2} \tau^{(130)} \zeta^\a_{\eta_1, \eta_2, \eta_3} + {D_{a} f_2 \over 2 f_2} (\tau^{(11a)} \zeta^\a)_{\eta_1, \eta_2, \eta_3} \right ]
\eea
where we have pulled a factor of $\g_M$ out front for equation (m) and (p).
To calculate the term $\phi \ep$, we observe that the 
only non-vanishing direction is along $e^a$,
\bea
\label{B10}
(a) &\qquad& -{1\over 4} \sum_{\eta_1 ,\eta_2, \eta_3}  \chi^{\eta_1, \eta_2, \eta_3} \otimes
q_a^{ij} (\Gamma_{ij})^{\alpha}_{\;\beta}\zeta^{\b}_{\eta_1, \eta_2, \eta_3}
\eea
Finally, to calculate the term $\theta \ep$, the equations along various directions are,
\bea
\label{B11}
(m) &\qquad&
\gamma_m \sum_{\eta_1 ,\eta_2, \eta_3} \chi^{\eta_1, \eta_2,\eta_3 } \otimes
\half g^i_a  (\Gamma^{i})^{\alpha}_{\;\; \beta} (\tau^{(01a)}\zeta^{\beta})_{\eta_1, \eta_2,\eta_3}
\\
(p) &\qquad&
\gamma_p \sum_{\eta_1 ,\eta_2,\eta_3} \chi^{\eta_1, \eta_2,\eta_3} \otimes
\left ( - {i\over 2} \right ) h^i_a  (\Gamma^{i})^{\alpha}_{\;\; \beta}(\tau^{(10a)} \zeta^{\beta})_{\eta_1, \eta_2,\eta_3}
\no \\
(a) &\qquad&
\sum_{\eta_1 ,\eta_2, \eta_3} \chi^{\eta_1, \eta_2, \eta_3} \otimes
\left({1\over 2} g^i_a (\Gamma^{i})^{\alpha}_{\;\; \beta}(\tau^{(100)} \zeta^{\beta})_{\eta_1, \eta_2, \eta_3}
- {i\over 2} h^i_a (\Gamma^{i})^{\alpha}_{\;\; \beta}(\tau^{(010)}\zeta^{\beta})_{\eta_1, \eta_2, \eta_3}\right)
\no 
\eea
Combining all the terms, we recover the reduced gravitino BPS equations of (\ref{3c2}),
\bea
\label{B12}
(m) \quad 0 &=&
 { 1\over  f_1} \t^{(300)} \zeta^\a+ {D_{a} f_1 \over  f_1} \t^{(11a)} \zeta^\a
+  g^i_a \t^{(01a)}(\Gamma^{i})^{\alpha}_{\;\; \beta}\zeta^{\beta}
 \\
(p) \quad 0&=&
{ i\over  f_2} \t^{(130)}\zeta^\a + {D_{a} f_2 \over  f_2}  \t^{(11a)} \zeta^\a
- i h^i_a  \t^{(10a)} (\Gamma^{i})^{\alpha}_{\;\; \beta}\zeta^{\beta}
\no \\
(a) \quad 0&=&
(D_a + {i\over 2} \hat \omega_a \tau^{(003)})\zeta^\a
-{1\over 4} q_a^{ij} (\Gamma_{ij})^{\alpha}_{\;\beta}\zeta^{\b}
+{1\over 2} g^i_a \t^{(100)} (\Gamma^{i})^{\alpha}_{\;\; \beta}\zeta^{\beta}
- {i\over 2} h^i_a \t^{(010)} (\Gamma^{i})^{\alpha}_{\;\; \beta}\zeta^{\beta}
\no
\eea

\subsection{Reduction of the $\chi ^{r \a}$ BPS equation}

The $\chi^{r \alpha}$ equation is
\bea
\label{B13}
0={1\over \sqrt{2}} \gamma^M P_M^{ir} (\Gamma^{i})^{\alpha}_{\;\; \beta} \epsilon^{\beta}
 +{1\over 12} \gamma^{MNP}H^{r}_{MNP}\epsilon^{\alpha}
\eea
The first term gives,
\bea
\label{B14}
&& \sum_{\eta_1 ,\eta_2, \eta_3} \chi^{\eta_1, \eta_2, \eta_3} \otimes
{1\over \sqrt{2}} p_{a}^{ir} (\Gamma^{i})^{\alpha}_{\;\; \beta} (\t^{(11a)} \zeta^{\beta})_{\eta_1, \eta_2, \eta_3}
\no \\
\eea
The second term gives,
\bea
\label{B15}
\sum_{\eta_1 ,\eta_2, \eta_3} \chi^{\eta_1, \eta_2, \eta_3} \otimes
\left( -{1\over 2} \tilde g_{a}^{r} (\t^{(01a)} \zeta^{\a})_{\eta_1, \eta_2, \eta_3}
+ {i\over 2} \tilde h_{a}^{r}  (\t^{(10a)} \zeta^{\a})_{\eta_1, \eta_2, \eta_3}
\right)
\eea
Combining all contributions, we recover the reduced spinor BPS equations of (\ref{3c3}),
\bea
\label{B16}
0 ={1\over \sqrt{2}} p_{a}^{ir} \t^{(11a)}(\Gamma^{i})^{\alpha}_{\;\; \beta} \zeta^{\beta}
-{1\over 2} \tilde g_{a}^{r} \t^{(01a)} \zeta^{\a}
+ {i\over 2} \tilde h_{a}^{r} \t^{(10a)} \zeta^{\a}
\eea

\end{document}